\documentclass[12pt]{article}
\usepackage{epsfig}
\usepackage{wrapfig}
\usepackage{amssymb}
\usepackage{latexsym}
\newcommand{\psla}{\mbox{$\not{\! p}$}}

\newcommand{\calst}{\mbox{$\cal S$}}
\newcommand{\cals}{\mbox{${\cal S}$}}
\newcommand{\invcals}{\mbox{${\cal S}^{-1}$}}
\newcommand{\icalst}[1]{\mbox{${\cal S}^{-1}_{#1}$}}
\newcommand{\micals}[1]{\mbox{${\cal S}^{\{#1\}-1}$}}
\newcommand{\mcals}[1]{\mbox{${\cal S}^{\{#1\}}$}}
\newcommand{\caltf}{\mbox{${\cal T}_{[4]}$}}
\newcommand{\mcaltf}[1]{\mbox{${\cal T}_{[4]}^{\{#1\}}$}}

\newcommand{\calt}{\mbox{${\cal T}$}}

\newcommand{\calv}{\mbox{${\cal V}$}}
\newcommand{\calc}{\mbox{${\cal C}$}}
\newcommand{\mcalv}[1]{\mbox{${\cal V}^{\{#1\}}$}}
\newcommand{\mcalc}[1]{\mbox{${\cal C}^{\{#1\}}$}}

\newcommand{\caly}{\mbox{${\cal Y}$}}

\newcommand{\dbar}{\mbox{$\bar{\delta}$}}
\newcommand{\diffz}{\mbox{$d \vec{Z}$}}
\newcommand{\diffk}{\mbox{$d \bar{k}$}}
\newcommand{\diffl}{\mbox{$d \bar{l}$}}

\newcommand{\be}{\begin{equation}}
\newcommand{\ee}{\end{equation}\noindent}
\newcommand{\bea}{\begin{eqnarray}}
\newcommand{\eea}{\end{eqnarray}\noindent}

\newcommand{\nn}{\nonumber}
\def\eps{\epsilon}
\def\tr{{\rm tr}\,}
\def\Eins{\mathord{1\hskip -1.5pt
\vrule width .5pt height 7.75pt depth -.2pt \hskip -1.2pt
\vrule width 2.5pt height .3pt depth -.05pt \hskip 1.5pt}}
\def\half{{1\over 2}}
\def\mn{\mu\nu}

\begin{document}

\setlength{\unitlength}{1mm}

\begin{titlepage}

\begin{flushright}
LAPTH-1097/05\\
WUE-ITP-2005-003\\
ZU-TH-08/05\\
April 2005
\end{flushright}
\vspace{1.cm}

\begin{center}
{\LARGE\bf An algebraic/numerical formalism for one-loop multi-leg amplitudes}\\[1cm]

{T.~Binoth$^{a}$, J.-Ph.~Guillet$^{b}$, G.~Heinrich$^{c}$, 
E.~Pilon$^{b}$, C.~Schubert$^{d}$}\\[1cm]%

{\em $^{a}$Institut f\"ur Theoretische Physik, 
Universit\"at W\"urzburg,\\ Am Hubland, 
D-97074 W\"urzburg, Germany}\\[.5cm]%

{\em $^{b}$Laboratoire d'Annecy-le-Vieux de Physique Th\'eorique\\
LAPTH, B.P. 110, F-74941 Annecy-le-Vieux Cedex, France}\\[.5cm]
    
{\em $^{c}$Institut f\"ur Theoretische Physik, 
Universit\"at Z\"urich,\\ Winterthurerstrasse 190, 
8057 Z\"urich, Switzerland}\\[.5cm]

{\em $^{d}$Department of Physics and Geology,\\
University of Texas Pan American,
1201 West University Drive,\\
Edinburg, Texas 78541, USA}\\[.5cm]
      
\end{center}
\normalsize

\vspace{7mm}

\begin{abstract}
We present a formalism for the calculation of multi-particle 
one-loop amplitudes, valid for an arbitrary number $N$ of 
external legs, and for massive as well as massless particles. 
A new method for the tensor reduction is suggested
which naturally isolates infrared divergences by construction.
We prove that for $N\geq 5$, higher dimensional integrals 
can be avoided. We derive many useful relations which allow for
algebraic simplifications of one-loop amplitudes. 
We introduce a form factor representation of tensor integrals
which contains no inverse Gram determinants by choosing 
a convenient set of basis integrals.
For the evaluation of these basis integrals we propose two
methods: An evaluation based on the analytical representation, 
which is fast and accurate away
from exceptional kinematical configurations, 
and a robust numerical one, 
based on multi-dimensional contour deformation.  
The formalism can be implemented straightforwardly 
into a computer program to calculate next-to-leading order 
corrections to multi-particle processes in a largely automated way.
\end{abstract}

\end{titlepage}

\tableofcontents

\newpage

\section{Introduction}\label{s1}

The quest for new physics at TeV colliders like the Fermilab Tevatron,
the upcoming Large Hadron Collider at CERN, as well as the
International Linear Collider (ILC) project, requires the quantitative calculation
of many hard multi-particle processes. Direct searches rely on the
proper identification of multi-particle signals and a precise understanding
and determination of the corresponding multi-particle/multi-jet backgrounds. The
latter, in particular those for multi-jet or vector boson(s) plus multi-jet 
production, are considerable, as can be estimated from leading order 
studies~\cite{lhcrep,Draggiotis:2002hm}. 
Indirect searches at high luminosity machines
will also involve high precision measurements of multi-particle observables of
the Standard Model, comprising six-point processes like $e^+e^-\to 4$ fermions
at the ILC, so that accurate predictions for these cross sections will be
mandatory. 
High energy physics is thus entering a new era where the
quantitative description of hard multi-particle production is becoming a topic
of prominent importance, whereas the lowest order estimates for such processes
are plagued by the well-known deficiencies of large renormalisation and
factorisation scale dependences, poor multi-jet modelling and large sensibility
to kinematic cuts. Therefore the calculation of next-to-leading-order (NLO)
corrections to such cross sections is a necessary step forward. However, this task
involves an enormous technical complexity.

\vspace{0.3cm}

To perform an NLO calculation with $N$ hard external particles, the following
points have to be addressed:
\begin{enumerate}
\item Generation of tree amplitudes with $N+1$ external particles
\item Subtraction of soft and collinear singularities 
      due to single unresolved real radiation
\item Generation of the one-loop amplitude with $N$ external particles
\item Evaluation of the loop diagrams, UV renormalisation, extraction of
      soft and collinear singularities 
\item Combination of the contributions above, cancellation of soft
      singularities, cancellation of collinear singularities or absorption 
      into distribution functions
\item Numerical evaluation of the finite amplitude 
\end{enumerate}

For step 1, efficient, highly automated tools and algorithms are available.
A similar statement can be made
for step 2. Although automatisation is less trivial for this point, the
available algorithms 
\cite{Giele:1991vf,Frixione:1995ms,Nagy:1996bz,Catani:1996vz} are well tested 
and in principle valid for an arbitrary number of legs.   
The same is not true for loop amplitudes. Although efficient programs like
FeynArts\,\cite{Hahn:2000kx},  GRACE\,\cite{Belanger:2003sd,Kaneko:1994fd} or 
QGRAF\,\cite{Nogueira:1991ex} exist which reliably deal with the combinatorial complexity of
generating multi-leg one-loop Feynman diagrams, the evaluation of these 
diagrams, in particular in the presence of infrared divergences and for more
than four external legs, is still far from being automated. 
So the bottleneck in constructing an automated program package for NLO cross 
sections is step 3, the evaluation of the loop graphs. Although a number of 
five-parton processes, see for 
example\,\cite{Bern:1993mq}--\cite{Binoth:2003xk}, 
have been calculated already, and very recently 
even physical $2\to 4$ results have 
become available\,\cite{Boudjema:2004id,Denner:2005es},
these calculations all required a tedious individual treatment, and most of 
them do not allow to obtain fully differential results. Therefore it is 
desirable to have tools which allow the calculation of NLO cross sections in a 
largely automated way. These tools should be able to handle massless as well as 
massive particles, and should be numerically reliable and fast.
Ideally, they should also allow to be interfaced with a parton shower in a 
universal way, using for example the formalism 
proposed in\,\cite{Frixione:2005gz} or \cite{Nagy:2005aa}.

\vspace{0.3cm}

Several approaches to this aim have been suggested in the literature so far, 
from purely numerical ones to ones where the emphasis is on algebraic 
manipulations.
A completely numerical approach has been worked out by  D.~Soper et
al.\,\cite{Soper:1999xk,Kramer:2002cd}, where the sum over cuts for a given
graph is performed  before the numerical integration over the loop momenta. 
In this way  unitarity is exploited to cancel soft and collinear divergences
before they show up as explicit poles. 

\vspace{0.3cm}

However, the conventional method of calculating the virtual (loop) and real
(radiation) parts separately, thus generating infrared poles which cancel in
the sum, is still the most widely and successfully used approach so far. Of
course, within this approach, there are still many different ways to proceed,
in particular in what concerns the evaluation of the one-loop amplitude.
The most straightforward procedure -- and historically the first one -- relies on
the use of recursion relations to reduce the tensor integrals occurring in the
one-loop amplitude to a set of known basis 
integrals\,\cite{melrose}--\cite{vanHameren:2005ed}.
In the recent work on this subject, the emphasis is primarily 
on methods which are suitable for an efficient numerical evaluation 
of multi-leg amplitudes.
For the massless case, a formalism has been proposed recently in
\cite{Giele:2004iy}, which produces spurious inverse Gram determinants,  
but in \cite{Giele:2004ub} a method is proposed how to 
deal with them. 
The formalism given in \cite{Denner:2002ii} avoids 
inverse Gram determinants in the reduction of  pentagon
integrals, but deals with massive particles only.
In \cite{delAguila:2004nf}, another algorithm is presented, 
using spinor helicity methods. 
Based on the formalism of~\cite{delAguila:2004nf},  an evaluation of
one-loop integrals in massless gauge theories for up to 12 
external legs has been given recently in~\cite{vanHameren:2005ed}.

\vspace{0.3cm}

A numerical approach to the one-loop integrals 
is the one of~\cite{passarino}, where
various concepts like the Bernstein-Tkachov theorem, Mellin-Barnes
representation and sector decomposition are combined to get a 
stable numerical
behaviour in all regions of configuration space.
A fully numerical approach to
the calculation of loop integrals by contour integration
also has been elaborated in\,\cite{Kurihara:2005ja}.
A semi-numerical approach, where
a subtraction formalism for the UV and soft/collinear divergences 
of the one-loop graphs has been worked out, is presented in\,\cite{Nagy:2003qn}. 
The idea is to integrate the remaining
finite part in loop momentum space without performing any tensor 
reduction. 
Another semi-numerical approach is the one described 
in\,\cite{nikolas}. It relies on  the fact that {\it every} one-loop amplitude
can be represented in terms of building blocks which are one- and two-dimensional 
parameter integrals in a form which is suitable for numerical integration.

\vspace{0.3cm}
 
An alternative method is to obtain loop amplitudes by using unitarity to sew
together tree amplitudes\,\cite{Neq4Oneloop,Neq1Oneloop,UnitarityMachinery}. A
difficulty of this approach has been to determine  ambiguities
of rational functions which are present when calculating QCD amplitudes. However, the application of
twistor-space inspired methods to one-loop amplitudes
\cite{Cachazo:2004zb} led
to new insights in this context~\cite{BBKR}--\cite{Badger:2005zh}, 
and a rapid development in this direction may be expected in the future.
 
\vspace{0.3cm}

Nevertheless, to calculate one-loop amplitudes involving massive particles, as
well as to handle infrared divergences due to massless particles, we still have
to rely on more conventional methods. As the size of the expressions for such
amplitudes increases factorially with the number of legs, efficient methods of
tensor reduction become more and more important.
Although many in principle viable approaches exist, computations which
rely on conventional reduction methods may get stuck due to the combinatorial growth of
intermediate expressions for moderate values of $N\,(N \sim 6)$ already.
It is the sheer size of the expressions, together with spurious denominators,
which in the end hampers a successful, i.e. numerically stable, evaluation of
the amplitude. Reduction algorithms in momentum space generically lead to
so-called inverse Gram determinants which vanish if an exceptional kinematical
configuration is approached. Reduction algorithms in Feynman parameter space
in principle overcome this problem, but other kinematical
determinants are still present in the denominator and one has to deal with
scalar integrals in higher 
dimensions\,\cite{tarasov,davydychev,Bern:1992em,Bern:1993kr,Giele:2004iy}. 

\vspace{0.3cm}

In this article, we propose an algorithm which is similar to the one given
in\,\cite{tenred1}, but improved in several respects. First, the new algorithm
is designed to restrict and control the occurrence of inverse Gram
determinants. Second, the formalism is valid for massive as well as massless
particles, the soft and collinear divergences being regulated by dimensional
regularisation. 
Our method is valid
for arbitrary $N$, and we give a constructive recipe how to deal with the 
cases where  kinematic matrices are not invertible. In addition, we prove 
explicitly in this formalism how $N$-point integrals with $N\geq 5$ in more
than $n=4-2\eps$ dimensions drop out of any physical one-loop amplitude. 
Moreover, we elaborate on the numerical evaluation of the basis integrals. 
Further, the
new method is formulated in a manifestly shift invariant way  
and thus avoids a proliferation of terms due to shifts of the loop momentum 
when the tensor reduction is applied iteratively.

\vspace{0.3cm} 

The outline of the paper is as follows. In section \ref{overview}, we
present a non-technical overview of our approach, which serves to 
point out its main features. 
The following sections contain a detailed description of the formalism. In
section \ref{secnotation}, we define our notation and the general setup. The
method of tensor reduction by subtraction is described in section \ref{method}. 
In section
\ref{algebraic-red} we elaborate the algebraic evaluation of the building
blocks of our reduction. 
The case $N=5$ is particularly interesting, and in section \ref{formfact_fivep}
we give form factors for $N=5$ which do not contain higher 
dimensional 5-point functions {\em and} are free from inverse Gram determinants.
The explicit proof that these integrals drop out and how the inverse Gram 
determinants cancel for $N=5$ is rather technical and is provided in appendix 
\ref{absence-N-eqn5}.
In section \ref{basicblocks} we deal with the numerical evaluation of the basis
integrals, by means of multi-dimensional contour deformation, and we present
explicit checks of the numerical stability near exceptional kinematical situations. 
Section \ref{practitioner} contains guidelines for the practitioner 
who is less interested in the mathematical details on 
how to implement the formalism directly into a computer code, 
before we conclude in section \ref{concl}. 
In the appendices, we provide explicit formulae 
and useful relations  for 
the direct application of our algorithm to multi-leg calculations.

\section{A brief overview of the method}\label{overview}

Before entering into the mathematical details of our formalism we
would like to give a short overview of the method. 
 
\vspace{0.3cm}

\noindent
We consider one-loop $N$-point diagrams with external momenta 
$p_1,\dots,p_N$.   They are typically expressed
in terms of integrals in momentum space, with and without
loop momenta in the numerator.
Any algebraic approach to evaluate these diagrams  starts by reducing 
these tensor and scalar integrals to
simpler objects, with the price to pay that the number of terms increases
at each reduction step. We will distinguish reduction formulae for 
{\em tensor} and {\em scalar} integrals in the following.   

\vspace{0.3cm}

\noindent
Before applying tensor reduction formulae 
which typically increase the complexity 
of an expression, {\em reducible} terms might be cancelled.
Numerators of Feynman integrals which contain scalar products 
between loop momenta and external momenta  
are called reducible if they can be expressed by differences of inverse
propagators of the given Feynman diagram and by kinematical invariants.
The momentum representations of the remaining  
irreducible tensor integrals are converted 
to linear combinations of form factors and Lorentz structures. 
In our approach this is done in a {\em non-standard} way, 
as we express the numerators of the 
tensor integrals in terms of  propagator momenta instead of the 
loop momentum solely. Thus we define a generalised rank $r$ tensor integral by
\begin{eqnarray}
I^{n,\,\mu_1\ldots\mu_r}_N(a_1,\ldots,a_r) = 
\int \frac{d^n k}{i \, \pi^{n/2}}
\; \frac{q_{a_1}^{\mu_1}\,\dots  q_{a_r}^{\mu_r}}{
(q_1^2-m_1^2+i\delta)\dots (q_N^2-m_N^2+i\delta)}
\label{eq0}
\end{eqnarray} 
where $q_a=k+r_a$, and $r_a$ is a combination of external momenta. 
The method is defined in $n=4-2\epsilon$ dimensions and thus is
applicable to general scattering processes with arbitrary 
propagator masses.
Taking integrals of the form (\ref{eq0}) 
as building blocks has two advantages: 1) combinations of
loop and external momenta appear naturally  in Feynman rules,   
2) it allows for a formulation of the tensor reduction which 
manifestly maintains 
the invariance of the integral under a shift $k \to k+r_0$  in the loop momentum. 
Such a shift can be
absorbed into a redefinition of the $r_j, \,r_j \to r_j - r_0$. 
The Lorentz structure of the integral (\ref{eq0}) is carried by
tensor products of the  metric $g^{\mu\nu}$ and the difference vectors $\Delta_{ij}^\mu=
r_i^\mu - r_j^\mu$, which are invariant under  such a  shift. 
The fact that the sums $q_a=k+r_a$, which generically appear in loop diagrams, 
are not split
into loop and external momenta, as well as the explicit shift invariance,
have the virtue of leading to
a reduced number of terms in the expressions for the loop graphs.

\vspace{0.3cm}

\noindent
A key point of our method is to reduce these tensor integrals 
by adding and subtracting terms such that a reduction into 
infrared (IR) {\em finite} and {\em reduced} (or pinched) integrals is achieved.
By applying the reduction again to the potentially IR divergent 
reduced terms one generates iteratively a separation into IR finite and 
IR divergent terms. The latter are IR divergent 3-point functions
which can be evaluated analytically in a closed form and separated
from the finite part of the amplitude.
This immediately provides a starting point for
a numerical evaluation. 

\vspace{0.3cm}

\noindent
We further show that our reduction
of $N$-point tensor integrals for $N\geq 6$ trivially maps to
5-point tensor integrals. The 5-point case needs special care, as
was noted earlier\,\cite{Bern:1993kr}. 
We will show that all tensor 5-point functions can 
be reduced to some basis integrals  without generating higher dimensional 5-point 
functions nor  inverse Gram determinants.   
The form factors for all nontrivial tensor structures for up to rank five 5-point functions
will be provided explicitly in terms of our basis integrals. 

\vspace{0.3cm}

\noindent
These basis integrals, i.e. the endpoints of our reduction, 
are  4-point functions in 6 dimensions
$I_4^6$, which are IR and UV finite, UV divergent 4-point functions in
$n+4$ dimensions, and various 3-point functions, some of
them with Feynman parameters in the numerator. This provides us with a very
convenient  separation of IR/UV divergences, as the IR poles are 
exclusively contained in
the triangle functions. Explicitly, our reduction 
basis is given by integrals of the type 
\begin{eqnarray}\label{basic_integral}
I^{n}_3(j_1, \ldots ,j_r) &=& 
-\Gamma \left(3-\frac{n}{2} \right) \, \int_{0}^{1} 
\prod_{i=1}^{3} \, d z_i \, \delta(1-\sum_{l=1}^{3} z_l) 
\, \frac{z_{j_1} \ldots z_{j_r}}{ 
(-\frac{1}{2}\, z \cdot \calst
\cdot z-i\delta)^{3-n/2}}\;,\nn\\
I^{n+2}_3(j_1) &=& 
-\Gamma \left(2-\frac{n}{2} \right) \, \int_{0}^{1} 
\prod_{i=1}^{3} \, d z_i \, \delta(1-\sum_{l=1}^{3} z_l) 
\, \frac{z_{j_1}}{ 
(-\frac{1}{2}\, z \cdot \calst
\cdot z-i\delta)^{2-n/2}}\nn\\
I^{n+2}_4(j_1, \ldots ,j_r) &=& 
\Gamma \left(3-\frac{n}{2} \right) \, \int_{0}^{1} 
\prod_{i=1}^{4} \, d z_i \, \delta(1-\sum_{l=1}^{4} z_l) 
\, \frac{z_{j_1} \ldots z_{j_r}}{ 
(-\frac{1}{2}\, z \cdot \calst
\cdot z-i\delta)^{3-n/2}}\;,\nn\\
I^{n+4}_4(j_1) &=& 
\Gamma \left(2-\frac{n}{2} \right) \, \int_{0}^{1} 
\prod_{i=1}^{4} \, d z_i \, \delta(1-\sum_{l=1}^{4} z_l) 
\, \frac{z_{j_1}}{ 
(-\frac{1}{2}\, z \cdot \calst
\cdot z-i\delta)^{2-n/2}}\;,
\end{eqnarray}
and $I^{n}_3,I^{n+2}_3,I^{n+2}_4,I^{n+4}_4$ with no Feynman parameters in the numerator.
Of course, 2-point functions also have to be considered. 

\vspace{0.3cm}

\noindent
It turns out that for an arbitrary $N$-point amplitude, calculated in a
gauge\footnote{Here we have in mind a gauge fixing for which gauge boson 
propagators are
proportional to $g_{\mu \nu}$. In more general covariant gauges where
the tensor structure of gauge boson propagators depends on the vector boson
momentum, the extra gauge dependent terms might lead to higher 
dimensional integrals with non-trivial numerators. However, using the pinch 
technique \cite{watson}, it can be shown that a reorganisation of the 
integrands allows to cancel the gauge dependent contributions at the level of
the integrands before the loop momentum integration is performed, so that this
complication can be avoided in these cases, too.}
where the rank can only be less or equal to the number of external legs, it
is sufficient to consider $r\leq 3$ in the integrals above. 
Note that the $(n+2)$-dimensional box integrals, being neither IR nor
UV divergent, can be evaluated using $n=4$. The IR divergent three-point
integrals are easy to handle, and we will give a complete list of all required
3-point integrals for the case of massless propagators in appendix \ref{3pointIR}. 

\vspace{0.3cm}

\noindent
Our reduction   formalism is designed such that 
any $N$-point amplitude can be written as a linear
combination of the basis integrals without encountering inverse Gram
determinants.  
We would like to emphasize that we 
do not only avoid Gram determinants of rank {\it four} matrices
in the reduction of five-point functions,
but obtain form factors which do not have inverse Gram 
determinants  from {\it lower rank} matrices either. Our form factors are 
completely free from {\it any} inverse Gram determinants.
In addition, we avoid the proliferation of higher dimensional integrals
by choosing a convenient set of basic functions, 
while controlling at the same time the occurrence of arbitrary 
inverse Gram determinants. 
These two restrictions define our form factor representation
and lead to the function basis we use. To our best knowledge 
no explicit form factor representations with these special
properties have been derived in the literature before.
To evaluate the basis functions, 
we propose here two complementary approaches: a purely numerical one 
and an algebraic one. 

\vspace{0.3cm}

\noindent
In the algebraic approach 
our basic buildings blocks, given in eq.\,(\ref{basic_integral}), 
are further reduced to scalar integrals using recursion formulae.
However, this  introduces
scalar integrals of dimensions higher than $n+4$ 
\cite{tarasov,davydychev,Giele:2004iy}. 
Applying scalar reduction formulae,  
the latter can be remapped to $(n+2)$-dimensional scalar box integrals and 
$n$-dimensional scalar two- and three-point functions. Doing so, the price to 
pay is the occurrence of inverse Gram determinants. 
These dangerous denominators are spurious: the expressions can be organised in such a way
that inverse Gram determinants are multiplied by linear combinations of scalar
integrals which also vanish in the case of exceptional kinematics, 
as has been done for example in\,\cite{Campbell:1997tv}. 
Without such a grouping of terms, cancellations between singular pieces  
in general pose numerical problems, and
even respecting such a grouping of terms does not guarantee
numerical stability.

\vspace{0.3cm}

\noindent
If one aims for compact algebraic expressions of
loop amplitudes, it is a good strategy to
express the amplitude in terms of simple  scalar integrals 
and to enforce compensations of Gram determinants algebraically.
Experience shows that compact expressions can be achieved 
in this way (for examples where algebraic reduction was 
successfully applied, see \cite{Binoth:2003xk,Binoth:2002qh,Binoth:2005ua}).

\vspace{0.3cm}

\noindent
In order to simplify the  amplitude 
representation in terms of finite basis
functions and to enforce explicit cancellations of Gram determinants, 
it can be useful to exploit
relations between determinants and sub-determinants (minors)
of the kinematic matrix $\cal S$ present in eq.~(\ref{basic_integral}). 
We provide  a collection of 
relations useful for this purpose in appendix \ref{relations}.
Many of these relations also served to obtain compact and convenient 
form factors for the tensor integrals.

\vspace{0.3cm}

\noindent
If the amplitude is too complex,  the purely algebraic treatment
becomes intractable and it is advantageous to avoid the introduction
of Gram determinants from the start.
For this case we  give a prescription how to evaluate the integrals
in eq.~(\ref{basic_integral}) without further reduction.
The IR divergent integrals, being only 
3-point integrals in our approach, can easily be handled analytically,
as they are simple enough to allow for explicit 
representations of all possible cases. 
For the finite 3-point and the 4-point functions we 
propose a new numerical method which allows for a
direct numerical evaluation. By analytic continuation
in Feynman parameter space 
and an adequate multidimensional contour deformation we find 
a numerically stable
integral representation of the basis functions.
The method is described in detail in section \ref{numeric}.
In section \ref{compare}, we compare the two approaches of either using 
eqs.\,(\ref{basic_integral}) as endpoints of the reduction and then proceed 
numerically, or reducing algebraically until  only scalar integrals are reached.
The comparison shows that
near exceptional momentum configurations the purely numerical evaluation is
stable, whereas the analytic implementation of the basis functions,
containing inverse Gram determinants, is not. This is true although 
the terms were grouped such that the coefficients of  inverse Gram determinants 
also vanish in the limit of exceptional kinematics.
We are lead to the conclusion that 
if the compensation of inverse Gram determinants is not possible algebraically,
a tensor reduction scheme which avoids them from the start is preferable.

\vspace{0.3cm}

\noindent
Thus we propose a method where the  form factors are such that 
their numerical evaluation in the kinematically dangerous phase
space regions poses no problem. 
Away from exceptional phase space regions, 
analytical representations can be used safely.
These two complementary approaches should guarantee
a successful evaluation of very complex multi-leg processes.

\section{Definitions and notation}\label{secnotation}

\subsection{Feynman parameter representations}

\begin{figure}[ht]
\unitlength=1mm%
\begin{picture}(150,60)
\put(55,5){\includegraphics[width=5cm, height=5cm]{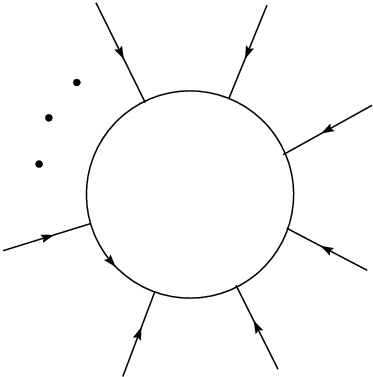}}
\put(50,15){$p_{N-2}$}
\put(60,5){$p_{N-1}$}
\put(85,5){$p_{N}$}
\put(100,15){$p_{1}$}
\put(102,34){$p_{2}$}
\put(92,50){$p_{3}$}
\put(70, 54){$p_{4}$}
\put(87,21){\footnotesize $N$}
\put(90,28){\footnotesize $1$}
\put(87,35){\footnotesize $2$}
\put(79,38){\footnotesize $3$}
\end{picture}
\caption{General $N$-point one-loop graph with momentum and propagator labelling.}
\label{fig1}
\end{figure}
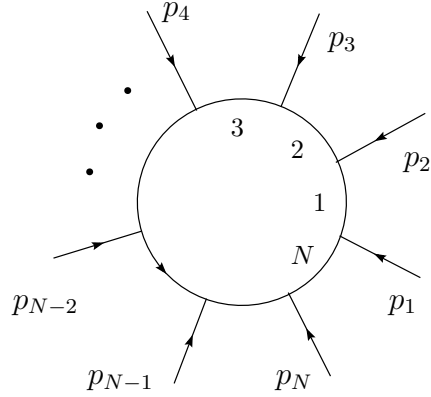

We consider general one-loop $N$-point graphs as the one shown in 
Fig. \ref{fig1}. 
All external momenta $p_i$ are defined as incoming. Momentum
conservation implies
\begin{equation}
\sum_{i=1}^N  p_i = 0 \;.
\label{eqEMC}
\end{equation}
For future reference we label each propagator $q_i$ by the number 
$i$  as shown in 
Fig.~\ref{fig1}. The {\em ordered} set containing the propagator labels
is denoted by $S$. In Fig. \ref{fig1} one has $S=\{1,2,\dots,N\}$.

\vspace{0.3cm}

\noindent
The propagator or internal momenta are labelled accordingly by $q_i = k + r_i$,  
where $k$ is the momentum running in the loop, and  the momenta $r_i$ are defined such that
$p_{i}=r_{i}-r_{i-1}$,  $(i=1,\ldots,N), \;,r_0=r_N$.  Thus one has
$q_{i}=p_{i}+q_{i-1}\; (q_0=q_N)$. By momentum conservation, one can choose one
of the vectors  $r_i$ 
to be zero. Most reduction algorithms specify either $r_N$ or $r_1$ to be zero.

\vspace{0.3cm}

\noindent
The momentum representation of the  scalar $N$-point integral 
in $n$ dimensions is denoted by 
\bea
I^n_N(S) &=& \int \frac{d^n k}{i \pi^{n/2}} \, 
\frac{1}{\prod_{i=1}^N (q_i^2-m_i^2+i\delta)} \;.
\label{isca1}
\eea
In the following, we will use the shorthand notation 
$d\bar k=d^n k/i \pi^{n/2}$ for the integration measure.
The  ordered set $S$ appearing here as an argument uniquely defines the 
one-loop integral. We will use the set $S$ as a basic 
object throughout the paper.

\vspace{0.3cm}

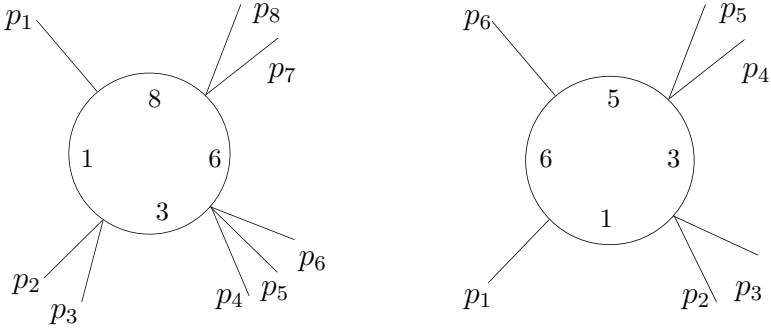
\begin{figure}[ht]
\unitlength=1mm
\begin{picture}(130,50)
\put(25,3){\epsfig{file=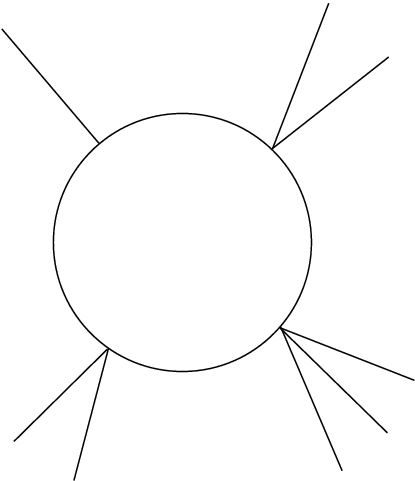,height=4.0cm}}
\put(21,40){$p_1$}
\put(22, 5){$p_2$}
\put(27, 1){$p_3$}
\put(49, 3){$p_4$}
\put(55, 4){$p_5$}
\put(60,8){$p_6$}
\put(56,33){$p_7$}
\put(54,41){$p_8$}
\put(41,14){\footnotesize 3}
\put(48,21){\footnotesize 6}
\put(40,29){\footnotesize 8}
\put(31,21){\footnotesize 1}
\put(85,3){\epsfig{file=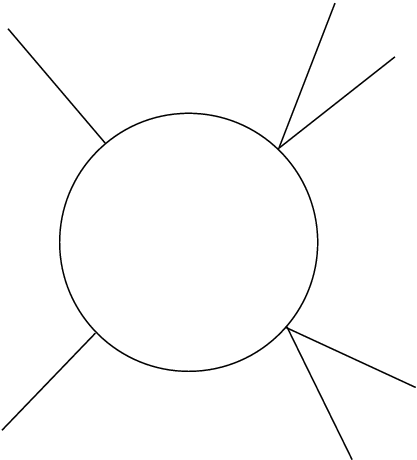,height=4.0cm}}
\put(82, 3){$p_1$}
\put(111, 3){$p_2$}
\put(118, 4){$p_3$}
\put(119,33){$p_4$}
\put(116,41){$p_5$}
\put(82,40){$p_6$}
\put(100,13){\footnotesize 1}
\put(109,21){\footnotesize 3}
\put(101,29){\footnotesize 5}
\put(92,21){\footnotesize 6}
\end{picture}
\caption{Graphical representation of pinch integrals. 
Each topology defines an ordered set $S$. The two diagrams
correspond to 
$N=8$, $M=4$, $S=\{1,3,6,8\}$ (left), and 
$N=6$, $M=4$, $S=\{1,3,5,6\}$ (right).}
\label{fig2}
\end{figure}

\noindent
After having introduced  Feynman parameters and performed the momentum integration, 
$I^n_N(S)$ can be written as
\bea
I^n_N(S) &=& (-1)^N\Gamma(N-\frac{n}{2})\int \prod_{i=1}^N dz_i\,
\delta(1-\sum_{l=1}^N z_l)\,\left(R^2\right)^{\frac{n}{2}-N}\nn\\
&& R^2 = -\frac{1}{2}\,z\cdot\calst\cdot z-i\delta= 
-\frac{1}{2} \sum\limits_{i,j=1}^N z_i\,\calst_{ij}  z_j\,\,-i\delta
\;.
\label{isca2}
\eea
The kinematic matrix  $\calst$ is  defined by
\begin{equation}
\calst_{ij} =  (r_i-r_j)^2-m_i^2-m_j^2\;.
\label{eqDEFS}
\end{equation}
In general, a one-loop $N$-point amplitude will contain  
$N$-point integrals as well as $(N-1),(N-2),\ldots, (N-M)$-point integrals
with tree graphs attached to some of the external legs of the loop integral. 
The latter are characterised by the omission of some propagators 
(say $j_1,\ldots,j_m$) of the ``maximal" one loop $N$-point graph. 
They consist of $N$ external particles and $M < N$ internal lines, where $M$ 
denotes the number of elements in the set
$S\setminus\{j_1,\ldots,j_m\}$. We give some examples in Fig. \ref{fig2}. 
The corresponding kinematic  matrix is denoted by
\bea
{\cal S}^{\{j_1 \cdots j_m\}}\equiv {\cal S}(S\setminus\{j_1, \dots ,j_m\}) \,\, .
\eea
It is obtained from $\calst$ 
by replacing the entries of the rows and columns $j_1,\dots, j_m$ by zero. 
In this way one can keep track of
the pinching of propagators in the iterative application of reduction formulae
without  changing the labels of the rows and columns
of reduced matrices ${\cal S}^{\{j_1 \cdots j_m\}}$ with respect
to the maximal set $S$. 

\vspace{0.3cm}

\noindent
For example, for $S=\{1,2,3,4\}$ one has, with $s_j = p_j^2$ and $s_{ij}=(p_i+p_j)^2$:
\bea
{\cal S} &=& \left( \begin{array}{cccc} 
-2\,m_1^2 & s_2-m_1^2 -m_2^2 & s_{23}-m_1^2 -m_3^2 &  s_{1}-m_1^2 -m_4^2 \\
                                       s_2-m_1^2 -m_2^2 & -2\,m_2^2 & s_3-m_2^2 -m_3^2 & s_{12}-m_2^2 -m_4^2\\
				       s_{23}-m_1^2 -m_3^2 & s_3-m_2^2 -m_3^2 & -2\,m_3^2 &  s_4-m_3^2 -m_4^2\\
				       s_{1}-m_1^2 -m_4^2 & s_{12}-m_2^2 -m_4^2 &  s_4-m_3^2 -m_4^2& -2\,m_4^2
				       \end{array}\right)
\eea
The symmetric $(4\times 4)$ matrix ${\cal S}^{\{2,4\}}$, which corresponds to the pinching of propagators 2 and 4, 
is now defined by
\bea
{\cal S}^{\{2,4\}} &=&\left( \begin{array}{cccc} -2\,m_1^2 &  0& s_{23}-m_1^2 -m_3^2 & 0 \\
                                                0 & 0 & 0& 0 \\
                                               s_{23}-m_1^2 -m_3^2 &  0&  -2\,m_3^2 &  0\\
					       0 & 0 & 0& 0\end{array}\right)
\eea
Inverse matrices are labelled analogously. 
Although ${\cal S}^{\{j_1 \cdots j_m\}}$ in our definition
is not a regular matrix it has a uniquely defined pseudo-inverse.
We recall that the so-called Moore-Penrose generalised inverse ${\cal P}$ to 
a symmetric matrix ${\cal S}$ is uniquely defined by the 
properties \cite{barnett,swedesbook}
\bea\label{DEFpseudo_inverse}
{\cal P}{\cal S}{\cal P} = {\cal P}\,,\, {\cal S}{\cal P}{\cal S} = {\cal S}\,,\, 
{\cal P}{\cal S} = {\cal S}{\cal P}.
\eea
This concept will also be used below.
To construct the pseudo inverse here one simply has to invert the sub-matrix
of  ${\cal S}^{\{j_1 \cdots j_m\}}$ with the zero rows and columns
omitted and promote the result back to an $N\times N$ matrix by inserting zeros
for the rows and columns $\{j_1, \dots ,j_m\}$. In our example
one finds, with $\lambda(x,y,z)=x^2+y^2+z^2-2xy-2yz-2xz$:
\bea
{\cal S}^{\{2,4\}-1} &=&
\frac{1}{\lambda(s_{23},m_1^2,m_3^2)}\left( \begin{array}{cccc} 2\,m_3^2 &  0& s_{23}-m_1^2 -m_3^2 & 0 \\
                                                0 & 0 & 0& 0 \\
                                               s_{23}-m_1^2 -m_3^2 &  0&  2\,m_1^2 &  0\\
					       0 & 0 & 0& 0\end{array}\right)
\eea
In the following, ${\cal S}^{-1}(S\setminus\{j_1, \dots ,j_m\})={\cal S}^{\{j_1, \dots ,j_m\}-1}$ 
has to be understood in this sense. 

\vspace{0.3cm}

\noindent
Using these conventions,  
Feynman parameter integrals with propagator pinches and Feynman 
parameters $z_{l_1}\dots z_{l_r}$ in the numerators can be defined as
\bea
&& I^n_N(l_1,\dots,l_r;S\setminus \{j_1,\dots,j_m\}) = \nn\\
&& 
(-1)^N\Gamma(N-\frac{n}{2})\int \prod_{i=1}^N dz_i\,
\delta(1-\sum_{l=1}^N z_l)\,
\delta(z_{j_1})\dots \delta(z_{j_m})z_{l_1}\dots z_{l_r}\left(R^2\right)^{n/2-N}
\label{isca_pinch}
\eea
Whenever one index of the pinch set ${\cal Q} = \{j_1,\dots,j_m\}$ 
coincides with
one numerator index in the set ${\cal N} = \{l_1,\dots,l_r\}$ 
this integral is trivially zero:
\be\label{isca_pinch_property}
\mbox{If }\;{\cal N} \cap {\cal Q} \neq \{\} \quad \Rightarrow\quad 
I^n_N( {\cal N};S\setminus {\cal Q}) = 0\,\, .
\ee
The above conventions lead to simple expressions in our
formalism and are designed for iteration purposes, as will become clear below.

\subsection{Definition of form factors}

Algebraic expressions of an amplitude typically consist
of spinors and Dirac chains of momenta and polarisation vectors, depending on
the process under consideration. The loop momenta which appear in the
numerator of a Feynman diagram often come in the combination $q_i=k+r_i$, 
like for instance in fermion propagators. 
We keep this natural combination in our tensor reduction formalism. 
Therefore we define  tensor integrals as 
\bea
I^{n,\,\mu_1\ldots\mu_r}_N(a_1,\ldots,a_r;S) = 
\int d\bar{k}\,
\frac{q^{\mu_1}_{a_1}\ldots q^{\mu_r}_{a_r}}{\prod_{i \in S}
(q_i^2-m_i^2+i\delta)}\;.
\label{iten}
\eea
By setting $a_1,\ldots, a_r=N$, and using momentum conservation to set $r_N=0$, 
we can always retrieve the commonly used form 
\be
I^{n,\,\mu_1\ldots\mu_r}_N(N,\dots,N,S) = 
\int d\bar{k}\,
\frac{k^{\mu_1}\ldots k^{\mu_r}}{\prod_{i \in S}
(q_i^2-m_i^2+i\delta)}\;.
\label{conventional}
\ee
In this more ``conventional" approach, 
one of the $q_i^\mu$ is specified to be $k^\mu$, which singles out one 
propagator and defines a standard form. 
After a reduction step one obtains integrals which are not of 
standard type, 
such that a shift operation  $k \to k + r_j$ is necessary to remap to the 
standard form, giving rise to  $2^r$ terms for a rank $r$ 
tensor integral.
Our  formulation, being manifestly translation invariant and 
thus making such shifts obsolete, avoids here a  
proliferation of terms.

\vspace{0.3cm}

\noindent
As pointed out above, shifts of the loop momentum can be absorbed into 
a redefinition of the $r_j^\mu$ vectors. To achieve a manifestly 
translation invariant formulation, we need
vectors which are invariant under shifts $r_j^\mu\to r_j^\mu + r_a^\mu$.
This motivates the definition of the shift-invariant vector $\Delta_{ij}^\mu$:
\begin{equation}
\Delta_{ij}^\mu = r_i^\mu-r_j^\mu = q_i^\mu-q_j^\mu 
\label{eqDEFDELTA}
\end{equation}
Apart from metric tensors  $g^{\mu\nu}$, the
Lorentz structure of  the integrals will be carried
by these vectors.

\vspace{0.3cm}

\noindent
As will become clear below, 
we have to distinguish  the cases $N\leq 5$ and $N\geq 6$.
In the case $N\leq 5$ we will express the different tensor structures in terms
of metric tensors and difference vectors, $\Delta_{ij}^\mu$.
Tensor integrals are expressible by linear combinations of such Lorentz tensors and 
form factors  denoted by $A^{N,r}_{l_1\cdots l_r}(S)$,
$B^{N,r}_{l_1\cdots }(S)$, $C^{N,r}_{\cdots }(S)$. 
$A^{N,r}$ is the coefficient of the Lorentz
structure containing only difference vectors. 
$B^{N,r}$ belongs to exactly  one metric tensor and $(r-2)$
$\Delta_{ij}^{\mu}$ vectors, and $C^{N,r}$ is the coefficient of the Lorentz
structure containing products of two metric tensors.  
Thus our form factors for $N\leq 5$ are defined by the  formula 
\bea 
\lefteqn{I^{n,\,\mu_1\ldots\mu_r}_N(a_1,\ldots,a_r;\,S) =}  
\nonumber \\
& &  
\; 
\sum_{j_1\cdots j_{r}\in S} \;\;\;
\left[ 
 \Delta_{j_1\cdot}^{\cdot} \cdots \Delta_{j_r\cdot}^{\cdot} 
\right]^{\{\mu_1\cdots\mu_r\}}_{\{a_1\cdots a_r\}} \, A_{j_1 \ldots ,j_{r}}^{N,r}(S) 
\nn\\ 
&+& 
\sum_{j_1\cdots j_{r-2}\in S} \, 
\left[
 g^{\cdot\cdot} \Delta_{j_1\cdot}^{\cdot} \cdots \Delta_{j_{r-2}\cdot}^{\cdot}
 \right]^{\{\mu_1\cdots\mu_r\}}_{\{a_1\cdots a_{r}\}}\,  B_{j_1 \ldots,j_{r-2}}^{N,r}(S) 
\nn\\ 
&+& 
\sum_{j_1\cdots j_{r-4}\in S} \, 
\left[
 g^{\cdot\cdot}g^{\cdot\cdot}  \Delta_{j_1\cdot}^{\cdot} \cdots
                               \Delta_{j_{r-4}\cdot}^{\cdot} 
\right]^{\{\mu_1\cdots\mu_r\}}_{\{a_1\cdots a_{r}\}}\, C_{j_1 \ldots ,j_{r-4}}^{N,r}(S) 
\label{fofageneral} 
\eea
where the distribution of the $r$ Lorentz indices $\mu_i$, and momentum 
labels $a_i$ to the vectors $\Delta_{j\,a_i}^{\mu_i}$, denoted by 
$[\cdots]^{\{\mu_1\cdots\mu_r\}}_{\{a_1\cdots a_r\}}$ in eq.~(\ref{fofageneral}),  
is  illustrated in the following equations (\ref{eqNpr1}) to (\ref{eqNpr4}).
\begin{eqnarray} 
&& I^n_N(S) 
=  A^{N,0}(S)
\label{eqNpr0}\\ 
&&
\nn\\ 
&&I_N^{n,\mu_1}(a_1;S) 
=  \sum_{l \in S} \, \Delta^{\mu_1}_{l\, a_1} \, A^{N,1}_{l}(S) 
\label{eqNpr1}\\ 
&&
\nn\\ 
&&I_N^{n,\mu_1 \mu_2}(a_1,a_2;S) 
=  
\sum_{l_1,l_2 \in S}  \;
\Delta^{\mu_1}_{l_1 \, a_1} \;  \Delta^{\mu_2}_{l_2 \, a_2} \, 
A^{N,2}_{l_1 \, l_2}(S) + g^{\mu_1 \, \mu_2} \, B^{N,2}(S)  
\label{eqNpr2}\\ 
&&
\nn\\ 
&&I_N^{n,\mu_1 \mu_2 \mu_3}(a_1,a_2,a_3;S) 
= 
\sum_{l_1,l_2,l_3 \in S} \; 
  \Delta^{\mu_1}_{l_1 \, a_1} \;  
  \Delta^{\mu_2}_{l_2 \, a_2} \;
  \Delta^{\mu_3}_{l_3 \, a_3} \,  A^{N,3}_{l_1 l_2 l_3}(S) 
\nonumber \\
&& \qquad + 
\sum_{l \in S} \; 
\left(
 g^{\mu_1 \mu_2} \, \Delta^{\mu_3}_{l \, a_3} + 
 g^{\mu_1 \mu_3} \, \Delta^{\mu_2}_{l \, a_2} + 
 g^{\mu_2 \mu_3} \,  \Delta^{\mu_1}_{l \, a_1}
\right) \, B^{N,3}_l(S)   
\label{eqNpr3}\\
&&
\nn\\ 
&&I_N^{n,\mu_1 \mu_2 \mu_3 \mu_4}(a_1,a_2,a_3,a_4;S)
= 
\sum_{l_1\ldots l_4 \in S}  \; 
\Delta^{\mu_1}_{l_1 \, a_1}  \; \Delta^{\mu_2}_{l_2 \, a_2} \;
\Delta^{\mu_3}_{l_3 \, a_3} \;  \Delta^{\mu_4}_{l_4 \, a_4} \, 
A^{N,4}_{l_1 l_2 l_3 l_4}(S) 
\nonumber \\ 
&&\; +   
\; 
\sum_{l_1,l_2 \in S}\;  
\left(  
  g^{\mu_1 \mu_2} \, \Delta^{\mu_3}_{l_1 \, a_3} \, \Delta^{\mu_4}_{l_2 \, a_4} 
+ g^{\mu_1 \mu_3} \, \Delta^{\mu_2}_{l_1 \, a_2} \, \Delta^{\mu_4}_{l_2 \, a_4} 
\right. 
+
  g^{\mu_1 \mu_4} \, \Delta^{\mu_2}_{l_1 \, a_2} \, \Delta^{\mu_3}_{l_2 \, a_3} 
+ g^{\mu_2 \mu_3} \, \Delta^{\mu_1}_{l_1 \, a_1} \, \Delta^{\mu_4}_{l_2 \, a_4}  
\nonumber \\ 
& &  \qquad\qquad\qquad
\left.
+
  g^{\mu_2 \mu_4} \, \Delta^{\mu_1}_{l_1 \, a_1} \, \Delta^{\mu_3}_{l_2 \, a_3}
+ g^{\mu_3 \mu_4} \, \Delta^{\mu_1}_{l_1 \, a_1} \, \Delta^{\mu_2}_{l_2 \, a_2} 
\right) \, B^{N,4}_{l_1 l_2}(S)  
\nonumber \\  &&\nonumber \\ 
&& \qquad\qquad\quad+   
\left(  
 g^{\mu_1 \mu_2} \, g^{\mu_3 \mu_4} + 
 g^{\mu_1 \mu_3} \, g^{\mu_2 \mu_4} +   
 g^{\mu_2 \mu_3} \, g^{\mu_1 \mu_4}  
\right) \, C^{N,4}(S)  
\label{eqNpr4} 
\end{eqnarray} 
We recall that  standard form factor representations can be simply obtained
by replacing  $a_j=N$ for all $j$, together with $r_N=0$.
This also shows that the form factors do not depend on the introduction
of the difference vector. The form factors are shift invariant by themselves.
One main result of the paper will be the explicit representation of all
these form factors in terms of higher dimensional 4-point parameter integrals
and $n=4-2\epsilon$ dimensional  3-point parameter integrals with nontrivial numerators,
see eq. (\ref{basic_integral}).

\vspace{0.3cm}

\noindent
For $N\geq 6$, the tensor reduction 
will be done in such a way that only the form factors for $N=5$ appear.
The  Lorentz structure of $N$-point rank $r$ tensor integrals
does not require the introduction of additional factors of $g^{\mu\nu}$ 
as compared to the $N=5$ case, only additional difference vectors  
$\Delta_{ij}^\mu$ appear. This is due to the fact that for $N\geq 5$, 
four linearly independent external vectors form a basis
of Minkowski space. We note that for $N=5$, 
one could already  express 
the metric by external momenta, but this would introduce 
inverse Gram determinants.

\vspace{0.3cm}

\noindent
Before closing this section we would like to note that  
momentum integrals, Feynman parameter integrals and form factors
are naturally related by \cite{tarasov,davydychev,tenred1}:
\bea
I^{n,\,\mu_1\ldots\mu_r}_N(a_1,\ldots,a_r\,;S) 
& = & 
(-1)^r \sum_{m=0}^{[r/2]} \left( -\frac{1}{2} \right)^m
\sum_{j_1\cdots j_{r-2m}=1}^N \, 
\left[ 
 (g^{..})^{\otimes m}\,\Delta_{j_1\cdot}^{\cdot} \cdots \Delta_{j_r\cdot}^{\cdot}
\right]^{\{\mu_1\cdots\mu_r\}}_{\{a_1\cdots a_r\}}
\nn\\
& & \;\;\;\;\;\;\;\;
\times I_N^{n+2m}(j_1 \ldots ,j_{r-2m}\,;S)
\label{eq32}
\eea 
where $I_N^{n+2m}(j_1 \ldots ,j_{r-2m}\,;S)$ is defined in eq.~(\ref{isca_pinch}).
In eq.~(\ref{eq32}), $[r/2]$ stands for the nearest integer less or equal to $r/2$ and the symbol 
$\otimes m$ indicates
that $m$ powers of the metric tensor are present.
It is obvious from this formula that the $g^{\mu\nu}$-terms are always 
associated to integrals in more than $n$ dimensions.

\vspace{0.3cm}

\noindent
In the following sections \ref{method} and \ref{algebraic-red}, we will 
formulate a reduction formalism for  
tensor integrals of the type (\ref{iten}) and give convenient
representations for the form factors defined in eq.~(\ref{fofageneral}).
In section \ref{absence-N-eqn5} we will show that higher than 
$n=4-2\eps$ dimensional integrals for 
$N\geq 5$ can be avoided completely if the external kinematics is 
defined in 4-dimensions, a fact which was already carefully investigated
elsewhere \cite{tarasov,Bern:1993kr,tenred1}.

\section{Tensor reduction by subtraction}\label{method}

The formalism described in this section   
naturally leads to a separation of IR divergent and finite
expressions and does not produce
spurious Gram determinants. The reduction is based on a  
subtraction technique which is analogous to the one
used in~\cite{tenred1} for the scalar case. 
Before we come to the tensorial case, 
let us recall the procedure of \cite{tenred1} for the scalar case,  
recast into the notation of this article. 

\subsection{Subtraction for scalar integrals}\label{subscalar}

Our aim is to split a scalar $N$-point integral as defined in 
eq.~(\ref{isca1}) into an IR finite part and a possibly IR divergent, 
but simpler part. Therefore we make the ansatz
\begin{eqnarray}
I_N^n(S) & = & I_{div}(S) + I_{fin}(S) \nn\\
&=&\sum_{i\in S} b_i(S) \, \int d\bar k \; 
\frac{ (q_i^2-m_i^2)}{\prod_{j\in S} (q_j^2-m_j^2+i\,\delta)} +
 \int d\bar k \; 
\frac{ 1 - \sum_{i\in S}  b_i(S) \, (q_i^2-m_i^2) 
}{\prod_{j\in S} (q_j^2-m_j^2+i\,\delta)}
\label{EQdivetfin}
\end{eqnarray}
We can see that $I_{div}$ is a sum of reduced integrals where 
one propagator has been pinched.
Now let us consider  $I_{fin}$ after having introduced Feynman
parameters. In order to arrive at a 
quadratic form in the loop momentum, we perform the shift
\begin{eqnarray}
k & = & l - \sum_{i\in S} z_i \, r_{i} \;.
\label{eqTRANSL} 
\end{eqnarray}
This shift transforms the denominator to the form $l^2-R^2$, 
where $R^2$ is defined in eq.~(\ref{isca2}).
For the numerator, we use 
\bea
\Delta_{ij}\cdot\Delta_{kl}&=&
\frac{1}{2}({\cal S}_{il}+{\cal S}_{jk}-{\cal S}_{ik}-
{\cal S}_{jl})\;,\nn\\
\left( \sum_{i\in S} z_i \, \Delta_{j i} \right)^2 &=& 
\sum_{i\in S} z_i \, \cals_{i j}\,+m_j^2 + R^2
\label{trafo}
\eea 
to obtain
\bea
1-\sum_{i\in S}  b_i(S) \, (q_i^2-m_i^2)  &=&
- (l^2+R^2 ) \sum\limits_{i\in S} b_i(S) + 
\sum\limits_{j\in S} z_j \Biggl[ 1- 
\sum\limits_{i\in S} b_i(S)\,\left\{
{\cal S}_{ij} +2\,l\cdot\Delta_{ij} \right\}\Biggr]\;.\nn\\
\label{ifin}
\eea
The term linear in the loop momentum $l$ vanishes due to symmetric integration
and we conclude that if the equation 
\be
 \sum\limits_{i\in S} b_i(S)\,
{\cal S}_{ij}=1\;\;,\;\;j=1,\ldots,N
\label{bdef}
\ee
is fulfilled, the term in square brackets in eq.~(\ref{ifin}) vanishes.
Then $I_{fin}$ is given by
\begin{eqnarray}
I_{fin}(S) & = & - B(S) \,\Gamma(N) \, 
 \int^1_0  \prod_{i\in S} \, d z_i \, 
\delta(1-\sum_{l\in S} z_l)  \;
\int \frac{d^n l}{i\pi^{n/2}} \; 
\frac{l^2+R^2}{(l^2-R^2)^N} \nn\\
&=& -B(S)\, (N-n-1)\, I_N^{n+2}(S)
\label{EQifin2}\\
&&\nn\\
B(S)&=&\sum_{i\in S} b_i(S)\;.\label{bigB}
\end{eqnarray}
If it is clear from the context which set $S$ we refer to, 
the argument $(S)$  is omitted in $B(S)$ and $b_i(S)$. 
If the $b_i$ belong to a reduced kinematic matrix 
${\cal S}^{\{j\}}$ where the $j^{\rm{th}}$ row and column is zero, 
associated with the set $S\setminus\{j\}$, 
one has $b_i(S\setminus\{j\})=
\sum_{k\in S\setminus\{j\}}({\cal S}^{\{j\}})^{-1}_{ki}$. 
For simplicity of notation, we introduce the shorthand
$b_i(S\setminus\{j\})=b_i^{\{j\}}$, 
and correspondingly $B^{\{j\}}$ is defined as
\be
B^{\{j\}}=\sum_{i \in S\setminus\{j\}} b^{\{j\}}_i \;.
\ee
All that remains to be shown now  is that eq.~(\ref{bdef})
indeed has a solution for the reduction coefficients $b_i$ 
for arbitrary $N$. 

\vspace{0.3cm}

\noindent
In the case of non-exceptional 4-dimensional 
kinematics\footnote{We call a kinematic configuration defined by $N$ external momentum vectors 
{\em exceptional} if the Gram determinant built from a subset of 
four of these vectors vanishes.}, 
$rank({\cal S}) = min(N,6)$ holds. Thus 
${\cal S}$ is invertible for $N \leq 6$, and eq.~(\ref{bdef})
has the unique solution 
\be
b_i=\sum_{k\in S}{\cal S}_{ki}^{-1}\;.\label{bsinv}
\ee
If ${\cal S}$ is not invertible, we proceed as follows.
First we single out the $a^{\mathrm{th}}$ row and column 
in ${\cal S}$, where  
$a$ is an arbitrary element of the set $S$,  
to write ${\cal S}_{ij}$ as:
\begin{equation}\label{e3}
{\cal S}_{ij} = - G_{ij}^{(a)} + V_{i}^{(a)} + V_{j}^{(a)}
\end{equation}
with 
\begin{equation}\label{e4}
G_{ij}^{(a)} = 2 \Delta_{ia}\cdot\Delta_{ja}, \;\;\;\; 
V_{i}^{(a)}  = \Delta_{ia}^{2} - m_{i}^{2}\;.
\end{equation}
The Gram matrix $G_{ij}^{(a)}$ is understood as an $N\times N$ matrix. By definition 
its entries are zero whenever $i=a$ or $j=a$.
For ease of notation, we will omit the superscript $(a)$ in the
following if it is clear from the context.
Using the above relations and distinguishing the cases
$j=a$ and $j\neq a$, 
eq.~(\ref{bdef}) is equivalent to the two equations 
\bea
\sum_{i\in S\setminus\{a\}}b_i\, G_{ij}&=&B\,(V_j-V_a)\label{bdef1}\\
\sum_{i\in S\setminus\{a\}}b_i\,\,(V_i-V_a)&=&1-2\,B\,V_a\;.
\label{bdef2}
\eea
Eqs.\,(\ref{bdef1}),(\ref{bdef2}) may be solved in the following way.  As
$G$ is not a regular matrix we first
construct the generalised
inverse $H$ of $G$, defined as in eq.~(\ref{DEFpseudo_inverse}).
To this end we introduce four linearly independent 4-vectors $E_{l = 1, \cdots, 4}^{\mu}$
forming a basis of the physical Minkowski space, and a $4 \times N$ coefficient matrix ${\cal R}$, such that
\begin{equation}\label{e7}
\Delta_{i\,a}^{\mu} = \sum_{m = 1}^{4} {\cal R}_{mi}^{(a)}\, E_{m}^{\mu}, \;\;\;\;
\tilde{G}_{mn} = 2 E_{m}\cdot E_{n}\;.
\end{equation}
Here ${\cal R}_{ma}^{(a)}=0$ for all $m$.
The Gram matrix $G$ is expressible as\footnote{Note that the pseudo-inverse 
of $G$ and the singular value decomposition of $G$ given in ref.\,\cite{Giele:2004iy}
are related in a simple way\,\cite{swedesbook}:
Defining an orthogonal $4\times 4$ matrix $O$ which 
diagonalises $\tilde{G}_{mn}$ (such a matrix $O$ always exists for 
a symmetric matrix $\tilde{G}$), i.e. 
$(O^T\tilde{G}O)_{mn}=\omega_m\delta_{mn}$, one has
$G_{ij}=({\cal R}^T\,O\,O^T\,\tilde{G}\,O\,O^T\,{\cal R})_{ij}=
\sum_{k=1}^4\,({\cal R}^T\,O)_{ik}\,\omega_k\,(O^T\,{\cal R})_{kj}=
\sum_{k=1}^4\,u_{ik}\,\omega_k\,v^T_{kj}$, 
which means that the $u_{ik}$ and $v^T_{kj}$ 
used in ref.\,\cite{Giele:2004iy} are related to the matrix ${\cal R}$ 
given here by $u_{ik}=({\cal R}^T O)_{ik}$ and $v^T_{kj}=(O^T {\cal R})_{kj}$.
} 
$G = {\cal R}^{T} \tilde{G} {\cal R}$.
The matrix ${\cal R}$ is of rank 4, and thus the $4 \times 4$ 
matrix ${\cal R} {\cal R}^{T}$
is invertible. The sought matrix $H$ is then uniquely defined by
\begin{equation}\label{e8}
H = {\cal R}^{T} ({\cal R} {\cal R}^{T})^{-1} \tilde{G}^{-1} 
({\cal R} {\cal R}^{T})^{-1} {\cal R}\;.
\end{equation}
Before going on we note that
the case of exceptional kinematics is now easily dealt with. 
If the external vectors only form a space of dimension
$k$ less then four, one may choose $k$ basis vectors 
$E_{l = 1, \cdots, k}^{\mu}$
defining ${\cal R}$ to be a $k \times N$ matrix.
This is the only change to be made.

\vspace{0.3cm}

\noindent
The system of equations (\ref{bdef1}),(\ref{bdef2}) admits 
solutions if and only if the consistency condition
\begin{eqnarray}\label{consistency}
B\sum_{j\in S\setminus\{a\}}(\Eins_{N-1} - GH)_{ij}\, \delta V_j^{(a)}
 &=& 0
\end{eqnarray}
is fulfilled, where we have defined 
\be
\delta V^{(a)}_j=V^{(a)}_j-V^{(a)}_a \quad,\quad j \in S\setminus \{a\}\;.
\label{vvec}
\ee
We will not always write  the sum 
$\sum_{j\in S\setminus \{a\}}$ explicitly in the following, but 
denote products like $\sum_{j\in S\setminus \{a\}} K_{ij}\,V_j$
by $(K\cdot V)_i$.

\vspace{0.3cm}

\noindent
Since, for $N>5$, in general $GH\cdot\delta  V \not= 
\delta V$, a solution of 
(\ref{consistency}) for $N>5$ exists if and only if 
$B = 0$. 
The solution  spans an $(N-5)$-dimensional
space which is just the kernel of the Gram matrix. 
It can be parametrized by $(N-5)$  vectors
$U^{(1,\dots,N-5)}$. 
Let us note that 
\begin{equation}
GH = HG = {\cal R}^{T} ({\cal R} {\cal R}^{T})^{-1} {\cal R}
\end{equation}
and that the projector onto $\mathrm{Ker}(G)$ is given by
\begin{eqnarray}
K & = & 
\Eins_{N-1} -{\cal R}^{T} ({\cal R} {\cal R}^{T})^{-1} {\cal R}\;.
\end{eqnarray}
It follows from the definition of $H$ that $K\cdot\delta V \in \mathrm{Ker}(G)$.  
Now one can choose
$U^{(N-5)}=
K\cdot \delta V/(\delta V\cdot K\cdot \delta V)$ 
parallel to $\delta V$ and
the other $U$-vectors orthogonal, $\delta V\cdot U^{(k=1,\dots,N-6)}=0$.
A general vector in $\mathrm{Ker}(G)$ is then parametrised by
$U=\sum_{k=1}^{N-6} \beta_k  U^{(k)} +\alpha U^{(N-5)}$ and 
a solution to eqs.~(\ref{bdef1}),(\ref{bdef2}) is given by
\begin{eqnarray}\label{bgeneral}
b_i  &=& \frac{(K\cdot \delta V)_i +
\sum_{k=1}^{N-6} \beta_k  U^{(k)}_i }{\delta V \cdot K \cdot 
\delta V} \, , \quad i \in S\setminus \{a\} \\ 
b_a&=& -\sum_{k\in S\setminus \{a\}} b_k
\;.\label{bzero_allN}
\end{eqnarray}
We recall that this solution is valid for all $N\geq 6$.
For exceptional
kinematics and $N< 6$ an analogous solution can be  derived.
In the other cases one can simply use eq.~(\ref{bsinv}).
\vspace{0.3cm}

\noindent
Near some special momentum configurations the denominators
of the reduction coefficients, $\delta V \cdot K \cdot 
\delta V$ or $\det({\cal S})$, may cause numerical instabilities in realistic 
applications. In contrast to inverse Gram determinants which can
be viewed as technical artefacts of the tensor reduction down to scalar integrals
in momentum space, the singular behaviour of the latter is  due to  physical 
singularities like soft/collinear configurations or 
thresholds and therefore they are unavoidable.  
We would like to point out that this is not a problem related to the 
use of the pseudo-inverse. If one would use the singular value 
decomposition, this problem would occur as well.
Experimental cuts very often exclude these 
numerically dangerous phase space regions.

\vspace{0.3cm}

\noindent
Finally, we quote the interesting relation between the kinematic determinants
\bea\label{detSGrel}
B\;\det{\cal S}= (-1)^{N+1}\det G^{(a)}\;, 
\eea
where $G_{ij}^{(a)}$ is the Gram  matrix defined above. The relation
is valid for all $a\in S$ and arbitrary $N$.

\vspace{0.3cm}

\noindent
We conclude that in the case $N\geq 6$ one always has $B=0$, if the external kinematics is defined
in four dimensions. This means that $I_{fin}(S)=0$ in eq. (\ref{EQifin2}).
In the case $N=5$ one finds $I_{fin}(S)= B\,(n-4)\, I_5^{n+2}(S)$.
As $I_5^{n+2}(S)$ is IR and UV finite, the whole term is of ${\cal O}(\epsilon)$
and can be dropped in phenomenological applications.
For $N=4$ we see that 4-point functions can be represented in terms
of 3-point functions and $(n+2)$-dimensional 4-point functions.
As the latter are IR finite, a splitting into IR divergent
and IR finite integrals is achieved.
We would like to emphasise that by iteration of 
eqs.~(\ref{EQdivetfin}) and (\ref{EQifin2})
an {\em arbitrary} scalar $N$-point function can be algebraically
reduced to 3-point functions and $(n+2)$-dimensional 4-point functions.
In this representation IR divergent integrals are naturally separated from
finite contributions. 

\vspace{0.3cm}

\noindent
As an example we give here the 6-point function
in terms of $(n+2)$-dimensional 4-point functions and 3-point functions.
In our notation, up to ${\cal O}(\eps)$ terms:
\bea
I^{n}_{6}(S) & = & \sum_{j \in S} \, b_j\sum_{k \in S\setminus \{j\}}\,b_k^{\{j\}}
 \,\left[ B^{\{j,k\}} \, 
I^{n+2}_4(S\setminus\{j,k\}) +  \sum_{l \in S\setminus\{j,k\}} 
b_l^{\{j,k\}} \,I_3^{n}(S\setminus\{j,k,l\}) \right]\nn
\eea
From this representation, IR divergences can be trivially isolated. 

\subsection{Subtraction for tensor integrals}\label{subtensor}

We now extend the above reasoning to  the tensorial case, i.e. we will 
split a general tensor
integral into an infrared finite part and a part which contains possible
IR divergences, the latter having one rank less and one propagator less. We will
give a constructive algorithm how to proceed for  any number $N$ of external
legs.

\vspace{0.3cm}

\noindent
Let us write  eq.~(\ref{iten}) as follows:
\begin{eqnarray}
I_N^{n,\mu_1\ldots\mu_r}(a_1,\ldots,a_r;S) & = & 
\int d\bar k \, \frac{ \left[\,q_{a_1}^{\mu_{1}} + 
\sum_{j \in S} \calc_{ja_{1}}^{\mu_{1}}\, (q_j^2 -m_j^2)\right] \; 
 q_{a_2}^{\mu_{2}}\ldots q_{a_r}^{\mu_{r}}}{\prod_{i \in S} (q_i^2-m_i^2+i\delta)} 
\nn\\
& & \mbox{} - \sum_{j \in S}\calc_{ja_1}^{\mu_{1}} \,  
\int d\bar k \;\frac{ (q_j^2 -m_j^2)\;
q_{a_2}^{\mu_{2}}\ldots q_{a_r}^{\mu_{a_r}}}{
\prod_{i \in S} (q_i^2-m_i^2+i\delta)}\;.
\label{eqNEWIDEF2}
\end{eqnarray}
The last line corresponds to $(N-1)$-point tensor integrals
of rank $(r-1)$. The coefficients $\calc_{ja_1}^{\mu_{1}}$ 
will be determined such that the first term in eq.\,(\ref{eqNEWIDEF2}) 
is an IR finite expression.  

\vspace{0.3cm}

\noindent
As in the scalar case, one introduces $N$ 
Feynman parameters $z_i$ and makes the substitution (\ref{eqTRANSL})
to obtain the form $l^2-R^2$ for the denominator.
Under the shift~(\ref{eqTRANSL}), the momenta $q_{a}$ 
become
\begin{equation}
q_{a} = l + \sum_{i\in S} z_i \, \Delta_{a\,i}\;.
\label{eqSHIFT1}
\end{equation}
Now let us consider the vector $A_{a_1}^{\mu}$, 
given by the square bracket in the first line of 
eq.~(\ref{eqNEWIDEF2})
\begin{equation}
A_{a_1}^{\mu_1} = q_{a_1}^{\mu_1} + \sum_{j\in S} \calc_{ja_1}^{\mu_1}\,(q_j^2-m_j^2)
\label{eq1}
\end{equation}
where
$\calc^{\mu}_{j a}$ is defined by the following equation, which will 
be the cornerstone of our derivation for general $N$\,:
\be
\sum_{j \in S} {\cal S}_{ij}\, {\cal C}_{j\,a}^{\mu} = 
\Delta_{i\, a}^{\mu}
\;, \;\;\;\; a \in S
\;.
\label{eqCDEF}
\ee
Note that this equation is analogous to eq.~(\ref{bdef}) for the scalar case.
We will solve equation (\ref{eqCDEF}) in the next subsection for arbitrary kinematics and arbitrary $N$.

\vspace{3mm}

\noindent
The vector $A^{\mu}_a$ 
 transforms under the shift~(\ref{eqTRANSL}) as:
\begin{eqnarray}
A_a^{\mu} & = & l^{\mu} +\sum_{i\in S} z_i \, 
\Delta_{a i}^{\mu} + \sum_{j \in S}  \calc_{ja}^{\mu} \, 
\left[ l^2 -m_j^2 + \left( \sum_{i\in S} z_i\, 
\Delta_{j i} \right)^2 - 2 \, l\cdot\sum_{i\in S} z_i \, 
\Delta_{ij} \right] \nn\\
&=& l^{\mu} +\left( l^2+R^2 \right) \, \calv_{a_1}^{\mu}+\sum_{i\in S} z_i\,\left[
\sum_{j \in S} \calc_{ja}^{\mu}\,\left(
\cals_{ij}-2\,l\cdot\Delta_{ij}\right) - \Delta_{ia}^{\mu}\right]\;,
\label{eqTRANSApre}
\end{eqnarray}
where eq.~(\ref{trafo}) and the definition 
\bea
\calv_a^{\mu} & = &
\sum_{j \in S} \, \calc_{ja}^{\mu}
=\sum_{k \in S} \, b_k \, \Delta^{\mu}_{k \, a}
\label{eqVDEF}
\eea
have been used.
Thus we see that if eq.~(\ref{eqCDEF}) is fulfilled, 
all the terms contained in $A_a^{\mu}$ are either proportional to 
the loop momentum $l$ or to $R^2$. 


\vspace{0.3cm}

\noindent
Provided that eq.~(\ref{eqCDEF}) is fulfilled, and using  
$\Delta_{ij}=\Delta_{ia_2}+\Delta_{a_2j}$, where $a_2\in S$  
is arbitrary, eq.~(\ref{eqTRANSApre}) can also be written in the
following form\,: 
\begin{eqnarray}
A_{a_1}^{\mu} & = & l_{\nu} \, \left( \calt^{\mu  \nu}_{a_1  a_2} + 
2 \, \calv_{a_1}^{\mu}\,\sum_{i\in S} z_i \Delta_{a_2 i}^{\nu}  
\right) + \left( l^2+R^2 \right) \, \calv_{a_1}^{\mu}
\label{eqTRANSA}\;,
\end{eqnarray}
where
\bea
\calt^{\mu  \nu}_{a_1 \, a_2} & = & g^{\mu  \nu} + 
2 \, \sum_{j \in S} \, 
 \calc_{j \, a_1}^{\mu} \, \Delta^{\nu}_{j \, a_2}
\label{eqTDEFN} \;.
\eea
From the expression (\ref{eqTRANSA}) 
one can see immediately that 
the first term on the right-hand side in eq.~(\ref{eqNEWIDEF2})  
has no infrared divergences:  
After the shift~(\ref{eqTRANSL}), 
the integral in the first line is proportional to $l$ or
$l^2+R^2$ and, after integration over the loop momentum, will 
give some higher-dimensional integrals (with Feynman parameters 
in the numerator for $r>1$). 
The  term in the second line of eq.~(\ref{eqNEWIDEF2}) is divergent 
but has one tensor rank less and one propagator less.
Thus, if eq.~(\ref{eqCDEF}) is fulfilled, 
the ansatz (\ref{eqNEWIDEF2}) leads to the desired splitting of 
a rank $r$ $N$-point integral into IR finite terms and 
$(N-1)$-point integrals of rank $r-1$.

\subsection{Solving the defining equation for arbitrary $N$}

It remains to be shown that eq.~(\ref{eqCDEF}) has a solution 
for general $N$, and to construct such a solution explicitly. 

\vspace*{3mm}

\noindent
As in the scalar case, the solution is simple if  ${\cal S}$ is 
invertible, i.e. 
in the case of non-exceptional  kinematics for $N \leq 6$. 
In this case, eq.~(\ref{eqCDEF}) has the unique solution
\begin{equation}\label{e2}
{\cal C}_{j a}^{\mu} = \sum_{k \in S} 
\left( {\cal S}^{-1} \right)_{jk} \Delta_{k a}^{\mu}\;.
\end{equation}
On the other hand, if $N \geq 7$ or in the case of 
exceptional kinematics, ${\cal S}$ is not invertible, so 
eq.~(\ref{eqCDEF}) does not have a unique solution. 
However, an explicit solution can be constructed 
in the same way as has been done in section \ref{subscalar}
for the scalar case. 
To this end,  we again write the matrix ${\cal S}$ as 
in eqs.~(\ref{e3}),(\ref{e4}).
Using the definition (\ref{vvec}) for 
 $\delta V_j^{(a)}$, eq.~(\ref{eqCDEF}) may be rewritten as
\begin{eqnarray}
\sum_{i \in S} {\cal S}_{ij}\, {\cal C}_{i\,b}^{\mu} &=& 
\Delta_{j\, b}^{\mu}
\;, \;\;\; b \in S \qquad \Leftrightarrow\nn\\
\sum_{i\in S\setminus \{a\}} G^{(a)}_{ij}\,{\cal C}_{i\,b}^{\mu} 
& = &
- \Delta_{j\,a}^{\mu} + \delta V_j^{(a)} \, {\cal V}_{b}^{\mu}, \;\;\;\; 
\label{e6a} \\
\sum_{i\in S\setminus \{a\}}\delta V_i^{(a)}\,{\cal C}_{i\, b}^{\mu}  
& = & 
\Delta_{a \,b}^{\mu} - 2 V^{(a)}_{a} \, {\cal V}_{b}^{\mu} 
\label{e6b}
\end{eqnarray}
where also $\sum_{i \in S} {\cal C}_{i\,b}^{\mu}={\cal V}_{b}^{\mu}$ has been used.
Eqs.~(\ref{e6a}),(\ref{e6b}) can be solved using the same 
pseudo-inverse as already constructed in section \ref{subscalar}.
The system of equations (\ref{e6a}),(\ref{e6b}) admits solutions 
if and only if the consistency condition
\begin{equation}\label{e8p}
\sum_{j\in S\setminus \{a\}} (\Eins_{N-1} - GH)_{ij}\, 
\left( 
 \Delta_{j\,a}^{\mu} - \delta  V_j^{(a)} \, {\cal V}_{b}^{\mu} 
\right) = 0
\end{equation}
is fulfilled. Since $\sum_{j\in S\setminus \{a\}}
(\Eins_{N-1} - GH)_{ij}\Delta_{j a}^{\mu} = 0$ 
whereas, for $N>5$, in general $GH \cdot \delta V \not= \delta  V$, 
a solution of 
(\ref{e6a}) for $N>5$ exists if and only if ${\cal V}_{b}^{\mu} = 0$. 
The general solution of (\ref{e6a}) for $N>5$ is thus given by
\begin{eqnarray}\label{e9}
{\cal C}_{i\, b}^{\mu} &=& -\sum_{j\in S\setminus \{a\}} H_{ij}\,
\Delta_{j\, a}^{\mu} + W_i^{\mu}\quad,\quad i\in S\setminus \{a\}\\
{\cal C}_{a\, b}^{\mu} &=& 
-\sum_{j\in S\setminus \{a\}}{\cal C}_{j\, b}^{\mu}\quad,\quad
{\cal V}_{b}^{\mu}=0\;.\nn
\end{eqnarray}
The vectors $W_i^{\mu}$ span the $(N-5)$-dimensional vector 
space $\mathrm{Ker}(G)$, the kernel of $G$. 
To construct a basis of $\mathrm{Ker}(G)$ 
we again use the vectors  $U_i^{(l= 1, \cdots N-5)}$
introduced in section \ref{subscalar}.
The vectors $W_i^{\mu}$ are 
then parametrised  as
\begin{equation}\label{e10}
W_i^{\mu} = \beta^{\mu}_{N-5} (K\cdot\delta V)_i + 
\sum_{l = 1}^{N-6} \beta^{ \mu}_{l} \,U_i^{(l)}\;.
\end{equation}
Substituting the parametrisation (\ref{e10}) into eq.~(\ref{e6b}) 
yields:
\begin{eqnarray}
\beta^{\mu}_{N-5} 
& = & 
\frac{1}{\delta V\cdot K \cdot\delta V}
\left( 
 \Delta_{ab}^{\mu} + \sum_{i,j\in S\setminus \{a\}}
 \delta  V_i\, H_{ij}\,\Delta_{j\,a}^{\mu}
\right)
\label{e11} \\
\beta^{ \mu}_{l=1, \cdots, N-6} 
& = & 
\mbox{arbitrary 4-vectors}\nn \;.
\end{eqnarray}
\noindent
If $N =6$, $G$ is not invertible,  
yet the solution (\ref{e9})-(\ref{e11}) is still unique, as it has to be since ${\cal S}$
is invertible. If $N \geq 7$, neither $G$ nor ${\cal S}$ are invertible. In
this case (\ref{eqCDEF}) still admits solutions, which are however no more
unique, but span the $(N-6)$-dimensional affine space defined by
(\ref{e9})-(\ref{e11}). It also has to be emphasised that the above construction
is equally valid if the external momenta become linearly dependent not due to
the fact that $N\geq 6$, but due to an exceptional kinematic configuration. 


\vspace{0.3cm}

\noindent
With relations (\ref{e7}) and (\ref{e9}) at hand, it is now easy to see
that for $N\geq 6$ one has 
\bea
\sum_{j\in S}\,{\cal C}_{j\,b}^{\mu}\, \Delta_{j\,a}^{\nu}&=&
-\frac{1}{2}\,g^{\mu\nu}_{[4]}\;\mbox{ and thus }\;
\caltf^{\mu \, \nu}_{a \, b}  =  0\;. \label{eqEQUATION1}
\eea
The subscript $``[4]"$ in $\caltf^{\mu \, \nu}_{a \, b}$ 
indicates that $\calt^{\mu \, \nu}_{a \, b}$ is 4-dimensional  
and not $n$-dimensional here.
Further, we already saw that 
\be
{\cal V}_{b}^{\mu} = 0 \;\mbox{ for } \; N\geq 6\;. 
\label{eqEQUATION2}
\ee 
Relations (\ref{eqEQUATION1}) and (\ref{eqEQUATION2}) have an
important consequence. They mean that $A^\mu_a$ in eq.\,(\ref{eqTRANSA}) 
is zero and thus  
they imply that no higher 
than $n$-dimensional $N$-point functions with $N\geq 6$ 
can be generated by the 
reduction: In eq.~(\ref{eqNEWIDEF2}), only the pinched terms
survive, leading to   
\begin{eqnarray}
I_N^{n,\mu_1\ldots\mu_r}(a_1,\ldots,a_r;S) & = & 
 - \sum_{j \in S}\calc_{ja_1}^{\mu_{1}} \, 
 I_{N-1}^{n,\mu_2\ldots\mu_r}(a_2,\ldots,a_r;S\setminus \{j\})\quad (N\geq 6)\;.
\label{TenRedNgeq6}
\end{eqnarray}
In this sense the tensor reduction of $N$-point integrals with $N\geq 6$ 
is trivial: Integrals with $N\geq 6$ can be reduced iteratively to 
5-point integrals, without generating higher dimensional remainders.
Therefore form factors for $N>5$ are not needed.

\vspace{0.3cm}

\noindent
The case $N=5$ needs special attention and was already
carefully analysed in the Feynman parameter approach \cite{Bern:1993kr}. 
We will give a full analysis of the 5-point case 
within our formalism in section \ref{absence-N-eqn5}. 

\vspace{3mm}

\noindent
In order to make contact to ref.~\cite{tenred1}, 
we note that the reduction is equivalent to applying first 
eq.~(32) and then eq.~(B.7) of \cite{tenred1}, but the method 
advocated here gives very naturally certain 
algebraic relations between reduction coefficients, 
which have been exploited in order 
to avoid Gram determinants (i.e. factors of $1/B$) as far as possible.

\section{Form factors for $N=3,4$ and algebraic representation of basis integrals}\label{algebraic-red}

In the following we will apply the formalism explained  
above to provide  explicit formulae for
all form factors present in the reduction of 3- and
4-point integrals. The form factors for $N=5$ 
will be given in section \ref{formfact_fivep} and the 2-point 
functions will be listed in appendix \ref{formfacsti12}.
The form factors will contain scalar integrals with non-trivial 
numerators.
In this representation no inverse Gram determinants are
present, which serves as an ideal  starting point for 
a numerical approach\footnote{
Parameter integrals with non-trivial numerators, corresponding to tensor integrals, 
are  also dealt with in~\cite{Kurihara:2005at}, where 
$n$-dimensional tensor 3-point and 4-point functions  with massless propagators
are expressed by hypergeometric series expansions.}.

\vspace{0.3cm}

\noindent
Nonetheless we will also provide an {\em analytical} 
representation
for the basis integrals in terms of scalar integrals 
with trivial numerators only. The price to pay in this case are  
inverse Gram determinants. 
For the calculation of a cross section this representation
is numerically stable in most parts of the phase space.
However, numerical problems can arise 
at the kinematical boundaries.
In section \ref{basicblocks} we will propose a numerical 
evaluation of the  basis integrals which also works for exceptional 
kinematical configurations.

\subsection{Three-point integrals}\label{i3mass}

The form factors for the 3-point tensor integrals, 
following directly from eqs.~(\ref{fofageneral}) and 
(\ref{eq32}), are given by
\begin{eqnarray}
A^{3,0}(S) & = & I_3^{n}(S) \nn\\
A^{3,1}_{l}(S) & = & -I_3^{n}(l;S) \nn\\
B^{3,2}(S) & = & - \frac{1}{2} \, I_3^{n+2}(S) \nn\\
A^{3,2}_{l_1 l_2}(S) & = & I_3^{n}(l_1,l_2;S) \nn\\
B^{3,3}_l(S) & = & \frac{1}{2} \, I_3^{n+2}(l;S) \nn\\
A^{3,3}_{l_1 l_2 l_3}(S) & = & 
- I_3^{n}(l_1,l_2,l_3;S) 
\label{eqA33}
\end{eqnarray}
If one would like to express the integrals with 
nontrivial numerators above in terms of scalar integrals only,
the  formulae given below in eqs.~(\ref{i3n1}) to (\ref{i3np23r2}) 
can be used if ${\cal S}$ is regular. 
The singular cases occur if either the 
3-point function is IR divergent or one hits an anomalous threshold.
Anomalous thresholds appear in scattering processes with unstable 
external particles. In very special kinematic situations all
internal particles in a given diagram can go on-shell, which can cause
integrable singularities inside the phase space \cite{bjorkendrell}.
One can show that in the 3-point case $\det({\cal S})=0$ is a necessary 
condition for an anomalous threshold.
For the case where IR divergences are present, 
we give a complete list
of 3-point integrals with massless propagators in appendix \ref{3pointIR}. 
\bea
I_3^{n}(l_1;S)&=&\frac{b_{l_1}}{B}\left[
I_3^{n}(S)-\sum_{j\in S}
b_j\,I_2^{n}(S\setminus\{j\})\right]+
\sum_{j\in S}{\cal S}^{-1}_{l_1 j}I_{2}^{n}(S\setminus\{j\})\label{i3n1}\\
I_3^{n}(l_1,l_2;S)&=&-{\cal S}^{-1}_{l_1 l_2}
I_{3}^{n+2}(S)+
b_{l_1}(n-1)I_3^{n+2}(l_2;S)+
\sum_{j\in S}{\cal S}^{-1}_{l_1 j}\,I_2^{n}(l_2;S\setminus\{j\})\label{i3n2}\\
I_3^{n}(l_1,l_2,l_3;S)&=&
-{\cal S}^{-1}_{l_1 l_2}
I_{3}^{n+2}(l_3;S)-{\cal S}^{-1}_{l_1 l_3}
I_{3}^{n+2}(l_2;S)+
n\,b_{l_1}\,I_3^{n+2}(l_2,l_3;S)\nonumber\\
&&+
\sum_{j\in S}{\cal S}^{-1}_{l_1 j}
\,I_2^{n}(l_2,l_3;S\setminus\{j\})\label{eqA33ints}\\
&&\nn\\
I_3^{n+2}(S)&=&\frac{1}{B}\frac{1}{(n-2)}\,\Bigl[I_3^{n}(S)-\sum_{l\in S}
b_l\,I_2^{n}(S\setminus\{l\})\Bigr] \label{i3np2} \\
I_3^{n+2}(l_1;S) & = & \frac{1}{B} \; \Bigl[ b_{l_1} \, 
I_3^{n+2}(S)  + \frac{1}{n-1} \, \sum_{j \in S}  
\icalst{j \, l_1} \, 
I_2^{n}(S\setminus\{j\}) \nonumber \\
& & \mbox{} - \frac{1}{n-1} \, \sum_{j \in S} b_j \, 
I_2^{n}(l_1;S\setminus\{j\}) \Bigr]
\label{i3np23r1}\\
I_3^{n+2}(l_1,l_2;S) & = & \frac{1}{n \, B} \; 
\Bigl[ b_{l_1} \, 
I_3^{n+2}(l_2;S) + b_{l_2} \, I_3^{n+2}(l_1;S)   +
I_3^{n}(l_1,l_2;S)\nn \\
& & \mbox{} - \sum_{j \in S} b_j \,
I_2^{n}(l_1,l_2;S\setminus\{j\}) \Bigr]\label{i3np23r2}
\eea
The two-point functions $I_2^n$ are given in 
appendix \ref{formfacsti12}. The scalar three-point function 
$I_3^{n}(S)$ is well known, 
see for example \cite{Bern:1993kr,Davydychev:1999mq,Davydychev:2000na}.

\vspace{0.3cm}

\noindent
By iterating the above formulae 
one arrives at a representation in terms of scalar integrals 
with  trivial numerator.
As a result of the iteration,  inverse Gram determinants 
up to the third power occur. 
To improve the numerical stability, the 
bracketing of the terms as given by 
eqs.~(\ref{i3n1}) and (\ref{i3np2}) to (\ref{i3np23r2}) should be respected.

\subsection{Four-point integrals}\label{formfacbox}

All form factors of the 4-point tensor integrals are expressed 
in terms 
of $(n+2)$- and $(n+4)$-dimensional scalar box integrals and 
$n$- and $(n+2)$-dimensional triangle integrals,  
with up to three Feynman parameters in the numerator.
\begin{eqnarray}
A^{4,0}(S) & = & B \, I^{n+2}_4(S) + \sum_{j \in S} \, 
b_j \,I_3^{n}(S\setminus\{j\}) \label{eqA40}\\
A^{4,1}_{l}(S) & = & - b_l \, I^{n+2}_4(S) - \sum_{j \in S} \, 
\icalst{j \, l} \,I_3^{n}(S\setminus\{j\}) \label{eqA41}\\
B^{4,2}(S) & = & - \frac{1}{2} \, I_4^{n+2}(S) \\
A^{4,2}_{l_1 l_2}(S) & = & b_{l_1} \, I^{n+2}_4(l_2;S) + 
b_{l_2} \, 
I^{n+2}_4(l_1;S) - \icalst{l_1 \, l_2} \, 
I^{n+2}_4(S)\nn \\
& & \mbox{} + \frac{1}{2} \, \sum_{j \in S} \, 
\left[ \icalst{j \, l_2} \, 
I_3^{n}(l_1;S\setminus\{j\}) + \icalst{j \, l_1} \, 
I_3^{n}(l_2;S\setminus\{j\}) \right] 
\label{eqA42}\\
&&\nn\\
&&\nn\\
B^{4,3}_l(S) & = & \frac{1}{2} \, I^{n+2}_4(l;S)  \\
A^{4,3}_{l_1 l_2 l_3}(S) & = & \frac{2}{3} \, \left[ \icalst{l_2 \, l_3} \, 
I^{n+2}_4(l_1;S) + \icalst{l_1 \, l_3} \, I^{n+2}_4(l_2;S) 
+ \icalst{l_1 \, l_2} \, I^{n+2}_4(l_3;S) \right] \nn \\
& & \mbox{} - \left[ b_{l_1} \, I^{n+2}_4(l_2,l_3;S) + b_{l_2} \, 
I^{n+2}_4(l_1,l_3;S)  + b_{l_3} \, I^{n+2}_4(l_1,l_2;S) \right] 
\nonumber \\
& & - \frac{1}{3} \, \sum_{j \in S} \, \left[ \icalst{j \, l_1} 
\,  I_{3}^{n}(l_2,l_3;S\setminus\{j\}) 
 + \icalst{j \, l_2} \,
I_{3}^{n}(l_1,l_3;S\setminus\{j\})\right.\nn\\
&& \left. \qquad\quad + \, \icalst{j \, l_3}  
\, I_{3}^{n}(l_1,l_2;S\setminus\{j\}) \right] \\
&&\nn\\
C^{4,4}(S) & = & \frac{1}{4} \, I_4^{n+4}(S) \\
B^{4,4}_{l_1 l_2}(S) & = & - \frac{1}{2}\,I^{n+2}_4(l_1,l_2;S) \\
A^{4,4}_{l_1 l_2 l_3 l_4}(S) & = & 
f^{4,4}(l_1, l_2; l_3, l_4)+
f^{4,4}(l_1, l_3; l_2, l_4)+
f^{4,4}(l_1, l_4; l_3, l_2)\nn\\
&&+
f^{4,4}(l_2, l_3; l_1, l_4)+
f^{4,4}(l_2, l_4; l_3, l_1)+
f^{4,4}(l_3, l_4; l_1, l_2)\nn\\
&&+g^{4,4}(l_1; l_2, l_3, l_4)+
g^{4,4}(l_2; l_1, l_3, l_4)
\nn\\
&&+
g^{4,4}(l_3; l_2, l_1, l_4)+
g^{4,4}(l_4; l_2, l_3, l_1)\nn\\
&&\nn\\
f^{4,4}(l_1, l_2; l_3, l_4)&=&
- \frac{1}{2} \,  \icalst{l_1 \, l_2} 
\, I^{n+2}_4(l_3,l_4;S)  \nonumber \\
&&\nn\\
g^{4,4}(l_1; l_2, l_3, l_4)& =& 
 b_{l_1} \, I^{n+2}_4(l_2,l_3,l_4;S) 
+ \frac{1}{4}\sum_{j \in S} \,  \icalst{j \, l_1} \,  
I_{3}^{n}(l_2,l_3,l_4;S\setminus\{j\}) 
\label{eqA44}
\end{eqnarray}
There are six different combinations of  terms 
$f^{4,4}(\ldots)$ and four different 
combinations of  terms  $g^{4,4}(\ldots)$.
The combinations are imposed by the symmetry of 
these objects and   
represent all different {\em distinguishable} index distributions.
The  object $f^{4,4}$ is symmetric in the first  and last
two indices. The symmetry in the first two indices is manifest,
as $\icalst{l_1 \, l_2}$ is symmetric. The symmetry in the last two indices
is only induced when the form factors are combined with 
formulae (\ref{fofageneral}), because the summations symmetrize 
these indices. The  object $g^{4,4}$ is symmetric in the  last
three indices in the same sense.

\vspace{0.3cm}

\noindent
The form factors given above are free from any inverse Gram determinant
and contain only $n$-dimensional 3-point  integrals with
up to three Feynman parameters in the numerator, 
$(n+2)$-dimensional 3-point  integrals with
maximally one Feynman parameter in the numerator, and
$(n+2)$- and $(n+4)$-dimensional 4-point  functions with up to
three respectively one Feynman parameter in the numerator.
These integrals form our {\it basis set}, i.e. the endpoints of 
the reduction. We will see that for $N>4$, no additional 
basis integrals are required. 
This basis set, being free from inverse Gram determinants, 
serves as an ideal starting point for a numerical evaluation.

\vspace{0.3cm}

\noindent
Of course, further algebraic reduction down to integrals 
with no Feynman parameters in the numerator is also possible. 
The following formulae and the results for the 3-point integrals
given above allow to achieve an algebraic representation
of the  form factors in terms of {\em scalar} integrals, 
i.e. integrals with no Feynman parameters in the numerator, only.
Again, such a representation produces $1/B$ terms, 
which is equivalent to the presence of inverse Gram determinants.
Exploiting the relations given in appendix \ref{formrel}, 
the following results can be derived:
\begin{eqnarray}
I^{n+2}_4(l;S) & = & \frac{1}{B}  \Biggl\{  b_{l} \, I^{n+2}_4(S) 
+ \frac{1}{2}  \sum_{j \in S} \icalst{j \, l}  I^{n}_3(S\setminus\{j\}) 
- \frac{1}{2}  \sum_{j \in  S} b_j \,
I^{n}_3(l;S\setminus\{j\})  \Biggr\}\;
\label{eq1RELA}
\end{eqnarray}
\begin{eqnarray}
I^{n+2}_4(l_1,l_2;S)&=&
\frac{2}{3 \, B} \, \Biggl\{ \, b_{l_1} \, I^{n+2}_4(l_2;S) + 
b_{l_2} \, I^{n+2}_4(l_1;S) \label{eq2RELA}
 \\& &  
- \frac{1}{2} \, \icalst{l_1 \, l_2} \, I^{n+2}_4(S) 
+ \frac{1}{4} \, \sum_{j \in S} \icalst{j \, l_2} \, I^{n}_3(l_1;S\setminus\{j\}) 
   \nonumber \\& & \mbox{} 
+ \frac{1}{4} \, \sum_{j \in S} \icalst{j \, l_1} \, I^{n}_3(l_2;S\setminus\{j\}) 
- \frac{1}{2} \, \sum_{j \in  S} b_j \, I^{n}_3(l_1,l_2;S\setminus\{j\}) \, \Biggr\}
\nn
\end{eqnarray}
\begin{eqnarray}
I^{n+2}_4(l_1,l_2,l_3;S) & =&  \frac{1}{2 \, B} \, 
\Biggl\{ \, b_{l_3} \, I^{n+2}_4(l_1,l_2;S)   + b_{l_2} \, 
I^{n+2}_4(l_1,l_3;S)+ 
b_{l_1} \, I^{n+2}_4(l_2,l_3;S) \label{eq3RELA}\\
& &  - \frac{1}{3} \Bigl( \icalst{l_1 \, l_2} \, I^{n+2}_4(l_3;S) + 
\icalst{l_1 \, l_3} \, I^{n+2}_4(l_2;S) 
+ \icalst{l_2 \, l_3} \, I^{n+2}_4(l_1;S) \Bigr) \nonumber \\
& & + \frac{1}{6} \Bigl( \sum_{i \in S} 
\icalst{i \, l_3} \, I^{n}_3(l_1,l_2;S\setminus\{i\})  
+  \sum_{i \in S} \icalst{i \, l_2} \, 
I^{n}_3(l_1,l_3;S\setminus\{i\}) 
\nonumber \\& & 
+ \sum_{i \in S} \icalst{i \, l_1} \, 
I^{n}_3(l_2,l_3;S\setminus\{i\}) \Bigr)
- \frac{1}{2}  \sum_{i \in  S} b_i \, 
I^{n}_3(l_1,l_2,l_3,S\setminus\{i\}) \, \Biggr\}\nn\\
&&\nn\\
&&\nn\\
I^{n+4}_4(S) & = & \frac{1}{(n-1) \, B} \, \Biggl\{ 
\, I^{n+2}_4(S)  - 
\sum_{j \in  S} b_j \, I^{n+2}_3(S\setminus\{j\}) \Biggl\}
\label{eq4RELA}
\end{eqnarray}
\begin{eqnarray}
I^{n+4}_4(l;S) & = & \frac{1}{n \, B} \,
\Biggl\{  b_{l} \, 
I^{n+4}_4(S) +  I^{n+2}_4(l;S) 
-\sum_{j \in S} \, b_j \,I^{n+2}_3(l;S\setminus\{j\})  
\Biggl\}
\label{eq5RELA}
\end{eqnarray}
Applying these formulae and eq.~(\ref{i3np23r1}), 
an algebraic representation of the form 
factors for 4-point tensor integrals
in terms of 3-point functions and six-dimensional integrals without 
Feynman parameters in the numerator
is achieved. The six-dimensional scalar  box functions 
for massless internal lines are well known and can be found 
for example in \cite{Bern:1993kr,yukawa}. 
In section \ref{basicblocks}, 
the behaviour of the form factors for $B\to 0$
in this algebraic representation will be compared  to the one 
given by eqs.~(\ref{eqA41}) to (\ref{eqA44}), where inverse Gram determinants 
have been avoided.

\section{Form factors for $N = 5$}\label{formfact_fivep}

In this section we will give numerically stable form factor
representation for five point functions. 
We have already seen that the reduction of tensor integrals 
for $N\geq 6$ does not produce higher dimensional remainders, 
see eq.~(\ref{TenRedNgeq6}).
The case $N=5$ is, in a sense, the most complicated one. 
The derivation is rather technical and can be found in 
appendix~\ref{absence-N-eqn5}.
For five-point integrals up to rank three,
the absence of higher dimensional five-point integrals already 
has been demonstrated by explicit calculation in \cite{Bern:1993kr}. 
In appendix \ref{absence-N-eqn5}, we show that these integrals 
drop out for 5-point integrals of {\em arbitrary} rank 
and we derive representations which are free from  both
higher dimensional five-point integrals {\em and} inverse Gram 
determinants {\em at the same time}. To the best of our knowledge, 
such a representation has not been given in the literature before.


By  application of eq.~(\ref{eqONEOVERB15}) 
and reduction of $n$-dimensional box integrals $I_4^n$ to
3-point functions and higher dimensional box integrals, we obtain, 
dropping ${\cal O}(\eps)$ terms:
\begin{eqnarray}
A^{5,0}(S) & = & \sum_{j \in S} \, b_j \, B^{\{j\}} \, 
I^{n+2}_4(S\setminus\{j\}) + \sum_{j \in S} \, \sum_{k \in S\setminus\{j\}} 
\, b_j \, b_k^{\{j\}} \,I_3^{n}(S\setminus\{j,k\}) \label{eqA50}\\
A^{5,1}_{l}(S) & = & - \sum_{j \in S} \, \icalst{j \, l} \, 
B^{\{j\}} \, I^{n+2}_4(S\setminus\{j\}) - \sum_{j \in S} \, 
\sum_{k \in S\setminus\{j\}} \, \icalst{j \, l} \, b_k^{\{j\}} \, 
I_3^{n}(S\setminus\{j,k\}) \\
B^{5,2}(S) & = & - \frac{1}{2} \, \sum_{j \in S} \, b_j \, 
I_4^{n+2}(S\setminus\{j\}) \\
A^{5,2}_{l_1 l_2}(S) & = & \sum_{j \in S} \, 
\left( \, \icalst{j \, l_1} \, b_{l_2} + \icalst{j \, l_2} \, 
b_{l_1} - 2 \, \icalst{l_1 \, l_2} \, b_{j} + b_{j} \, 
\micals{j}_{l_1 \, l_2}  \right) \, I^{n+2}_4(S\setminus\{j\}) \nonumber \\
& & \mbox{} + \frac{1}{2} \, \sum_{j \in S} \, 
\sum_{k \in S\setminus\{j\}} \, \left[ \icalst{j \, l_2} \, 
\micals{j}_{k \, l_1}  + \icalst{j \, l_1} \, \micals{j}_{k \, l_2}  
\right]
I_3^{n}(S\setminus\{j,k\}) \\
&&\nn\\
B^{5,3}_l(S) & = & \frac{1}{3} \, \sum_{j \in S} \, \left( \, b_j \, 
I^{n+2}_4(l;S\setminus\{j\})  + \frac{1}{2} \, \icalst{j \, l} \, 
I^{n+2}_4(S\setminus\{j\}) \right) \\
A^{5,3}_{l_1 l_2 l_3}(S) & = & \frac{2}{3} \, \sum_{j \in S} \, 
\left[ \, I^{n+2}_4(l_3;S\setminus\{j\})\right. 
\nonumber \\
& & \mbox{} \times \left. \left( \, 2 \, \icalst{l_1 \, l_2} 
\, b_{j} - \icalst{j \, l_1} \, b_{l_2} - \icalst{j \, l_2} \, 
b_{l_1} - b_{j} \, \micals{j}_{l_1 \, l_2}  \right) + 
l_2 \leftrightarrow l_3 + l_1 
\leftrightarrow l_3 \right] \nonumber \\
& & \mbox{} + \frac{1}{3}\,\sum_{j \in S} \, I^{n+2}_4(S\setminus\{j\})\,
\left[
 \icalst{j \, l_3} \, \micals{j}_{l_1 \, l_2} 
  +  \icalst{j \, l_1} \, \micals{j}_{l_2 \, l_3}  
 + \icalst{j \, l_2} \, \micals{j}_{l_1 \, l_3}  \right] \nonumber \\
& & - \frac{1}{6} \, \sum_{j \in S} \, \sum_{k \in S\setminus\{j\}} \,
 \left[  
 \, I_{3}^{n}(l_1;S\setminus\{j,k\}) \,\left( \icalst{j \, l_3} \, 
\micals{j}_{k \, l_2}  
+ \icalst{j \, l_2} \, \micals{j}_{k \, l_3}  \right) \right.\nn\\
&& + \left. l_1 \leftrightarrow l_2 + l_1 \leftrightarrow l_3 \right] \\
&&\nn\\
C^{5,4}(S) & = & \frac{1}{4} \, \left( 1 + \frac{n-4}{3} \right) \, 
\sum_{j \in S} \, b_j\,I_4^{n+4}(S\setminus\{j\}) \\
B^{5,4}_{l_1 l_2}(S) & = & \frac{1}{4} \, \sum_{j \in S} \, \Biggl[ 
\, \left( 1 + \frac{n-4}{3} \right) \, \left( \, 
\icalst{l_1 \, l_2} \, b_{j} - \frac{1}{2}\icalst{j \, l_1} \, b_{l_2} - 
\frac{1}{2}\icalst{j \, l_2} \, b_{l_1} 
\right) \, I_4^{n+4}(S\setminus\{j\})  \nonumber\\
& &  -  b_j \,I^{n+2}_4(l_1,l_2;S\setminus\{j\})   \nonumber\\
& & \mbox{} - \frac{1}{2} \, \icalst{j \, l_1}  \, I^{n+2}_4(l_2;S\setminus\{j\})
- \frac{1}{2} \, \icalst{j \, l_2}  \, I^{n+2}_4(l_1;S\setminus\{j\})   \Biggr] \\
%
%
A^{5,4}_{l_1 l_2 l_3 l_4}(S) & = &  \frac{1}{4} \, 
\sum_{j \in S} \, \Biggl[
f^{5,4}(l_1, l_2; l_3, l_4)+
f^{5,4}(l_1, l_3; l_2, l_4)+
f^{5,4}(l_1, l_4; l_2, l_3)\nn\\
&&+
f^{5,4}(l_2, l_3; l_1, l_4)+
f^{5,4}(l_2, l_4; l_1, l_3)+
f^{5,4}(l_3, l_4; l_1, l_2)
\nn\\
&&+
g^{5,4}(l_1; l_2, l_3, l_4)+
g^{5,4}(l_2; l_1, l_3, l_4)\nn\\
&&+
g^{5,4}(l_3; l_1, l_2, l_4)+
g^{5,4}(l_4; l_1, l_2, l_3)
\Biggr]
\end{eqnarray}

\begin{eqnarray}
f^{5,4}(l_1, l_2; l_3, l_4)&=&- 2  \, 
I^{n+2}_4(l_1,l_2;S\setminus\{j\}) \, \left( \, 2 \, 
\icalst{l_3 \, l_4} \, b_{j} - 
\icalst{j \, l_3} \, b_{l_4} - \icalst{j \, l_4} \, b_{l_3} - 
b_{j} \, 
\micals{j}_{l_3 \, l_4}  \right) \nonumber \\
& & \mbox{} + \frac{1}{3} \, 
\sum_{k \in S\setminus\{j\}}  \, 
I_{3}^{n}(l_1,l_2;
S\setminus\{j,k\})\, \left( \icalst{j \, l_3} \, 
\micals{j}_{k \, l_4}  + \icalst{j \, l_4} \, \micals{j}_{k \, l_3} 
\right)  \nonumber \\
& &  \nonumber \\
g^{5,4}(l_1; l_2, l_3, l_4)&= &    - \frac{2}{3} \, 
I^{n+2}_4(l_1;S\setminus\{j\})\, \left( \, 
\icalst{j \, l_4} \, \micals{j}_{l_2 \, l_3} + \icalst{j \, l_3} \, 
\micals{j}_{l_2 \, l_4}  
+ \icalst{j \, l_2} \, \micals{j}_{l_3 \, l_4}  \right)  \nn\\&&
\end{eqnarray}
The combinations of $f^{5,4}(\ldots)$ and $g^{5,4}(\ldots)$
appearing in $A^{5,4}_{l_1 l_2 l_3 l_4}(S)$ are imposed by the 
symmetry of these objects and   
represent all different {\em distinguishable} index distributions.
The  object $f^{5,4}$ is symmetric in the first  and last
two indices and the  object $g^{5,4}$ is symmetric in the  last
three indices when combined with formulae (\ref{fofageneral}).
\begin{eqnarray}
C^{5,5}_l(S) & = & \frac{1}{5} \, \sum_{j \in S} \, \Biggl[
 \, -  \left( 1 + \frac{n-4}{4} \right) \, 
  b_j\,I_4^{n+4}(l;S\setminus\{j\})  - \frac{1}{4} \, 
 \icalst{j \, l} \, I_4^{n+4}(S\setminus\{j\}) \Biggr] \nn\\
B^{5,5}_{l_1 l_2 l_3}(S) & = & \frac{1}{5} \, \sum_{j \in S} \, \Biggl[ 
\, \Biggl\{  \frac{n}{4}\,I_4^{n+4}(l_1;S\setminus\{j\}) \nonumber \\
& & \mbox{} \times \left( \frac{1}{2} \icalst{j \, l_3} \, b_{l_2} + \frac{1}{2}\icalst{j \, l_2} \, b_{l_3} -  \icalst{l_2 \, l_3} \, b_{j}  \right) + 
 l_1 \leftrightarrow l_2 + l_1 \leftrightarrow l_3 \Biggr\} \nonumber \\
& & \mbox{} +  b_j\,I_4^{n+2}(l_1,l_2,l_3;S\setminus\{j\})  \nonumber \\
& & \mbox{} + \Biggl\{ \frac{1}{2} \, I_4^{n+2}(l_1,l_2;S\setminus\{j\}) \, \icalst{j \, l_3} + l_1 \leftrightarrow l_3 + l_2 \leftrightarrow l_3 \Biggr\}   \Biggr] \\
%
%
A^{5,5}_{l_1 l_2 l_3 l_4 l_5}(S) & = &  \frac{1}{5} \,\sum_{j\in S} \,\Biggl[ 
\left( f^{5,5}(l_1, l_2, l_3; l_4, l_5)+ 9 \,\,\mbox{combinations}\right)
\nn\\
&&\qquad + \left( g^{5,5}(l_1, l_2; l_3, l_4, l_5)+9 \,\,\mbox{combinations} \right)
\Biggl] 
\eea
\bea
&& f^{5,5}(l_1, l_2, l_3; l_4, l_5)=
- \frac{1}{4} \, \sum_{k \in S\setminus\{j\}} \,  I_{3}^{n}(l_1,l_2,l_3;S\setminus\{j,k\}) 
\, \left( \icalst{j \, l_5} \, \micals{j}_{k \, l_4} \, 
 + \icalst{j \, l_4} \, \micals{j}_{k \, l_5} \, 
\right)  \nonumber \\ &&
\qquad \qquad\qquad\qquad - 2 \,  
I^{n+2}_4(l_1,l_2,l_3;S\setminus\{j\})
\, \left( \,  \icalst{j \, l_5} \, b_{l_4} + 
\icalst{j \, l_4} \, b_{l_5} - 2 \, \icalst{l_4 \, l_5} \, b_{j} + 
b_{j} \, \micals{j}_{l_4 \, l_5}  \right)  \nonumber 
\eea

\bea
&&g^{5,5}(l_1, l_2; l_3, l_4, l_5)=
  \frac{1}{2} 
\, I^{n+2}_4(l_1,l_2;S\setminus\{j\})  \nonumber \\&& \qquad\qquad\qquad\qquad\qquad \times
\left( \, \icalst{j \, l_5} \, \micals{j}_{l_3 \, l_4} 
+  \icalst{j \, l_4} \, \micals{j}_{l_3 \, l_5} \, 
+ \icalst{j \, l_3} \, \micals{j}_{l_4 \, l_5} 
\,  \right)  
\label{eqA55}
\end{eqnarray}
The ten different combinations of the functions 
$f^{5,5}(\ldots)$ and 
$g^{5,5}(\ldots)$
are  all different  distinguishable index distributions.
The function $f^{5,5}$ is symmetric in the first three and last
two arguments and the function $g^{5,5}$ is symmetric in the first two and last
three arguments when combined with formulae (\ref{fofageneral}).

\section{Numerical evaluation of the basis integrals}
\label{basicblocks}

In this section we present a method to evaluate the basis integrals of
our reduction formalism numerically. 
The method is an alternative to an approach proposed 
earlier\,\cite{nikolas}, which would also be viable, 
but which needs more analytical input.
The  method presented here allows to deal easily with the 
case of complex masses, which is necessary if unstable
particles are present in the loop.

\vspace{0.3cm}

\noindent
First we will explain the mathematical details of the contour deformation.
Then we will show a comparison between the numerical and the algebraic
implementation of some basis integrals.

\subsection{Contour deformation of parameter integrals}\label{numeric}

The method which will be presented here is based on contour deformation 
in Feynman parameter space.
As explained above, in our formalism  it is 
sufficient to evaluate the following functions, which are the  
endpoints of our reduction  
\begin{eqnarray}\label{basic_scalar34}
I_3^{4}(j_1,\dots ,j_r) &=&   
-\int_{0}^{1} \prod_{i=1}^3 dz_i  \,\delta(1-\sum\limits_{l=1}^3 z_{l})
\frac{z_{j_1}\dots z_{j_r}}{(-z\cdot {\cal S}\cdot z/2 -i\delta)}\;,\nonumber \\
I_3^{n+2}(j_1) &=&   
-\Gamma(\epsilon)\int_{0}^{1} \prod_{i=1}^3 dz_i  \,\delta(1-\sum\limits_{l=1}^3 z_{l})
\frac{z_{j_1}}{(-z\cdot {\cal S}\cdot z/2 -i\delta)^\eps}\;,
\nonumber \\
I_4^{6}(j_1,\dots ,j_r) &=& 
\int_{0}^{1} \prod_{i=1}^4 dz_i \,\delta(1-\sum\limits_{l=1}^4 z_{l})
\frac{z_{j_1}\dots z_{j_r}}{(-z\cdot {\cal S}\cdot z/2 -i\delta)} \;, 
\nonumber \\
I_4^{n+4}(j_1,\dots ,j_r) &=& \Gamma(\epsilon)
\int_{0}^{1} \prod_{i=1}^4 dz_i \,\delta(1-\sum\limits_{l=1}^4 z_{l})
\frac{z_{j_1}\dots z_{j_r}}{(-z\cdot {\cal S}\cdot z/2
-i\delta)^\epsilon} \;, 
\end{eqnarray}
together with integrals of the same type, but with no Feynman 
parameters in the numerator, and two-point functions.
We will find  numerically stable representations of these integrals
as special cases of a completely general derivation which is valid 
for scalar integrals of the form 
\begin{eqnarray}\label{basic_scalar}
I_N^{D}(j_1,\dots ,j_r) &=&  (-1)^N\Gamma(N-\frac{D}{2}) 
\int_{0}^{1}  \prod_{i=1}^N dz_i\,\delta(1-\sum\limits_{l=1}^N z_{l})
\frac{z_{j_1}\dots z_{j_r}}{(-z\cdot {\cal S}\cdot z/2 -i\delta)^{N-D/2}} \nonumber \\&&
\end{eqnarray}
in the case when no IR/UV divergences are present. The 
IR/UV singular cases are discussed below.
\vspace{0.3cm}

\noindent
For loop calculations with unstable particles, it is necessary that internal 
propagators can be defined
with complex masses, ${\cal M}^2_j = M^2_j - i\,M_j\Gamma_j$. 
The denominators
of the integrands in eq.~(\ref{basic_scalar}) are defined accordingly 
by changing the quadratic form to
\begin{eqnarray}
-\frac{1}{2} z\cdot {\cal S}\cdot z - i\delta  \to -\frac{1}{2} z\cdot {\cal S}\cdot z 
- i \left(\sum\limits_{j=1}^{N} z_j\right)\left(\sum\limits_{j=1}^{N} 
z_j M_j\Gamma_j\right)
\end{eqnarray}
The finite width improves the stability of a numerical integration but is not
sufficient to guarantee a stable evaluation. We construct a contour
deformation in those parameter integrals which belong to propagators 
with zero width only.  This has technical reasons which will become clear below.

\vspace{0.3cm}

\noindent
In a first step we eliminate the delta function in eq.~(\ref{basic_scalar}) by 
decomposing the integration region into $N$ sectors where in each sector, one
of the Feynman parameters is larger than all the others. 
\begin{eqnarray}
1 = \sum\limits_{l=1}^N \theta( z_l > z_1,\dots,z_{l-1},z_{l+1},\dots,z_N )
\end{eqnarray}
Note that the splitting into $N$ sectors can be avoided by a clever choice
of Feynman parametrisation adapted to the special case at hand. However, as we 
are interested in a general, automated approach, we do not pursue this option.
The basic integral decays into $N$ sector integrals $J_l$
\begin{eqnarray}
I_N^D(j_1,\dots ,j_r) = (-1)^N \Gamma(N-D/2)\,\sum\limits_{l=1}^N  
J_l(N,D,j_1,\dots ,j_r)\;.
\end{eqnarray}
The latter can be written as integrals over the $(N-1)$-dimensional 
unit hypercube. Focusing on one sector, say sector $l$, and introducing the 
vector $\vec T=(t_1,\dots,t_{l-1},1,t_{l},\dots,t_{N-1})$ which defines the $N-1$ 
coordinates $t_1,\dots , t_{N-1}$ of the unit hypercube, one gets:
\begin{eqnarray}
J_l(N,D,j_1,\dots ,j_r) = \int\limits_0^1 \prod\limits_{l=1}^{N-1}dt_l \;
\Bigl( \sum\limits_{j=1}^N T_j \Bigr)^{N-D-r}
\;\frac{T_{j_1}\dots T_{j_r}}{ \Bigl( -T\cdot {\cal S}\cdot T/2 - i\delta \Bigr)^{N-D/2}}
\end{eqnarray}
The denominator becomes singular if the quadratic form 
$Q(\vec{t}\,)=- T\cdot {\cal S}\cdot T/2$ approaches zero inside the integration region, 
explicitly:
\begin{eqnarray}\label{sing_surface}
Q(\vec{t}\,) = \frac{1}{2} \sum\limits_{j,k=1}^{N-1} X_{jk} t_j t_k + 
\sum\limits_{j=1}^{N-1} Y_{j} t_j + Z = 0\;,
\end{eqnarray}
where $X_{jk}$, $Y_{j}$ and $Z$ are defined through ${\cal S}_{jk}$. 
The singularity is protected by the $i\delta$ prescription or a finite width
in some propagators. Viewing the 
integration volume as an $(N-1)$-dimensional hypercontour in a space with $N-1$ 
complex dimensions, we are looking for a contour deformation which leads to a 
smooth and bounded integrand without intersecting the singularity hypersurface 
defined by eq.~(\ref{sing_surface}). 
For the analytic continuation of $Q(\vec{t})\to Q(\vec{x})$
we make the ansatz\,\cite{Soper:1999xk} $\vec{x}= \vec{t} -i\; \vec{\tau}$. Now:
\begin{eqnarray}
Q(\vec{x}) = Q(\vec{t}\,) 
- \frac{1}{2} \sum\limits_{j,k=1}^{N-1} X_{jk} \tau_j \tau_k
- i\sum\limits_{k=1}^{N-1} \tau_k \sum\limits_{j=1}^{N-1} ( X_{jk} t_j +
Y_k  )
\end{eqnarray}
This suggests the following choice for the deformation vector $\vec\tau$:
\begin{eqnarray}\label{cont_def}
\vec{x}( \vec t) &=& \vec{t} - i\;  \vec{\tau}(\vec{t})\nonumber\\
\tau_k &=& \left\{
\begin{array}{ll}
\lambda t_k^\alpha (1-t_k)^\beta 
\sum\limits_{j=1}^{N-1} ( X_{jk} t_j +Y_k  ) & \mbox{if } \, \Gamma_k=0
\label{ansatz} \\
0 & \mbox{if } \,\Gamma_k\neq 0
\end{array}\right.
\end{eqnarray}

\vspace{0.3cm}

\noindent
Note that the implementation of such a contour deformation also for the 
parameters which correspond to a non-vanishing width would not lead to the 
desired result, as Im$(Q)<0$ can not be guaranteed then. 
For $\lambda$, $\alpha$, $\beta > 0$ the deformation moves the integration contour 
away from the poles, i.e. Im$(Q)<0$, without causing any harm at the boundaries.

\vspace{0.3cm}

\noindent
The invariance under diffeomorphisms of the contour, ${\cal C}_\lambda$, means
\begin{eqnarray}
\int\limits_{{\cal C}_0} \prod\limits_{l=1}^{N-1}dx_l\, f(x) = 
\int\limits_{{\cal C}_\lambda} \prod\limits_{l=1}^{N-1}dx_l \, f(x) 
\end{eqnarray}
or in the given parametrisation:
\begin{eqnarray}
\int\limits_{0}^1 \prod\limits_{l=1}^{N-1}dt_l \; f(\vec{t}) = 
\int\limits_{0}^1 \prod\limits_{l=1}^{N-1}dt_l 
\;\det\left( \frac{\partial x_i}{\partial t_j}\right)\; 
f(\vec{t} - i\vec{\tau}(\vec{t}\,)) 
\end{eqnarray}
The Jacobian is defined by
\begin{eqnarray}
\frac{\partial x_l}{\partial t_j} = \delta_{lj} - i \,
\lambda \,t_l^{\alpha-1} (1-t_l)^{\beta-1}
\Bigl[  \delta_{lj} [ \alpha (1-t_l) - \beta t_l ] 
\Bigl( \sum\limits_{k=1}^{N-1} X_{lk} t_k + Y_l \Bigr) + t_l  (1-t_l) X_{lj} \Bigr]
\bar\delta(\Gamma_l)\nn
\end{eqnarray}
Here $\bar \delta(\Gamma_l)$ is equal to one if $\Gamma_l=0$ 
and equal to zero else. To prove the invariance one can closely follow 
the derivation presented in the appendix of \cite{Soper:1999xk}, apart from the fact that
in our case a surface term is present which however turns out to be
zero for the proposed contour deformation (\ref{cont_def}). 
  
 \vspace{0.3cm}

\noindent
Some comments are in order. 
To simplify the  discussion let us 
assume that $\Gamma_l=0$ for all $l$.
\begin{itemize}
\item While $\lambda \nabla\cdot Q$ controls the size of the deformation, 
$\alpha,\beta$ control the
smoothness of the deformation at the integration boundaries. 
The vanishing of the gradient of $Q(\vec{x})$
{\em inside} the integration volume
is only critical if $Q\to 0$ at the same time. Using 
$\tilde X_{ij}^{-1} = X_{ij}^{-1} \det(X)$, 
this critical situation is defined by the equations 
\begin{eqnarray}\label{critical_points}
&&\det(X)\; Z = \frac{1}{2} \; \sum\limits_{l,j=1}^{N-1} \tilde 
X^{-1}_{lj}Y_l Y_j \nonumber \\
&&0 < - {\rm sgn}(\det(X))\sum\limits_{l=1}^{N-1} \tilde X^{-1}_{lj}Y_l < |\det(X)| \quad , 
\quad j\in \{1,\dots,N-1\}
\end{eqnarray} which are polynomial in the kinematical invariants.
This situation corresponds to an  anomalous threshold which is an 
exceptional kinematical configuration.
With respect to integration over the phase space of external particles 
this is an integrable
singularity. The critical surface in the integration regions 
is given by
\begin{eqnarray}\label{critical_ts}
t_j &=& t_j^{\textrm{crit.}} = - \sum\limits_{l=1}^{N-1}  X^{-1}_{jl} Y_l 
\end{eqnarray} 
All one has to do is to split the integration hypercube 
subsequently into $2^{N-1}$ parts. This maps the critical surface
to the integration boundary where adaptive integration routines 
can cope with the problem.  If $t_j^{\textrm{crit.}}=0$ or $1$
subleading singularities may be probed.

\item In the case of an UV divergent integral it is easy to 
explicitly isolate the UV pole. At one loop,
only overall UV divergences are present, which manifest themselves in terms 
of $\Gamma$-functions in front of 
the parameter integral. With $D= 4+2m-2\epsilon$ one has an UV divergence, 
if and only if $N \le 2+m$.
Then $\Gamma(N-D/2)$ has a single pole in $\epsilon$. 
In this case we need the sector integral $J_l$
to order $\epsilon$\,:
\begin{eqnarray}
&&J_l(N,D,j_1,\dots ,j_r) = \int\limits_0^1 \prod\limits_{l=1}^{N-1}dt_l \;
\Bigl( \sum\limits_{j=1}^N T_j \Bigr)^{N-4-2m-r} 
\Bigl( -\frac{1}{2}T\cdot {\cal S}\cdot T - i\delta\Bigr)^{2+m-N}\nonumber \\&&\qquad
\times\;\Bigl(T_{j_1}\dots T_{j_r} \Bigr)
\Bigl[ 1
    -\epsilon \log\Bigl( -\frac{1}{2} T\cdot {\cal S}\cdot T - i\delta \Bigr)  
  + 2\epsilon \log\Bigl(\sum\limits_{j=1}^N T_j  \Bigr)   + 
  {\cal O}(\epsilon^2) \Bigr]
\end{eqnarray}
If $N<2+m$, this integral is numerically stable without any modifications. The ${\cal O}(1)$
term is a real number multiplying the UV pole. If  $N=2+m$, 
there is a logarithmic singularity in the ${\cal O}(\epsilon)$ term 
which can be dealt with by using the same contour deformation
as defined in eq.~(\ref{cont_def}).
Note that no IR poles are present in the case $N \le 2+m$ (the UV case). The UV and IR problems
are thus  nicely separated. 
\item The case of IR divergent 3-point functions can be treated 
analytically\footnote{It is also possible to extract the infrared poles by the 
method of iterated sector decomposition
 \cite{secdec_papers} from the parameter representations. 
This can be done in a completely automated way. A Laurent series in 
$\epsilon$ is produced,
where the coefficients are again Feynman parameter integrals  with
slightly more complicated denominators. The contour deformation can 
then be applied to a reduced number of Feynman parameters.}. 
We give a complete list of  3-point functions for internal 
masses equal to zero in appendix \ref{3pointIR}. 
\end{itemize}
We note that the presented method should  also be applicable  for
the direct numerical computation of finite Feynman parameter integrals
without doing any algebraic tensor reduction.
This is presently under investigation.

\subsection{Comparison between numerical and algebraic approach}\label{compare}

We present now a comparison between the algebraical and 
the numerical approach of evaluating the basis functions
of our reduction formalism.
 \vspace{0.3cm}

\noindent
As explained above we have algebraic representations
of the higher dimensional 4-point functions where
inverse Gram determinants are present, see section \ref{algebraic-red}. 
In a realistic application one expects that these representations
will be numerically well-behaved in the bulk of the phase space 
under consideration.
However, when approaching exceptional kinematical situations, 
compensations of large numbers
will happen which will finally spoil a reliable evaluation.

 \vspace{0.3cm}

\noindent
On the other hand, one expects that the method where
integrals with Feynman parameters in the numerator are 
evaluated numerically from the start 
should not be sensitive to the presence of 
inverse Gram determinants, 
as there are no singular denominators in this representation.
We have implemented the method for $(n+2)$-dimensional box and 
 $n$-dimensional triangle functions 
for general kinematics 
and obtain good numerical behaviour 
using standard deterministic and Monte-Carlo methods.
The two- and three-dimensional integral representations allow for
a reliable direct evaluation of the required integrals.
Comparing our numerical implementation 
to the algebraic one, we found that
the algebraic implementation is much faster and accurate in 
the interior of the phase space, while 
the numerical one allows for the automated evaluation of the integrals 
near exceptional momentum configurations. This will be illustrated 
in the following by a simple example. 

\begin{figure}[htb]
\label{fignum3}
\includegraphics[width=11.0cm]{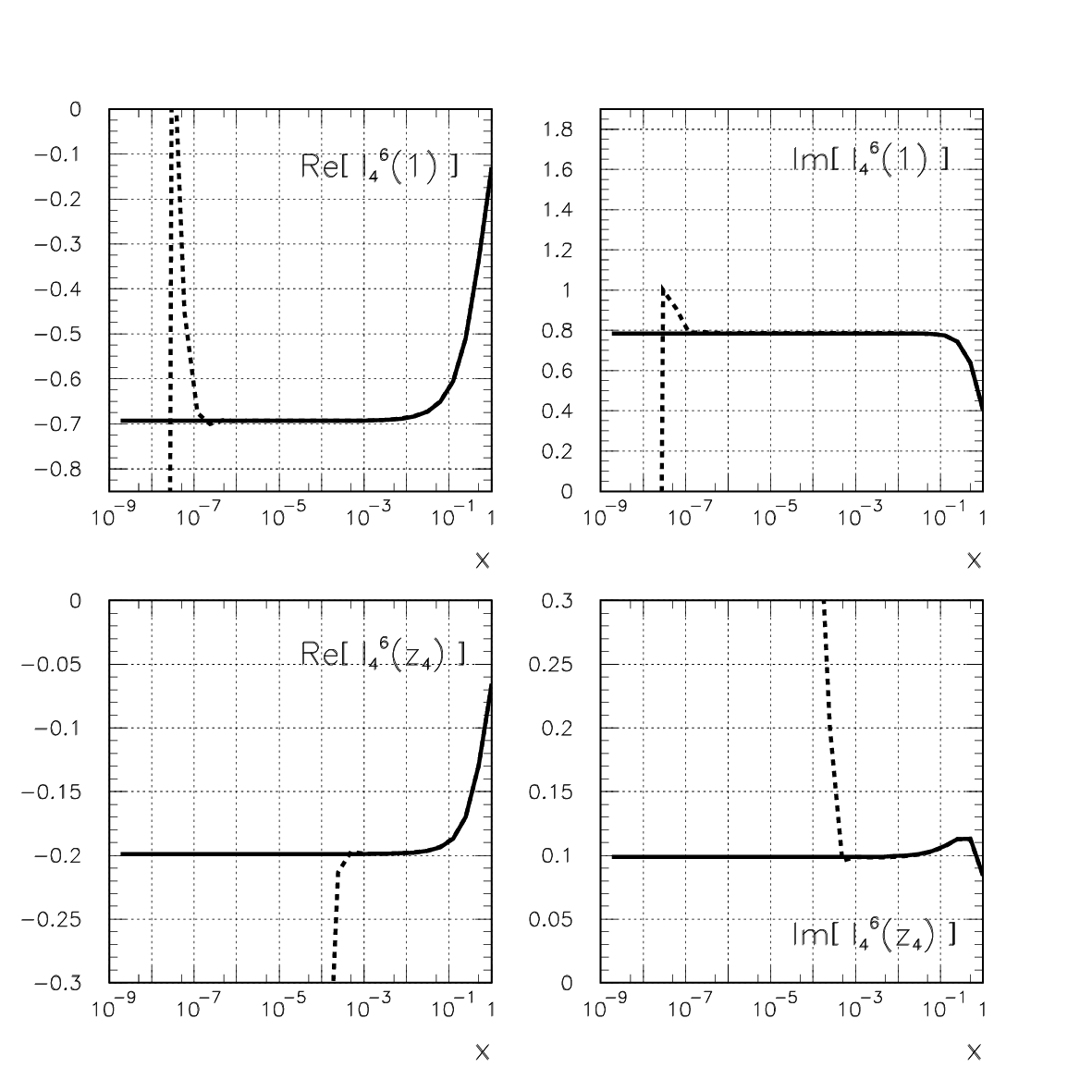} 
\caption{Real and imaginary parts of the basis integrals 
$I_4^6(1)$ and  $I_4^6(z_4)$, 
plotted versus the parameter $x$
which interpolates between exceptional and non-exceptional kinematics, as explained in the text.
The solid line stems from the numerical implementation, the dashed curves
show the numerical behaviour of the algebraic representation.}
\end{figure}
Let us consider a $2 \to 2 $ process 
$p_1+p_2\to p_3+p_4$ with $p_1^2=p_2^2=M^2$\,, where the momenta 
are parametrised as
\bea
p_1 = ( E(x),0,0, M\,x )\,,\qquad&&\;
p_2 = ( E(x),0,0,-M\,x ) \nn\\
p_3 = E(x) \, ( 1,0,\sin\theta,\cos\theta ), &&\;
p_4 = E(x) \, ( 1,0,-\sin\theta,-\cos\theta ) \nn\\
E(x) = M\,\sqrt{ 1+x^2 } \qquad\;\;\qquad&&\nn
\eea
The Gram determinant is given by 
$\det(G)=32 M^6\,(1+x^2)^2\,x^2\, \sin^2\theta$. 
Exceptional configurations are the forward/backward scattering region, 
$\theta=0,\pi$
and the case $x = 0$. 

In Fig. \ref{fignum3} we plot the real and imaginary
parts of the functions $I_4^6(1)$,  $I_4^6(z_4)$ with 
the parameter $x$ varied from 1 to 0. 
For the plots we have set $M=1$ and $\theta=7\pi/30$. 
The output for the plot was obtained by a Fortran code
working in double precision.
Whereas the 6-dimensional box function with numerator equal to one,
 $I_4^6(1)$, 
shows a perfect agreement with the result from the algebraic 
representation 
for $x$ values as low as $x \sim 10^{-6}$,  
the box with a numerator, $I_4^6(z_4)$,  
already starts to fluctuate severely 
for values of $x$ as large as  $x \sim 10^{-3}$. 
However, for both of these integrals the 
numerical instabilities 
in the algebraic implementation occur in a region where 
the result could be safely extrapolated to the boundary $x=0$
for our kinematical situation.
\begin{figure}[htb]
\label{fignum4}
\includegraphics[width=11.0cm]{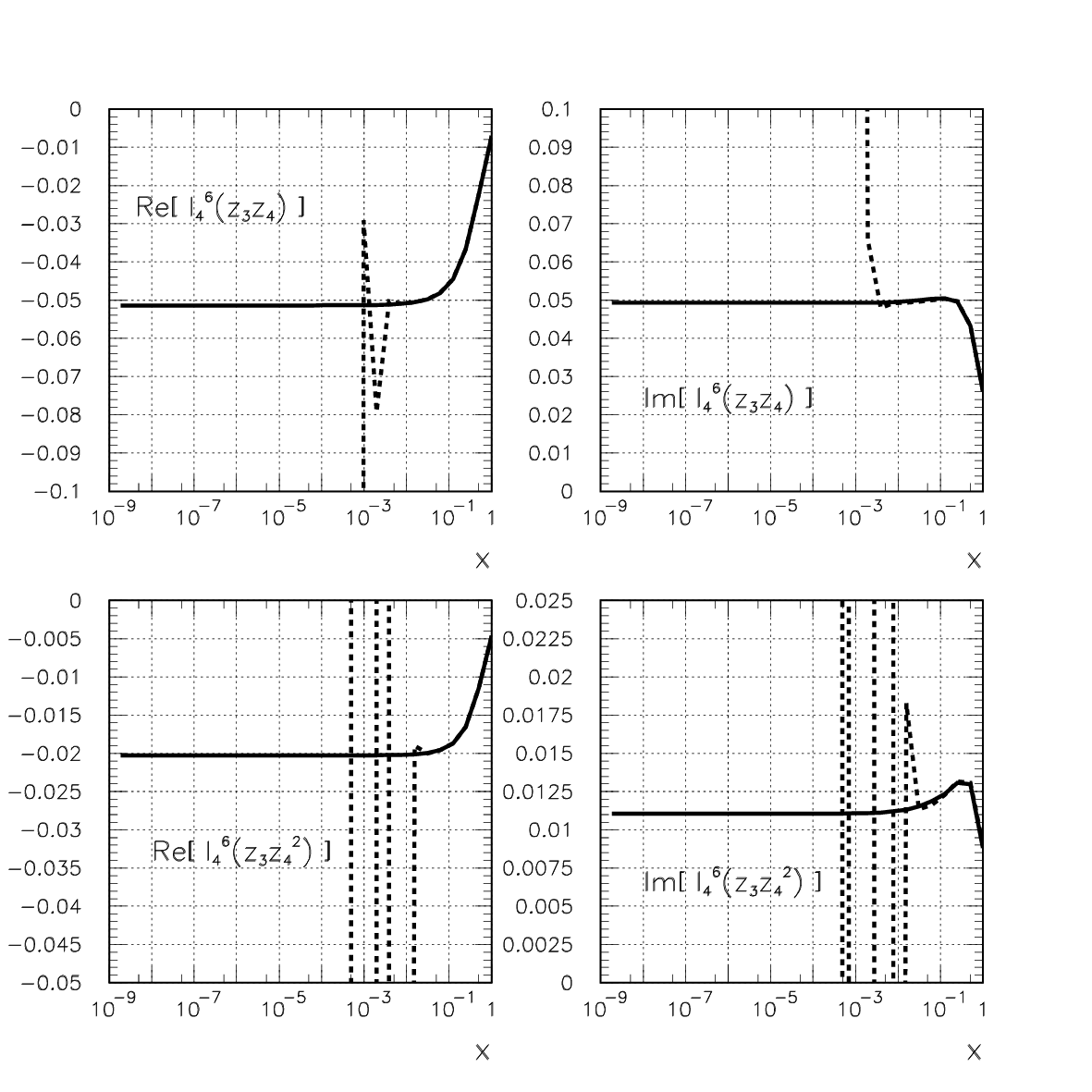} 
\caption{Same as Fig. \ref{fignum3} but for the basis 
integrals $I_4^6(z_3z_4)$ and  $I_4^6(z_3z_4^2)$.}
\end{figure}
In Fig.~\ref{fignum4}  the same plots are shown for the functions 
$I_4^6(z_3z_4)$ and  $I_4^6(z_3z_4^2)$.
The algebraic representations of these integrals 
have higher powers of inverse Gram determinants
and are thus less stable. It is important to note that the instability occurs 
already for values of $x$ where one cannot yet safely extrapolate 
to the integration boundary. We have checked that applying
quadruple precision for the evaluation of the discussed integrals
leads to an improved behaviour. The function
$I_4^6(1)$  can be evaluated correctly for the whole plotted $x$-range for the given kinematics.
The functions $I_4^6(z_4)$, $I_4^6(z_3z_4)$ and  
$I_4^6(z_3z_4^2)$ show instabilities below $x\sim 10^{-7}$, $x\sim 10^{-6}$ and $x\sim 10^{-4}$
respectively. 
The numerical problems are confined to a smaller
phase space region but are still present. 
Of course the analytical approach could be improved by
an algebraic expansion of the expression around critical regions, 
as has been done for example in \cite{Campbell:1996zw,Giele:2004ub}. 
Methods relying on such a Taylor expansion are faster, but require additional 
manual work, whereas the method suggested here is automated.

\vspace{0.3cm}

\noindent 
Note that in all cases the 
purely numerical implementation
is completely stable. It is actually possible to evaluate 
the given integrals numerically for all degenerate cases $x=0$ 
and $\theta=0,\pi$. 
 
\vspace{0.3cm}

\noindent The evaluation time for each plot point with an accuracy of better than one per cent
using the numerical method  is  of the order of seconds on a standard PC with a 
Pentium 4 processor if Monte Carlo methods are applied. A precision of one per cent 
for the higher order correction should be well sufficient for phenomenological applications.
The analytical evaluation is of the order of milliseconds and precise to
standard Fortran double precision. Given the fact that the numerical evaluation is only 
called in a very small fraction of the phase space the relative speed is compensated by the small
size of the critical regions.

\vspace{0.3cm}

\noindent 
We conclude from this study that the best procedure 
for practical applications is to use numerical
implementations near exceptional kinematical configurations 
and analytical ones in the interior phase space domains.  


\section{Recipe for the practitioner}\label{practitioner}

In this section we would like to summarise how to 
apply the main results of the article in a practical calculation.
Two steps have to be distinguished: first, expressing an amplitude
in terms of our basis integrals and second, the evaluation of
the basis integrals.

\subsection*{Expressing the amplitude in terms of basis functions}

After having generated the amplitude as a combination 
of Feynman diagrams, it has to be expressed 
in terms of tensor integrals as defined in  eq.~(\ref{iten}).
The further processing of these tensor integrals depends on the 
number $N$ of external legs.

\vspace{3mm}

\begin{figure}[htb]
\unitlength=1mm
\begin{picture}(130,150)
\put(10,0){\includegraphics[width=13.0cm]{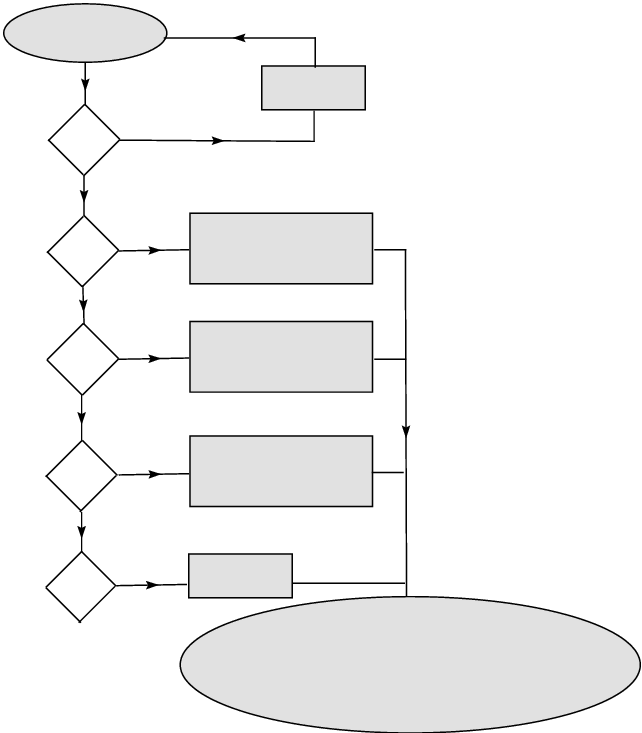}} 
\put(21,140){$I_N^{n,\mu_1\dots \mu_r}$}
\put(66,129){Eq. (\ref{TenRedNgeq6})}
\put(22,119){\small $N\geq6$}
\put(39,125){yes}
\put(20,109){no}
\put(21.4,96.1){\small $N=5$}
\put(20,86){no}
\put(21.5,74.5){\small $N=4$}
\put(20,65){no}
\put(21.3,51){\small $N=3$}
\put(20,40){no}
\put(21.2,28.5){\small $N\leq 2$}
\put(48,108){Form factors in eqs.:}
\put(53,97){(\ref{fofageneral}), (\ref{eqA50}--\ref{eqA55})}
\put(53,75){(\ref{fofageneral}), (\ref{eqA40}--\ref{eqA44})}
\put(53,52){(\ref{fofageneral}), (\ref{eqA33})}
\put(49,30){(\ref{eqFF2}--\ref{eqFF2end})}
\put(80,21){$I_2^n(1|j_1|j_1,j_2)$,}
\put(65,14){$I_3^n(1|j_1|j_1,j_2|j_1,j_2,j_3)$, $I_3^{n+2}(1|j_1)$}
\put(65,7){$I_4^{n+2}(1|j_1|j_1,j_2|j_1,j_2,j_3)$, $I_4^{n+4}(1|j_1)$}
\end{picture}
\caption{Reduction of $N$-point tensor integrals 
to basis integrals. The indicated equations 
can be implemented directly into an algebraic computer program.
$I_N^n(1|j_1|j_1,j_2|j_1,j_2,j_3)$ denotes the integral $I^n_N$ 
with zero, one, two or three Feynman parameters in the numerator.}\label{fig_prac}
\end{figure}

\noindent
In Fig.~\ref{fig_prac} we show the decision tree 
which indicates where to find the formulae  to reduce
an $N$-point tensor integral to our basis integrals. 
The latter are 2-point functions, $n$- and $(n+2)$-dimensional 
3-point functions and 
$(n+2)$- and $(n+4)$-dimensional 4-point functions with up to three
Feynman parameters in the numerator. To perform the reduction 
with algebraic programs it is sufficient to look up and code 
the given equation numbers.
Up to this point the representation is free 
from inverse Gram determinants. 
Algebraic simplifications might be applied after reduction 
for the coefficients of a given basis integral before 
proceeding to the numerical evaluation of the amplitude. 
We give a list of useful relations
in appendices \ref{relations} and \ref{chiral_hex_rels}.


\noindent
As an  illustration for a tensor reduction to basis integrals, 
let us show two examples. 
The first is the explicit expression for a 
rank two 5-point integral:
\begin{eqnarray}
I_5^{n,\,\mu_1 \mu_2}(a_1,a_2;S) & = & \int \diffk \; 
\frac{q_{a_1}^{\mu_1} \, q_{a_2}^{\mu_2}}{\prod_{i \in S}
(q_i^2-m_i^2+i\delta)} \nonumber \\
& = & g^{\mu_1 \, \mu_2} \, B^{5,2}(S) + \sum_{l_1,l_2 \in S}  \; 
\Delta^{\mu_1}_{l_1 \, a_1} \; \Delta^{\mu_2}_{l_2 \, a_2} \, 
A^{5,2}_{l_1 \, l_2}(S)\nonumber\\
&&\nn\\
B^{5,2}(S) & = & - \frac{1}{2} \, \sum_{j \in S} \, b_j \, 
I_4^{n+2}(S\setminus\{j\}) \nn\\
A^{5,2}_{l_1 l_2}(S) & = & \sum_{j \in S} \, \left( \, \icalst{j \, l_1} \, 
b_{l_2} + \icalst{j \, l_2} \, b_{l_1} - 2 \, \icalst{l_1 \, l_2} \, 
b_{j} + b_{j} \, 
\micals{j}_{l_1 \, l_2} \right) \, 
I^{n+2}_4(S\setminus\{j\}) \nonumber \\
& & \mbox{} + \frac{1}{2} \, \sum_{j \in S} \, \sum_{k \in S\setminus\{j\}} \, 
\left[ \icalst{j \, l_2} \, \micals{j}_{k \, l_1}  + 
\icalst{j \, l_1} \, \micals{j}_{k \, l_2}  \right]
I_3^{n}(S\setminus\{j,k\}) \nn\;.
\end{eqnarray}
As one can see, there is no inverse Gram determinant, 
only the inverse of the kinematic matrix $\cals$ is 
present\footnote{The determinant det\,$\cals$ is vanishing only if one hits an 
anomalous threshold\,\cite{bjorkendrell,edenlandshoff}, which
corresponds to an integrable singularity. The location of anomalous
thresholds depends on the external kinematics and the internal particle
masses. Note that there are no anomalous thresholds in the physical
region if all internal masses are zero. This is also true for the
massive case as long as the external particles are stable.}.


\noindent
A rank one 6-point integral has the form
\bea
&&I_6^{n,\,\mu}(a;S) = 
-\sum_{j \in S} \, \calc_{j\,a}^{\mu}\,I_5^n(S\setminus\{j\})\\
&& \quad =-\sum_{j,l \in S} \, \Delta_{la}^{\mu}\,
{\cal S}_{lj}^{-1}\,\sum_{k \in S}\,b_k^{\{j\}}
 \,\left[ B^{\{j,k\}} \, 
I^{n+2}_4(S\setminus\{j,k\}) +  \sum_{m \in S\setminus\{j,k\}} 
\,b_m^{\{j,k\}} \,I_3^{n}(S\setminus\{j,k,m\}) \right]\;.\nn
\eea
In both examples the basis integrals are already scalar integrals without
Feynman parameters in the numerator. For higher rank tensor integrals
this is not the case anymore.

\vspace{0.3cm}

\subsection*{Evaluation of the basis integrals}

\noindent
The case $N=2$ needs no extra discussion. All necessary formulae are 
gathered in appendix \ref{formfacsti12}.
For the evaluation of the remaining  basis integrals 
we  distinguish two cases, depending on 
whether the Gram determinants 
are numerically problematic or not. By eq.~(\ref{detSGrel})  Gram determinants
are related to the quantity $B$, the sum of reduction coefficients.
Technically, the splitting into safe and problematic regions
can be achieved  by introducing  an adequate energy scale 
$\Lambda$. This scale should be chosen such that
for  $B \Lambda^2\geq 1$ the evaluation of the 
{\em analytical} expression for   
the basis integrals in terms of purely scalar integrals  is 
numerically stable.
In this case the evaluation of the basis integrals for $N=3$ 
can be done 
by analytic reduction to scalar integrals by using 
eqs.~(\ref{i3np2})-(\ref{i3np23r2}) and eqs.~(\ref{eqFF2})-(\ref{eqFF2end}). 
If the 3-point functions are IR divergent, reduction formulae are
not applicable as $\det{\cal S}=0$. In this case one has to use 
 appendix \ref{3pointIR}, where explicit representations for all 
3-point functions
with and without Feynman parameters in the numerator
for  massless internal particles are provided. We do not quote the 
formulae for the IR 
divergent integrals where internal masses are present.
Note that {\em all} 3-point functions with IR poles are relatively simple 
functions which can always be treated analytically. 

\vspace{0.3cm}

\begin{figure}[ht]
\unitlength=1mm
\begin{picture}(150,140)
\put(30,0){\includegraphics[width=10.0cm]{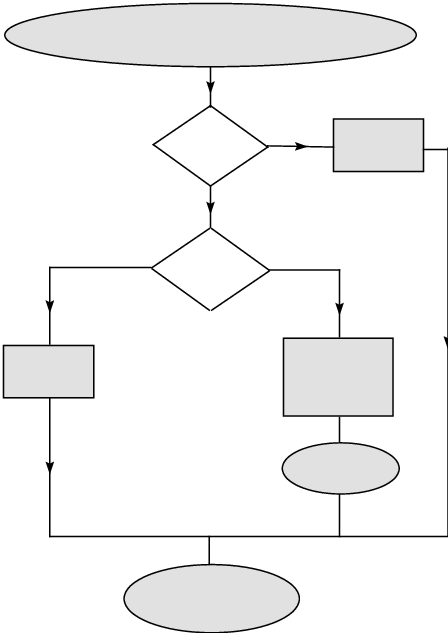}} 
\put(45,132){$I_3^n(1|j_1|j_1,j_2|j_1,j_2,j_3)$, $I_3^{n+2}(1|j_1)$}
\put(70,107){IR finite}
\put(90,117){Explicit functions:}
\put(107,108){App.~\ref{3pointIR}}
\put(90,102){no}
\put(68,98){yes}
\put(70,80){\small $B\,>\Lambda^{-2}$}
\put(90,85){analytic}
\put(87,74){yes}
\put(40,85){numerical}
\put(34,57){Sec.~\ref{numeric}}
\put(62,74){no}
\put(95,68){Eqs.}
\put(96,60){(\ref{i3n1}--\ref{i3np23r2})}
\put(96,53){(\ref{eqFF2}--\ref{eqFF2end})}
\put(96,35){$I_2^n(1),I_3^n(1)$}
\put(63,7){Numerical value}
\end{picture}
\caption{Evaluation of the basis integrals: the triangle case, $N=3$.}\label{prac_Neq3}
\end{figure}

\noindent
We illustrate our evaluation strategy for the 3-point functions in Fig.~\ref{prac_Neq3},
where we indicate all relevant sections or equation numbers needed for 
the implementation of our method.
\vspace{0.3cm}

\begin{figure}[ht]
\unitlength=1mm
\begin{picture}(150,130)
\put(30,0){\includegraphics[width=10.0cm]{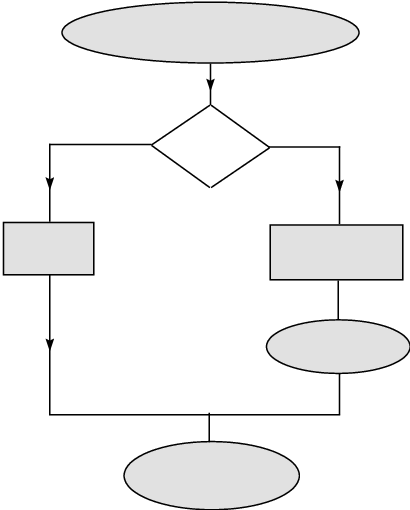}} 
\put(51,115){$I_4^{n+2}(1|j_1|j_1,j_2|j_1,j_2,j_3)$, $I_4^{n+4}(1|j_1)$}
\put(73,88){\small $B\,>\Lambda^{-2}$}
\put(100,95){analytic}
\put(95,82){yes}
\put(43,95){numerical}
\put(34,62){Sec.~\ref{numeric}}
\put(64,82){no}
\put(95,72){Eqs.}
\put(96,62){(\ref{eqA33}), (\ref{eq1RELA})-(\ref{eq5RELA})}
\put(96,38.5){$I_2^n(1)$, $I_3^n(1)$, $I_4^{6}(1)$}
\put(68,7){Numerical value}
\end{picture}
\caption{Evaluation of the basis integrals: the box case, $N=4$. For the analytical
evaluation of eqs.~(\ref{eqA33}) use Fig.~\ref{prac_Neq3}.
}\label{prac_Neq4}
\end{figure}

\noindent
The evaluation strategy for  the $N=4$ basis integrals is depicted in Fig.~\ref{prac_Neq4}.
There are no IR divergences in this case. Note that the analytic branch 
in Fig.~\ref{prac_Neq4} contains implicitly evaluations of 3-point functions
given in fig.~\ref{prac_Neq3} and an evaluation
of scalar box integrals in $6$-dimensions. For massless propagators
analytical representations can be found in \cite{Bern:1993kr,yukawa}. 
To our best knowledge  no complete list of these integrals 
for all combinations of internal and external masses is
available in the literature. However, 
we emphasise that the numerical evaluation,
using the contour deformation method introduced in 
section~\ref{numeric}, 
can always be used if the analytical representation is not known.
\noindent
Analytical and numerical representation are complementary to each other
in  the sense that the former are fast and accurate in the main part of the
phase space. The slower but robust numerical implementation also works
well  for exceptional kinematical configurations. 
As these dangerous phase space regions  
only cover a small part of the phase space, the speed issue does not 
pose a problem.  

\clearpage

\section{Conclusion}\label{concl}

In this paper, we have presented a complete method 
for the calculation of one-loop multi-leg amplitudes.
In principle it can be applied to arbitrary $N$-point problems,
the limitation coming only from computer power.  
We offer a new method for tensor reduction which has 
several advantages:

\begin{itemize}

\item The formalism, using dimensional regularisation, 
is valid for both massless and massive 
particles.
 
\item
Infrared divergences are easily isolated by construction. 
They appear only 
in terms of three-point integrals
which we list explicitly for the massless case.

\item The formalism is completely 
shift invariant, such that its iterated application 
does not require the redefinition of loop momenta.

\item Integrals in more than $n=4-2\eps$ dimensions do not   
have to be evaluated  for $N\geq 5$ external legs, as we have proven that 
they drop out.

\item By using  Feynman parameter integrals with non-trivial
numerators as basis functions, inverse Gram determinants 
can be completely avoided. 

\end{itemize}

\noindent
We present two methods  to compute the basis integrals which 
are the endpoints of our reduction.
First we  discuss the possibility of a 
``purely algebraic" approach, where all non-scalar integrals 
are reduced further to end up with scalar integrals only. 
This procedure re-introduces inverse Gram determinants,
which could spoil the subsequent numerical evaluation if 
an exceptional kinematic configuration is approached. 
We also propose a 
method  to compute the basis integrals completely 
numerically by multi-dimensional contour deformation in 
Feynman parameter space. 

\vspace{0.3cm}

\noindent
We have compared the two alternatives, with special emphasis
on the behaviour for exceptional kinematics. We show that 
the semi-numerical approach, where non-scalar integrals 
are used as basis integrals, is very stable if 
an exceptional kinematic configuration is approached.
On the other hand, 
the evaluation of the same integrals in a representation 
where they have been reduced algebraically 
down to scalar integrals, is stable only in the interior of the 
phase space. In this region however -- which is the bulk 
of the phase space -- their evaluation is of course faster
than evaluating the non-scalar form.  
In a program to calculate one-loop amplitudes, 
it is possible to combine the virtues of both alternatives
by using the algebraic representation in the interior phase space 
domains and switching to the semi-numerical one at the 
phase space boundaries.
\vspace{0.3cm}

\noindent
The paper contains a complete list of form factors 
for integrals of rank $r\leq N$, where $N=1,\ldots,5$, 
and it is shown that for $N>5$ it is not  necessary
to introduce new form factors. The formalism naturally 
maps tensor $N$-point integrals  with $N\geq 6$ 
to combinations of 5-point integrals and reduction 
coefficients.
\vspace{0.3cm}

\noindent
All the formulae 
needed for a direct implementation of the formalism are given 
in the paper, except some well-known 3-point and 4-point 
integrals, such that readers who are less interested 
in the technical details can straightforwardly use the method
by following the guidelines in the section ``recipe for the 
practitioner". 
\vspace{0.3cm}

\noindent
Further, we list several relations, between 
form factors as well as between reduction coefficients,
which can be very useful to achieve a more compact form 
of a given amplitude and/or to perform checks of the program. 
In particular, we give a helicity decomposition for the case 
of massless 6-point functions which leads to very compact expressions
for the reduction coefficients.
\vspace{0.3cm}

\noindent
In summary, the formalism presented here, as it can 
deal with massive particles as well as infrared divergences, 
has no restriction on the number of external legs and 
is suitable for numerical integration, can be used 
as the basis of a general program to 
calculate multi-leg one-loop amplitudes efficiently in a highly 
automated way.

\section*{Acknowledgements}
T.B. wants to thank D.~Soper for enlightening discussions concerning
multi-dimensional contour deformations. 
G.H. would like to thank Daniel Egli for useful discussions about 
the relations between reduction coefficients.
G.H., T.B. and C.S. thank the LAPTH for its 
hospitality while part of this article has been completed.
G.H. and J.Ph.G. are grateful to the Minami-Tateya group 
of KEK, Japan, for its warm hospitality 
and for interesting discussions while part this work has been carried out.\\
This research was supported in part (G.H.) by the Swiss National Science
Foundation (SNF) through grant no.\ 200021-101874.
The work of T.B. was supported by the Bundesministerium f\"ur 
Bildung und Forschung
(BMBF, Bonn, Germany) under the contract number 05HT4WWA2 and by the Deutsche 
Forschungsgemeinschaft (DFG) under the grant number, Bi1050/1-1.\\
LAPTH is a ``Unit\'e Mixte de Recherche (UMR 5108) associ\'ee au CNRS
et \`a l'Universit\'e de Savoie".

\begin{appendix}
\renewcommand{\theequation}{\Alph{section}.\arabic{equation}}
\setcounter{equation}{0}

\section{Form factors for $N=1,2$}\label{formfacsti12}

For completeness, we provide  the 
one- and two-point functions in this appendix.
We give the kinematical arguments  here in terms of invariants,
as  ${\cal S}$ is trivial in these cases. 

\bea
I_{1}^{n,\mu_1}(a_1;m_1^2) 
& = &  \Delta^{\mu_1}_{1\,a_1} I_1^n(m_1^2) = \Delta^{\mu_1}_{1\,a_1} \, m_1^2  \,I_2^n(0,0,m_1^2) \label{eqFF2}
\eea
Here we  use the fact that 1-point functions may be written as degenerate
2-point functions.

The kinematical matrix for the 2-point functions is
\bea
{\cal S} = -\left( \begin{array}{cc} 2\, m_1^2 & -s+ m_1^2+ m_2^2 \\
 -s+ m_1^2+ m_2^2 & 2\, m_2^2 \end{array} \right)
\eea
The Lorentz tensor decomposition for the tensor 2-point functions 
\bea
I_2^{n\,,\mu_1 }(a_1;s,m_1^2,m_2^2) 
& = & \sum_{l_1 \in S} \, \Delta^{\mu_1}_{l_1 \, a_1} \; A^{2,1}_{l_1}(S)
\eea
\bea
I_2^{n\,,\mu_1 \mu_2}(a_1,a_2;s,m_1^2,m_2^2) 
& = & g^{\mu_1 \, \mu_2} \, B^{2,2}(S) + \sum_{l_1,l_2 \in S}  \; 
\Delta^{\mu_1}_{l_1 \, a_1} \; \Delta^{\mu_2}_{l_2 \, a_2} \, 
A^{2,2}_{l_1 \, l_2}(S)
\eea
defines the form factors 
\bea
A_1^{2,1}(s,m_1^2,m_2^2)&=&-\frac{1}{2} I_2^n(s,m_1^2,m_2^2) \nonumber\\
   &&+ \frac{m_1^2-m_2^2}{2 \, s} [  I_2^n(s,m_1^2,m_2^2)-I_2^n(0,m_1^2,m_2^2) ]\nonumber\\
B^{2,2}(s,m_1^2,m_2^2)&=& \frac{1}{2 (n-1)}  \biggl[ 2\; m_2^2 I_2^n(s,m_1^2,m_2^2)+m_1^2 I_2^n(0,0,m_1^2)\nonumber\\
&& +\frac{-s+m_1^2-m_2^2}{2} \Bigl(  
I_2^n(s,m_1^2,m_2^2)\nonumber\\
&&  - \frac{m_1^2-m_2^2}{s} [ I_2^n(s,m_1^2,m_2^2)-I_2^n(0,m_1^2,m_2^2) ] \Bigr) \biggr] \nonumber\\
A^{2,2}_{11}(s,m_1^2,m_2^2)&=& \frac{1}{2 (n-1) s} \Bigl[ 
\frac{ n (s-m_1^2+m_2^2)}{2} \Bigl( 
I_2^n(s,m_1^2,m_2^2) \nonumber\\
 && - \frac{m_1^2-m_2^2}{s} [I_2^n(s,m_1^2,m_2^2)-I_2^n(0,m_1^2,m_2^2)] 
 \Bigr) \nonumber\\
 &&-2\, m_2^2\, I_2^n(s,m_1^2,m_2^2)+\, (n-2)\, m_1^2 I_2^n(0,0,m_1^2) \Bigr]
\eea
The other tensor coefficients are obtained by the following relations
\bea
A^{2,1}_1 + A^{2,1}_2 &=& -I_2^n \nonumber\\
A^{2,2}_{11} + A^{2,2}_{12} &=& -A^{2,1}_1 \nonumber\\
A^{2,2}_{21} + A^{2,2}_{22} &=& -A^{2,1}_2 \nonumber\\
A^{2,2}_{11} + A^{2,2}_{12}+A^{2,2}_{21} + A^{2,2}_{22} &=&  I_2^n
\eea
which follow directly from the fact that in one-loop 
parameter integrals,  
the sum of all Feynman parameters is equal to one.
The general 2-point scalar integral is 
well known\,\cite{Davydychev:2000na}. 
We  give here the integral representation for completeness.
\bea
I_2^n(s,m_1,m_2)&=& \Gamma(\epsilon) - \int\limits_{0}^1 dx 
\log( -s\, x\,(1-x) + x m_1^2 + (1-x) m_2^2 - i\delta )
+ \cal{O}(\epsilon)\eea
\noindent
If the external vector is light-like the formulae degenerate to
\bea
A^{2,1}_1(0,m_1^2,m_2^2)&=&+\frac{(4\,m_1^2-n\,m_1^2+n\,m_2^2)\,m_1^2}{2 n (m_2^2-m_1^2)^2}\,I_2^n(0,0,m_1^2)\nonumber\\
&& +\frac{(-4\,m_2^2-n\,m_1^2+n\,m_2^2)\,m_2^2}{2n(m_2^2-m_1^2)^2}\,I_2^n(0,0,m_2^2)\nonumber\\
&& -\frac{1}{2(m_1^2-m_2^2)}\,(m_1^2\,I_2^n(0,0,m_1^2)-m_2^2\,I_2^n(0,0,m_2^2)) \nonumber\\
B^{2,2}(0,m_1^2,m_2^2)&=&-\frac{m_1^4}{n(m_2^2-m_1^2)}\,I_2^n(0,0,m_1^2)
 +\frac{m_2^4}{n(m_2^2-m_1^2)}\,I_2^n(0,0,m_2^2) \nonumber\\
A^{2,2}_{11}(0,m_1^2,m_2^2)&=& \frac{1}{(n+2)n(m_1^2-m_2^2)^3}\, \Bigl( [(m_1^2-m_2^2)^2\,n\,(n+2) \nonumber\\
 &&-4\,m_1^2\,(n\,(m_1^2-m_2^2)-2\,m_2^2)]\,m_1^2\,I_2^n(0,0,m_1^2) \nonumber\\&&
 -8\,m_2^6\,I_2^n(0,0,m_2^2) \Bigr)
\eea
where
\bea
I_2^n(0,m_1^2,m_2^2)&=& \frac{m_2^2\, I_2^n(0,0,m_2^2) - m_1^2\,I_2^n(0,0,m_1^2)}{m_2^2-m_1^2} \nonumber\\
I_2^n(0,0,m_1^2) &=& \frac{I_1^n(m_1^2)}{m_1^2}=\frac{2}{n-2}I_2^n(0,m_1^2,m_1^2) =-\Gamma(1-n/2)\, (m_1^2)^{n/2-2}
\eea
In the case $m_1=m_2$  one finds
\bea
A^{2,1}_1(0,m_1^2,m_1^2)&=&-\frac{(n-2)}{4}\,I_2^n(0,0,m_1^2)\nonumber\\
B^{2,2}(0,m_1^2,m_1^2)&=&\frac{m_1^2}{2}\,I_2^n(0,0,m_1^2)\nonumber\\
A^{2,2}_{11}(0,m_1^2,m_1^2)&=&\frac{n-2}{6}\,I_2^n(0,0,m_1^2)\label{eqFF2end}
\eea
Two-point functions with no scale at all are defined as 
zero in dimensional regularisation.

\section{Divergent three--point functions with massless propagators}\label{3pointIR}

We consider here only three-point functions with massless propagators.
The parameter representation is given by
\bea
I^n_3(j_1, j_2 ,j_3;S) &=& - \Gamma(3-\frac{n}{2})\int_0^1 \prod_{i=1}^{3} 
\, d z_i \, \delta(1-\sum_{l=1}^{3} z_l) \, z_{j_1} z_{j_2} z_{j_3} \, 
(R^2)^{\frac{n}{2}-3} \\
R^2 &=& -z_1 \, z_2 \, \cals_{12} - z_2 \, z_3 \, \cals_{23} - 
z_1 \, z_3 \, \cals_{13}\,-i\,\delta\nn
\label{eqDEFJ3}
\eea
If one or two invariants out of the set $\{ \cals_{12} , \cals_{23} , \cals_{13} \}$
vanish, one gets an IR divergence. In this case one has 
$\det\cals=0$, such that the formulae given in section \ref{i3mass}
do not apply.
Therefore we provide analytic representations for all  3-point 
parameter integrals with massless propagators here. 
An overall coefficient $r_\Gamma$ is defined as
\begin{eqnarray}
r_\Gamma & = & \frac{\Gamma(1 +\eps) \,
\Gamma^2(1-\eps)}{\Gamma(1-2 \,\eps)} \nonumber \;.
\label{eqDEFGAMMAEPS}
\end{eqnarray}
For three-point functions with one non-zero invariant, denoted by $X$,
we labelled the internal propagators in such way that 
$\cals_{1 3}=X$ and $\cals_{1 2}=\cals_{2 3}=0$. 
For three-point functions with two non-zero invariants $X$ and $Y$, 
we set 
$\cals_{2 3} = X$ and $\cals_{1 3} = Y$. 
Thus the integrals $I^n_3(z_i,z_j,\ldots;0,0,X)$ 
are symmetric under exchange of 
$z_1\leftrightarrow z_3$ and the $I^n_3(z_i,z_j,\ldots;0,X,Y)$ 
are symmetric under 
simultaneous exchange of 
$z_1\leftrightarrow z_2$ and $X\leftrightarrow Y$. 
We obtain for integrals with no Feynman parameters in the numerator:
\begin{eqnarray}
I^n_3(0,0,X) & = & \frac{r_\Gamma}{\eps^2} \, H_0(X,-\eps) \\
I^n_3(0,X,Y) & = & \frac{r_\Gamma}{\eps^2} \, H_1(X,Y,-\eps) \end{eqnarray}
with one Feynman parameter:
\begin{eqnarray}
I^n_3(1;0,0,X) & = & -\frac{r_\Gamma}{\eps} \,  \frac{1}{1- 2 \,\eps} \, H_0(X,-\eps) \\
I^n_3(2;0,0,X) & = & \frac{r_\Gamma}{\eps^2} \, \frac{1}{1- 2 \,\eps} \, H_0(X,-\eps) \\
I^n_3(3;0,0,X) & = & -\frac{r_\Gamma}{\eps} \, \frac{1}{1- 2 \,\eps}\, H_0(X,-\eps) \\
&&\nonumber\\
I^n_3(1;0,X,Y) & = & \frac{r_\Gamma}{\eps^2} \,  \frac{1-\eps}{1- 2 \,\eps} \, H_2(X,Y,-\eps) \\
I^n_3(2;0,X,Y) & = & \frac{r_\Gamma}{\eps^2} \, \frac{1-\eps}{1- 2 \,\eps}  \, H_2(Y,X,-\eps) \\
I^n_3(3;0,X,Y) & = & -\frac{r_\Gamma}{\eps} \, \frac{1}{1- 2 \,\eps} \, H_1(X,Y,-\eps)
\end{eqnarray}
with two Feynman parameters:
\begin{eqnarray}
I^n_3(1,1;0,0,X) & = & -\frac{r_\Gamma}{\eps} \, \frac{1}{2(1-2\eps)} \, H_0(X,-\eps) \\
I^n_3(2,2;0,0,X) & = & \frac{r_\Gamma}{\eps^2} \, \frac{1}{(1-\eps)(1-2\eps)} \, H_0(X,-\eps) \\
I^n_3(3,3;0,0,X) & = & -\frac{r_\Gamma}{\eps} \, \frac{1}{2(1-2\eps)} \, H_0(X,-\eps) \\
I^n_3(1,2;0,0,X) & = & -\frac{r_\Gamma}{\eps} \, \frac{1}{2(1-\eps)(1-2\eps)} \, H_0(X,-\eps) \\
I^n_3(1,3;0,0,X) & = & \frac{r_\Gamma}{2(1-\eps)(1-2\eps)} \, H_0(X,-\eps) \\
I^n_3(2,3;0,0,X) & = & -\frac{r_\Gamma}{\eps} \, \frac{1}{2(1-\eps)(1-2\eps)} \, H_0(X,-\eps) \\
&&\nonumber\\
I^n_3(1,1;0,X,Y) & = & \frac{r_\Gamma}{\eps^2} \, \frac{2-\eps}{2(1-2\eps)} \, H_3(X,Y,-\eps) \\
I^n_3(2,2;0,X,Y) & = & \frac{r_\Gamma}{\eps^2} \, \frac{2-\eps}{2(1-2\eps)} \, H_3(Y,X,-\eps) \\
I^n_3(3,3;0,X,Y) & = & -\frac{r_\Gamma}{\eps} \, \frac{1}{2(1-2\eps)} \, H_1(X,Y,-\eps) \\
I^n_3(1,2;0,X,Y) & = & \frac{r_\Gamma}{\eps^2} \, \frac{2-\eps}{2(1-2\eps)} \, ( H_2(X,Y,-\eps) - H_3(X,Y,-\eps) )\\
I^n_3(1,3;0,X,Y) & = & -\frac{r_\Gamma}{\eps} \, \frac{1}{2(1-2\eps)} \, H_2(X,Y,-\eps) \\
I^n_3(2,3;0,X,Y) & = & -\frac{r_\Gamma}{\eps} \, \frac{1}{2(1-2\eps)} \, H_2(Y,X,-\eps) \\
\end{eqnarray}
with three Feynman parameters:
\begin{eqnarray}
I^n_3(1,1,1;0,0,X) & = &
	-\frac{r_\Gamma}{\eps} \, \frac{2-\eps}{2(3-2\eps)(1-2\eps)}  \, H_0(X,-\eps)\\
I^n_3(2,2,2;0,0,X) & = &
	\frac{r_\Gamma}{\eps^2} \, \frac{3}{(1-\eps)(3-2\eps)(1-2\eps)} \, 
	H_0(X,-\eps)\\
I^n_3(3,3,3;0,0,X) & = &
	-\frac{r_\Gamma}{\eps} \, \frac{2-\eps}{2(3-2\eps)(1-2\eps)}\, H_0(X,-\eps)\\
I^n_3(1,1,2;0,0,X) & = &
	-\frac{r_\Gamma}{\eps} \, \frac{1}{2(3-2\eps)(1-2\eps)} \, H_0(X,-\eps)\\
I^n_3(1,2,2;0,0,X) & = &
	-\frac{r_\Gamma}{\eps} \, \frac{1}{(1-\eps)(3-2\eps)(1-2\eps)}  \, H_0(X,-\eps)\\
I^n_3(1,1,3;0,0,X) & = &
	r_\Gamma \,\frac{1}{2(3-2\eps)(1-2\eps)}   \, H_0(X,-\eps)\\
I^n_3(2,2,3;0,0,X) & = &
	-\frac{r_\Gamma}{\eps} \,\frac{1}{(1-\eps)(3-2\eps)(1-2\eps)} \, H_0(X,-\eps)\\
I^n_3(1,3,3;0,0,X) & = &
	r_\Gamma \, \frac{1}{2(3-2\eps)(1-2\eps)}\, H_0(X,-\eps)\\
I^n_3(2,3,3;0,0,X) & = &
	-\frac{r_\Gamma}{\eps} \, \frac{1}{2(3-2\eps)(1-2\eps)} \, H_0(X,-\eps)\\
I^n_3(1,2,3;0,0,X) & = &
	r_\Gamma \, \frac{1}{2(3-2\eps)(1-2\eps)} \, H_0(X,-\eps)\\
	&&\nonumber\\
I^n_3(1,1,1;0,X,Y) & = &
	\frac{r_\Gamma}{\eps^2} \, \frac{(3-\eps)(2-\eps)}{2(3-2\eps)(1-2\eps)} \, 
	H_4(X,Y,-\eps)\\
I^n_3(2,2,2;0,X,Y) & = &
	\frac{r_\Gamma}{\eps^2} \,  \frac{(3-\eps)(2-\eps)}{2(3-2\eps)(1-2\eps)} \, 
	H_4(Y,X,-\eps)\\
I^n_3(3,3,3;0,X,Y) & = &
	-\frac{r_\Gamma}{\eps} \, \frac{(2-\eps)}{2(3-2\eps)(1-2\eps)} \, 
	H_1(X,Y,-\eps)\\
I^n_3(1,1,2;0,X,Y) & = &
	\frac{r_\Gamma}{\eps^2} \, \frac{(3-\eps)(2-\eps)}{2(3-2\eps)(1-2\eps)}  \, (
	H_3(X,Y,-\eps)-H_4(X,Y,-\eps))\\
I^n_3(1,2,2;0,X,Y) & = &
	\frac{r_\Gamma}{\eps^2} \, \frac{(3-\eps)(2-\eps)}{2(3-2\eps)(1-2\eps)} \, (
	H_3(Y,X,-\eps)-H_4(Y,X,-\eps))\\
I^n_3(1,1,3;0,X,Y) & = &
	-\frac{r_\Gamma}{\eps} \, \frac{(2-\eps)}{2(3-2\eps)(1-2\eps)} \, 
	H_3(X,Y,-\eps)\\
I^n_3(2,2,3;0,X,Y) & = &
	-\frac{r_\Gamma}{\eps} \, \frac{(2-\eps)}{2(3-2\eps)(1-2\eps)} \, 
	H_3(Y,X,-\eps)\\
I^n_3(1,3,3;0,X,Y) & = & 
	-\frac{r_\Gamma}{\eps} \, \frac{(1-\eps)}{2(3-2\eps)(1-2\eps)}  \, 
	H_2(X,Y,-\eps)\\ 
I^n_3(2,3,3;0,X,Y) & = &
	-\frac{r_\Gamma}{\eps} \, \frac{(1-\eps)}{2(3-2\eps)(1-2\eps)} \, 
	H_2(Y,X,-\eps)\\
I^n_3(1,2,3;0,X,Y) & = &
	-\frac{r_\Gamma}{\eps} \, \frac{(2-\eps)}{2(3-2\eps)(1-2\eps)} \, 
	(H_2(X,Y,-\eps)-H_3(X,Y,-\eps))
\label{}
\end{eqnarray}
Higher dimensional 3-point integrals:
\bea
I_3^{n+2}(0,0,X)=\frac{r_\Gamma}{\eps}
\frac{1}{2(1-\eps)
(1-2\eps)}\,H_0(X,1-\eps)\label{uv1}\\
I_3^{n+2}(0,X,Y)=\frac{r_\Gamma}{\eps} 
\frac{1}{2(1-\eps)(1-2\eps)}
\,H_1(X,Y,1-\eps)\\
&&\nonumber\\
I_3^{n+2}(1;0,0,X)=\frac{r_\Gamma}{\eps} 
\frac{1}{2(3-2\eps)(1-2\eps)}\,
H_0(X,1-\eps)\\
I_3^{n+2}(2;0,0,X)=\frac{r_\Gamma}{\eps} 
\frac{1}{2(1-\eps)(3-2\eps)(1-2\eps)}
\,H_0(X,1-\eps)\\
I_3^{n+2}(3;0,0,X)=\frac{r_\Gamma}{\eps} 
\frac{1}{2(3-2\eps)(1-2\eps)}\,
H_0(X,1-\eps)\\
&&\nonumber\\
I_3^{n+2}(1;0,X,Y)=\frac{r_\Gamma}{\eps}
\frac{(2-\eps)}{2(1-\eps)
(3-2\eps)(1-2\eps)}\,
H_2(X,Y,1-\eps)\\
I_3^{n+2}(2;0,X,Y)=\frac{r_\Gamma}{\eps}
\frac{(2-\eps)}{2(1-\eps)
(3-2\eps)(1-2\eps)}
\,H_2(Y,X,1-\eps)\\
I_3^{n+2}(3;0,X,Y)=\frac{r_\Gamma}{\eps} 
\frac{1}{2(3-2\eps)(1-2\eps)}\,
H_1(X,Y,1-\eps)\label{uv8}
\eea

The  functions $H_0$, $H_1$, $H_2$, $H_3$ and $H_4$ are given by:
\begin{eqnarray}
H_0(X,\alpha) & = & \frac{\bar{X}^{\alpha}}{X} \\
\label{eqH0}
H_1(X,Y,\alpha) & = & \frac{\bar{X}^{\alpha}-\bar{Y}^{\alpha}}{X-Y} \\
\label{eqH1}
H_2(X,Y,\alpha) & = & \frac{\bar{Y}^{\alpha}}{Y-X}+\frac{1}{1+\alpha} \, 
\frac{\bar{Y}^{1+\alpha}-\bar{X}^{1+\alpha}}{(Y-X)^2} \\
\label{eqH2}
H_3(X,Y,\alpha) & = & \frac{\bar{Y}^{\alpha}}{Y-X}+\frac{2}{1+\alpha} \, 
\frac{\bar{Y}^{1+\alpha}}{(Y-X)^2}+\frac{2}{(1+\alpha) \, (2+\alpha)} \, 
\frac{\bar{Y}^{2+\alpha}-\bar{X}^{2+\alpha}}{(Y-X)^3} \\
\label{eqH3}
H_4(X,Y,\alpha) & = & \frac{\bar{Y}^{\alpha}}{Y-X}+\frac{3}{1+\alpha} \, 
\frac{\bar{Y}^{1+\alpha}}{(Y-X)^2}+\frac{6}{(1+\alpha) \, (2+\alpha)} \, 
\frac{\bar{Y}^{2+\alpha}}{(Y-X)^3} \nonumber \\
& & \mbox{}+\frac{6}{(1+\alpha) \, (2+\alpha) \, (3+\alpha)} \, 
\frac{\bar{Y}^{3+\alpha}-\bar{X}^{3+\alpha}}{(Y-X)^4}\label{eqH4}\\
&&\nonumber\\
\bar{X}&=&-X-i\,\delta\nonumber
\end{eqnarray}
All these functions have a regular behaviour when $X = Y$. 
It is for this reason that we want to keep them as they are. 
If the coefficient in front of them is proportional to $(X-Y)$, 
then they can be reduced using the following properties:
\begin{eqnarray}
(Y-X) \, H_1(X,Y,\alpha) & = & \bar{Y}^{\alpha}-\bar{X}^{\alpha} \\
\label{eqPRO1H1}
(Y-X) \, H_2(X,Y,\alpha) & = & \frac{\alpha}{1+\alpha} \, \bar{Y}^{\alpha} - 
\frac{1}{1+\alpha} \, X \, H_1(X,Y,\alpha) \\
\label{eqPRO1H2}
(Y-X) \, H_3(X,Y,\alpha) & = & \frac{\alpha}{2+\alpha} \, \bar{Y}^{\alpha} - 
\frac{2}{2+\alpha} \, X \, H_2(X,Y,\alpha) \\
\label{eqPRO1H3}
(Y-X) \, H_4(X,Y,\alpha) & = & \frac{\alpha}{3+\alpha} \, \bar{Y}^{\alpha} - 
\frac{3}{3+\alpha} \, X \, H_3(X,Y,\alpha)
\label{eqPRO1H4}
\end{eqnarray}
\begin{eqnarray}
H_1(Y,X,\alpha) & = &  H_1(X,Y,\alpha)\nn\\
H_2(Y,X,\alpha) & = & H_1(X,Y,\alpha)-H_2(X,Y,\alpha) \nn\\
H_3(Y,X,\alpha) & = & H_3(X,Y,\alpha)-2 \, H_2(X,Y,\alpha) + 
H_1(X,Y,\alpha)\nn\\
H_4(Y,X,\alpha) & = & - H_4(X,Y,\alpha) + 3 \, H_3(X,Y,\alpha) - 
3 \, H_2(X,Y,\alpha) + H_1(X,Y,\alpha)
\label{eqPRO2H4}
\end{eqnarray}

We use $n=4-2\eps$ where $\eps<0$ in the infrared region.

\section{Proof of the absence of higher dimensional integrals and Gram determinants
for $N=5$}\label{absence-N-eqn5}

In the next subsection, we will show that the coefficient multiplying the 
higher dimensional five-point integrals is of order $\eps$. 
We will repeatedly use the fact that
for general 5-point kinematics, the metric tensor in 4 dimensions is 
expressible by 
a tensor product of  external vectors.  Many simplifications will
occur by neglecting terms of ``${\cal O}(\epsilon)$''. 
For scalar quantities it is clear what that means. 
For {\em tensors}, we say that a tensorial structure 
is of ${\cal O}(\epsilon)$ if differences of tensors defined
in $n$ and in 4 dimensions occur. Contracting such differences 
with kinematical objects like external momenta, polarisation vectors
or fermion currents will always lead finally to 
scalar quantities of ${\cal O}(\epsilon)$, 
which can be neglected in phenomenological applications at one loop.

The price to pay for the disappearance of the higher dimensional 
integrals is that inverse Gram 
determinants ($1/B$) are reintroduced explicitly. In subsection \ref{oneoverb}, 
we show how these spurious divergences cancel out analytically.
This will lead us to a representation of 5-point
functions which is free from higher dimensional 5-point integrals 
{\em and} $1/B$ terms. The corresponding
form factors are listed in section \ref{formfact_fivep} of the main text.

\subsection{The fate of higher dimensional integrals}\label{sec5point}

For $N=5$, the following relation, shown in appendix \ref{proof_of_eq3},
\begin{equation}
\caltf^{\mu \, \nu}_{a \, b} = 2 \, \frac{\calv_a^{\mu} \, 
\calv_b^{\nu}}{B}
\label{eqEQUATION3}
\end{equation}
will enable us to remove the higher dimensional 
five-point integrals. The tensors  
$\calv_a^{\mu}$ and $\calt^{\mu \, \nu}_{a \, b}$ 
are defined in eqs.~(\ref{eqVDEF}) and (\ref{eqTDEFN}). 
 
\vspace{0.3cm} 

\noindent
For rank zero and rank one, it is trivial to see that the coefficient 
of $I^{n+2}_5$ is of order $\eps$. 
For rank 0, one has, according to eqs.~(\ref{EQdivetfin}) 
and (\ref{EQifin2}): 
\begin{equation}
I^n_5(S)=\sum_{j\in S}b_j\,I^n_4(S\setminus\{j\})-(4-n)B\,
I^{n+2}_5(S)\;.
\end{equation}
For rank 1, one obtains by application of 
(\ref{eqNEWIDEF2})\,:
\begin{equation}
I^{n,\,\mu}_5(a,S)=-\sum_{j\in S}\,{\cal C}_{j\,a}^{\mu}\,
I^n_4(S\setminus\{j\})
+(4-n)\,{\cal V}^{\mu}_a\,I^{n+2}_5(S)\;.
\label{n5r1}
\end{equation}
\noindent
For higher rank ($r \ge 2$), we prove by induction on $r$ that the coefficient 
of the higher dimensional five-point integrals is of order $\eps$.
The higher dimensional five-point integrals $I^{n+2m}_5$ are UV finite for $m<3$. 
Therefore, the order $\eps$ terms can be dropped for pentagons of rank $r\leq 5$. 
Care has to be taken for higher dimensional pentagons $(r>5)$, which can occur for example 
in the presence of effective Higgs-gluon couplings, as the integrals $I^{n+2m}_5$
are UV divergent for $m\geq 3$ and therefore the order $\eps$ terms canot 
be dropped.

\vspace{0.3cm} 

\noindent
In step one we show that the assumption is true for $r=2$. 
Indeed, in the case of rank 2, direct application of (\ref{eqNEWIDEF2}) leads to
\bea
I_5^{n,\,\mu \nu}(a_1,a_2;S) & = & \int \diffk \; 
\frac{q_{a_1}^{\mu}\, q_{a_2}^{\nu}}{\prod_{i \in S} 
(q_i^2-m_i^2+i\delta)}\nonumber \\
& = & - \frac{1}{2} \, \calt^{\mu \nu}_{a_1a_2} \, 
I^{n+2}_{5}(S)  + (3-n) \, \sum_{i\in S} 
I^{n+2}_{5}(i\,;S) \, \Delta^{\mu}_{a_1 \,i} \, 
\calv^{\nu}_{a_2} \nonumber \\
& & \mbox{} - \sum_{j \in S} \, \calc^{\nu}_{j\,a_2} \, 
I^{n,\,\mu}_4(a_1\,;S\setminus\{j\})\;.
\label{eqRANK2}
\eea
The integral $I^{n+2}_{5}(i\,;S)$
with Feynman parameter $z_i$ in the numerator can be expressed in terms of 
$I^{n+2}_{5}(S)$. For this purpose, we 
contract both sides of eq.~(\ref{eqRANK2}) with $\Delta_{b\,c}^\nu$,
where $b$ and $c$ are arbitrary labels in $S$. 
Using 
\bea
q_a^{\mu}  \, \Delta_{b \, c\,\mu}  & = & \frac{1}{2} \left( \, q_b^2-m_b^2 -
[\,q_c^2 - m_c^2\,]-
\calst_{a \, b} + \calst_{a \, c} \right)\label{qdotdel}\\
{\cal T}^{\mu \, \nu}_{a \, b} \, \Delta_{c\,d\,\mu}  &=&
({\cal S}_{ac}-{\cal S}_{ad})\,\calv_b^{\nu}  \label{eq39}\\
\calv_a^{\mu} \, \Delta_{c\,d\,\mu}  &=& 
\frac{B}{2}\,({\cal S}_{ac}-{\cal S}_{ad})\label{eq40}\\
\calc^{\mu}_{j\,a}\, \Delta_{c\,d\,\mu}  &=& 
\frac{1}{2}\,\left(\delta_{jd}-\delta_{jc}+b_j\,[\,{\cal S}_{ac}-{\cal
S}_{ad}\,]\right)\label{eq41}
\eea
leads to
\begin{eqnarray}
\sum_{i\in S} \, I^{n+2}_5(i\,;S) \, 
\Delta^{\mu}_{a\,i} &=&  \frac{1}{B} \, 
\Biggl\{ - \calv_a^{\mu} \, I^{n+2}_5(S)  \nonumber \\ 
& & +\frac{1}{3-n}\,\sum_{i\in S} \,\left( 
b_i\, I^{n,\,\mu}_4(a,S\setminus\{i\})
+ \calc_{i \, a}^{\mu} \, I^{n}_4(S\setminus\{i\})\right)\Biggr\}\;.
\label{eqFIVERED1}
\end{eqnarray}
The insertion of the latter into eq.~(\ref{eqRANK2}) 
yields
\begin{eqnarray}
I_5^{n,\,\mu \nu}(a_1,a_2;S) & = &\left\{ - \frac{1}{2} \,\calt^{\mu \nu}_{a_1a_2}
 + (n-3) \, \frac{\calv_{a_1}^{\mu} \, \calv_{a_2}^{\nu}}{B} \right\} \, 
 I^{n+2}_{5}(S) \nonumber \\
& & \mbox{} + \sum_{i \in S} \, 
  \left( \frac{\calv_{a_2}^{\nu} }{B}\, b_i - 
\calc_{i \, a_2}^{\nu} \right) \,  I^{n,\,\mu}_4(a_1\,;S\setminus\{i\})
\nonumber \\
& & \mbox{} + \frac{\calv_{a_2}^{\nu}}{B}\,\sum_{i \in S} \, 
 \calc_{i \, a_1}^{\mu}\, 
I^{n}_4(S\setminus\{i\})\;.
\label{eqRANKpre}
\end{eqnarray}
Using  now eq.~(\ref{n5r1}) for the last term in eq.~(\ref{eqRANKpre}) results in 
\begin{eqnarray}
I_5^{n,\,\mu \nu}(a_1,a_2;S) & = & - \frac{1}{2} \,\left\{ \calt^{\mu \nu}_{a_1a_2}
 -  \frac{2 \, \calv_{a_1}^{\mu} \, \calv_{a_2}^{\nu}}{B} \right\} \, 
 I^{n+2}_{5}(S) \nonumber \\
& & \mbox{} + \sum_{i \in S} \, 
  \left( \frac{\calv_{a_2}^{\nu} }{B}\, b_i - 
\calc_{i \, a_2}^{\nu} \right) \,  I^{n,\,\mu}_4(a_1\,;S\setminus\{i\})
\nonumber \\
& & \mbox{} -
\frac{\calv_{a_2}^{\nu}}{B}\, 
I^{n,\,\mu}_5(a_1\,;S)\;.
\label{eqRANK2FIVEP}
\end{eqnarray}
Now we can use eq.~(\ref{eqEQUATION3}) to see that the coefficient 
of $I^{n+2}_{5}$ 
in eq.~(\ref{eqRANK2FIVEP}) is indeed of order $\eps$.

Let us now assume that for rank $r-1$, all the higher dimensional integrals involve an  
$\cal{O}(\eps)$ tensor of the type $ \calt^{\mu \nu}_{a_1a_2}
 -  2 \, \calv_{a_1}^{\mu} \, \calv_{a_2}^{\nu}/B $, and show that this is also the case 
 for rank $r$. Firstly, we carry out the momentum integration in the first term of 
eq.~(\ref{eqNEWIDEF2}), using the shift (\ref{eqTRANSL}) and 
eqs.~(\ref{eqSHIFT1}), (\ref{eqTRANSA}). 
This leads to\footnote{Here we keep $N$ arbitrary to show that the relations derived are 
not only valid for $N=5$.}:
\bea
&&I_N^{n,\,\mu_1\ldots\mu_r}(a_1,\ldots,a_r;S)  = 
-\sum_{j \in S} \, \calc_{j\,a_r}^{\mu_r}
\int \diffk \, \frac{ (q_j^2 -m_j^2) \; 
 q_{a_1}^{\mu_{1}}\ldots q_{a_{r-1}}^{\mu_{r-1}}}{
 \prod_{i \in S}(q_i^2-m_i^2+i\delta)} \label{eqdef}
\\
&&+ \int \diffz
\int \diffl \,
\frac{\left[l_{\nu}\,\left(\calt_{a_r d}^{\mu_r\nu}+
2\,\calv_{a_r}^{\mu_r}\sum_{i\in S} z_i\Delta_{d\,i}^\nu \right)+
\calv_{a_r}^{\mu_r}\,(l^2+R^2)\right]\,
\tilde{q}_{a_1}^{\mu_{1}}\ldots \tilde{q}_{a_{r-1}}^{\mu_{r-1}}}{(l^2-R^2)^N}\nn
\;,
\eea
where 
\begin{eqnarray*}
\diffz & = & \Gamma(N) \, \prod_{i\in S} dz_i\,
\delta(1-\sum_{l=1}^N z_l) 
\end{eqnarray*}
and the $\tilde{q}_{a}$ denote the $q_a$-vectors in terms 
of the shifted loop momentum $l$, given by eq.~(\ref{eqSHIFT1}).

Secondly, we contract eq.~(\ref{eqdef})  with 
$\Delta_{b\,c\,\mu_r}$, where $b,c\in S$ are arbitrary, 
using
(\ref{qdotdel}) for the 
left-hand side and (\ref{eq39}) to  (\ref{eq41}) for the right-hand side.
This yields
\begin{eqnarray}
&&\int \diffk \, \frac{  
 q_{a_1}^{\mu_{1}}\ldots q_{a_{r-1}}^{\mu_{r-1}}}{
 \prod_{i \in S}(q_i^2-m_i^2+i\delta)} = \sum_{j \in S} \, b_j \,
 \int \diffk \, 
 \frac{ (q_j^2 -m_j^2) \; 
 q_{a_1}^{\mu_{1}}\ldots q_{a_{r-1}}^{\mu_{r-1}}}{
 \prod_{i \in S}(q_i^2-m_i^2+i\delta)} \label{eqNEWRED} \\
&&- \int \diffz \, \int \diffl \, 
\frac{\left[ 2 \, l_{\nu} \left(\calv_{d}^{\nu} + B \,
\sum_{i\in S} z_i\,\Delta_{d\,i}^\nu \right) 
+ B\,(l^2+R^2) \right] \;
\tilde{q}_{a_1}^{\mu_{1}}\ldots 
\tilde{q}_{a_{r-1}}^{\mu_{r-1}}}{(l^2-R^2)^N} \;.\nonumber
\end{eqnarray}
Using now eq.~(\ref{eqNEWRED}) to replace the term containing 
$\sum_{i\in S} z_i\,\Delta_{d\,i}^\nu$ 
in (\ref{eqdef}), we obtain:
\bea
I_N^{n,\,\mu_1\ldots\mu_r}(a_1,\ldots,a_r\,;S) & = & 
\sum_{j \in S} \,\left( \frac{\calv_{a_r}^{\mu_r} \, b_j}{B} 
- \calc_{j \, a_r}^{\mu_r} \right) \, 
\int \diffk \, \frac{ (q_j^2 -m_j^2) \; 
 q_{a_1}^{\mu_{1}}\ldots q_{a_{r-1}}^{\mu_{r-1}}}{
 \prod_{i \in S}(q_i^2-m_i^2+i\delta)} 
\nn\\
& & \mbox{} - \frac{\calv_{a_r}^{\mu_{r}}}{B} \,  
I_N^{n,\mu_1\ldots\mu_{r-1}}(a_1,\ldots,a_{r-1}\,;S)\nn\\
&&+
\int \diffz \int \diffl \; \frac{l_{\nu}\,\left(\calt_{a_r d}^{\mu_r\nu}-
2\,\calv_{a_r}^{\mu_r}\calv_{d}^{\nu}/B \right)\,
\tilde{q}_{a_1}^{\mu_{1}}\ldots \tilde{q}_{a_{r-1}}^{\mu_{r-1}}}{(l^2-R^2)^N}\;.
\label{final}
\eea
Specifying $N=5$, we can see from eq.~(\ref{final}) that a rank $r$ 5-point integral 
can be written as the sum of  4-point integrals of rank $r-1$ 
plus a  5-point integral of rank $r-1$
plus a term 
which potentially generates higher dimensional 5-point integrals 
but is proportional to 
$ \calt^{\mu_r \mu_k}_{a_ra_k}
 - 2\,\calv_{a_r}^{\mu_r} \, \calv_{a_k}^{\mu_k}/B\,, 
$ 
which is of order $\eps$. \\
The last term, 
\be
I^{\mu_1\ldots\mu_r}_\eps=\int \diffz \int \diffl \; \frac{l_{\nu}\,\left(\calt_{a_r d}^{\mu_r\nu}-
2\,\calv_{a_r}^{\mu_r}\calv_{d}^{\nu}/B \right)\,
\tilde{q}_{a_1}^{\mu_{1}}\ldots \tilde{q}_{a_{r-1}}^{\mu_{r-1}}}{(l^2-R^2)^5}
\ee
can only lead to a finite contribution for rank $r\geq 6$, as 5-point integrals 
start to develop an ultraviolet pole for $r\geq 6$. 

\vspace*{3mm}

\noindent
A few comments are in order here.
\begin{itemize}
\item[(i)] Eq.~(\ref{eqNEWRED}) defines another way of reducing the tensor integrals, by which a rank 
$r$ $N$-point integral is expressed as an infrared finite part plus a sum of rank $r$ 
$(N-1)$\,-point 
integrals. Contrarily to eq.~(\ref{eqdef}), eq.~(\ref{eqNEWRED}) reduces the number of propagators but 
not the rank. For this  reason  the way used in 
section \ref{method} is preferable.
\item[(ii)] In the case where ${\cal S}$ is 
not invertible, the derivation is equally valid, as eq.~(\ref{eq41}) still 
holds in this case. This can be easily seen by contracting 
eq.~(\ref{eqdef}) for $r=1$ with $\Delta_{bc\,\mu}$. 
One obtains
$$
\sum_{j \in S} \,
\calc_{j a}^{\mu}\Delta_{bc\,\mu}\,I^n_{N-1}(S\setminus\{j\})=
\frac{1}{2}\left( I^n_{N-1}(S\setminus\{c\})-I^n_{N-1}(S\setminus\{b\}) 
+({\cal S}_{ab}-{\cal S}_{ac})\,I^n_{N}(S)\right) \;.
$$
Now we can use  $I^n_{N}(S)=\sum_{j\in S}\,b_j\,I^n_{N-1}(S\setminus\{j\})$,  
which is always valid if $B=\sum_{j\in S}\,b_j=0$, and fulfilled to order
$\eps$ for $N=5$.
The reduction coefficients 
$b_j$ can be constructed as in section \ref{subscalar}, if ${\cal S}$ is 
not invertible. Thus one obtains 
\be
\sum_{j \in S} \,
\calc_{j a}^{\mu}\Delta_{bc\,\mu}\,I^n_{N-1}(S\setminus\{j\})=
\frac{1}{2}\sum_{j \in S} \,\,I^n_{N-1}(S\setminus\{j\})
\Bigl[\delta_{jc}-\delta_{jb}
+ b_j\,({\cal S}_{ab}-{\cal S}_{ac})\Bigr]\;.
\ee
which leads to
eq.~(\ref{eq41}) without using the inverse of ${\cal S}$.
\item[(iii)] A similar proof is given in 
eqs.~(34) to (37) of \cite{tenred1}, 
but now we have a shift invariant formulation, and 
all coefficients are expressed in terms of 
quantities containing only ${\cal S}_{ij}^{-1}$ 
instead of $H_{ij}$ (the pseudo-inverse of the Gram matrix $G$).
\end{itemize}

\vspace{3mm}

\noindent
In summary, we have shown that
\begin{eqnarray}
\lefteqn{I_5^{n,\,\mu_1 \mu_2 \ldots \mu_r}(a_1,a_2, \cdots,a_r\,;S) =} \nonumber \\
& &
\mbox{} \frac{\calv_{a_r}^{\mu_r}}{B}\, \left[
-I_5^{n,\,\mu_1 \cdots \mu_{r-1}}(a_1, \ldots,a_{r-1}\,;S) + \sum_{j \in S} \, b_j \,  
I^{n,\,\mu_1\ldots\mu_{r-1}}_4(a_1,\ldots,a_{r-1}\,;S\setminus\{j\}) \right]
 \nonumber \\
& &  -\sum_{j \in S} \, 
\calc_{j \, a_r}^{\mu_r}  \, 
I^{n,\,\mu_1\cdots\mu_{r-1}}_4(a_1,\ldots,a_{r-1}\,;S\setminus\{j\})+
I^{\mu_1\ldots\mu_r}_{\eps}(S)\;.
\label{eqFIVERANKN}
\end{eqnarray}
For $r=6$, we obtain
\begin{eqnarray}
I^{\mu_1\ldots\mu_6}_{\eps}(S)&=&\int \diffz \int \diffl \; \frac{l_{\nu}\,\left(\calt_{a_6 d}^{\mu_6\nu}-
2\,\calv_{a_6}^{\mu_6}\calv_{d}^{\nu}/B \right)\,
\tilde{q}_{a_1}^{\mu_{1}}\ldots \tilde{q}_{a_{5}}^{\mu_{5}}}{(l^2-R^2)^5}\nonumber\\
&=&\left(\calt_{a_6 d}^{\mu_6\nu}-2\,\calv_{a_6}^{\mu_6}\calv_{d}^{\nu}/B \right)\,
\left(-\frac{1}{2}\right)^3\,I_5^{n+6}\,\left[
 g^{\cdot\cdot}g^{\cdot\cdot} g^{\cdot\cdot} \right]^{\{\nu\mu_1\cdots\mu_5\}}+\mbox{ terms leading to  }{\cal O}(\eps)\nonumber\\
&=&\left(\calt_{a_6 d}^{\mu_6\nu}-2\,\calv_{a_6}^{\mu_6}\calv_{d}^{\nu}/B \right)\,
\left\{-\frac{1}{8}\,\left(-\frac{1}{24\eps}\right)\,\left[
 g^{\cdot\cdot}g^{\cdot\cdot} g^{\cdot\cdot} \right]^{\{\nu\mu_1\cdots\mu_5\}} +\mbox{ finite }\right\}\nonumber\\
&=&
\frac{1}{8}\,\left(\frac{1}{24\eps}+\mbox{ finite }\right)\,\left[
 \tilde{g}^{\cdot\cdot}g^{\cdot\cdot} g^{\cdot\cdot} \right]^{\{\mu_1\cdots\mu_6\}} \;,
\end{eqnarray}
where $\tilde{g}^{\mu\nu}=g^{\mu\nu}_{[n]}-g^{\mu\nu}_{[4]}$.

In eq. (\ref{eqFIVERANKN}), inverse Gram determinants ($1/B$) have been explicitly 
reintroduced in the term
\begin{eqnarray}
\lefteqn{F^{\mu_1\ldots\mu_{r}}(a_1,\ldots,a_{r}) =} \nonumber \\
& &
\frac{\calv_{a_r}^{\mu_r}}{B}\, \left[
-I_5^{n,\,\mu_1 \cdots \mu_{r-1}}(a_1, \ldots,a_{r-1}\,;S) + \sum_{j \in S} \, b_j \,  
I^{n,\,\mu_1\cdots\mu_{r-1}}_4(a_1,\ldots,a_{r-1}\,;S\setminus\{j\})
\right]\;
\label{eqDEFF1}
\end{eqnarray}
We show in the next section how they drop out. 


\subsection{Cancellation of 1/B terms for $N=5$}\label{oneoverb}

In this section we will prove by induction that $F^{\mu_1\ldots\mu_{r}}(a_1,\ldots,a_{r})$, 
defined in eq.~(\ref{eqDEFF1}), is in fact free from $1/B$ terms 
if the factor $\calv_{a_r}^{\mu_r}/B$ is combined with the expressions inside the 
square bracket.

\vspace{0.3cm} 

\noindent
To this end, we first use 
eq.~(\ref{eqFIVERANKN}) for rank $r-1$
\begin{eqnarray}
\lefteqn{I_5^{n,\,\mu_1 \mu_2 \ldots \mu_{r-1}}(a_1,a_2, 
\cdots,a_{r-1}\,;S) =} \nonumber \\
& &
\mbox{} F^{\mu_1\ldots\mu_{r-1}}(a_1,\ldots,a_{r-1})  -\sum_{j \in S} \, 
\calc_{j \, a_{r-1}}^{\mu_{r-1}}  \, 
I^{n,\,\mu_1\cdots\mu_{r-2}}_4(a_1,\ldots,a_{r-2}\,;S\setminus\{j\})
\label{eqrm1}
\end{eqnarray}
and insert the above equation into eq.~(\ref{eqDEFF1}), leading to 
\begin{eqnarray}
\lefteqn{I_5^{n,\,\mu_1 \mu_2 \ldots \mu_r}(a_1,a_2, \cdots,a_r\,;S) 
=} \nonumber \\
& &
\mbox{} \frac{\calv_{a_r}^{\mu_r}}{B}\, \Biggl[
-F^{\,\mu_1 \cdots \mu_{r-1}}(a_1, \cdots,a_{r-1})  \nonumber \\
& & \mbox{} + \sum_{j \in S} \, 
\calc_{j \, a_{r-1}}^{\mu_{r-1}}  \, 
I^{n,\,\mu_1\cdots\mu_{r-2}}_4(a_1,\ldots,a_{r-2}\,;S\setminus\{j\})+
 \sum_{j \in S} \, b_j \,  I^{n,\,\mu_1\cdots\mu_{r-1}}_4(a_1,
 \ldots,a_{r-1}\,;S\setminus\{j\}) \Biggr]
 \nonumber \\
& &  -\sum_{j \in S} \, 
\calc_{j \, a_r}^{\mu_r}  \, 
I^{n,\,\mu_1\cdots\mu_{r-1}}_4(a_1,\ldots,a_{r-1}\,;S\setminus\{j\})
\label{eqONEOVERB0}
\end{eqnarray}
Actually, the iteration of eq.~(\ref{eqFIVERANKN}) performed in 
 eq.~(\ref{eqONEOVERB0}) singles out the pair of indices 
 $(a_{r-1},\mu_{r-1})$, which hides the manifest symmetry of 
 $I_5^{n,\,\mu_1 \cdots \mu_{r-1}}(a_1, \cdots,a_{r-1}\,;S)$ 
 with respect to all pairs $(a_1,\mu_1)$, $\cdots$, 
 $(a_{r-1},\mu_{r-1})$. However, the explicit cancellation of $1/B$ 
 terms relies on this symmetry.
 Therefore we introduce a symmetrisation operator $\Xi_s$ which acts as
 follows: If $W^{\mu_1\ldots\mu_{s}}(a_1,\ldots,a_{s})$ is a tensor 
 which already is symmetric with respect to the $s-1$ first indices, $\Xi_s$ is
 defined by
\begin{eqnarray}
\Xi_s[W^{\mu_1\ldots\mu_{s}}(a_1,\ldots,a_{s})] & = & 
\frac{1}{s} \, [ W^{\mu_1\ldots\mu_{s}}(a_1,\ldots,a_{s}) + 
\mbox{c. p. } ]\;,
\label{eqONEOVERB8C}
\end{eqnarray}
where ``c.~p." means the sum over cyclic permutations of 
$(a_{1},\mu_{1})$, $\cdots$, $(a_s,\mu_s)$.
Thus we can write eq.~(\ref{eqONEOVERB0}) as
\begin{eqnarray} 
I_5^{n,\,\mu_1 \mu_2 \ldots \mu_r}(a_1,a_2, \cdots,a_r\,;S) 
&=& 
\frac{\calv_{a_r}^{\mu_r}}{B}\, \Biggl[
-\Xi_{r-1}F^{\,\mu_1 \cdots \mu_{r-1}}(a_1, \cdots,a_{r-1}) +
Q^{\mu_1\ldots\mu_{r-1}} \Biggr]
 \nonumber \\
& &  -\sum_{j \in S} \, 
\calc_{j \, a_r}^{\mu_r}  \, 
I^{n,\,\mu_1\cdots\mu_{r-1}}_4(a_1,\ldots,a_{r-1}\,;S\setminus\{j\})\;,
\label{eqFQ}
\end{eqnarray}
where 
\begin{eqnarray}
Q^{\mu_1\ldots\mu_{r-1}} & = &  \sum_{j \in S} \, 
\Biggl\{ \, \Xi_{r-1}\left[
\calc_{j \,a_{r-1}}^{\mu_{r-1}} \, I_4^{n,\mu_1\ldots\mu_{r-2}}(a_1,
\ldots,a_{r-2}\,;S\setminus\{j\})\right] \nonumber \\ 
& &  \; \; \;  \;  \mbox{} + b_j \, 
I_4^{n,\mu_1\ldots\mu_{r-1}}(a_1,\ldots,a_{r-1}\,;S\setminus\{j\}) \, 
\Biggr\}
\label{eqONEOVERB6}
\end{eqnarray}
In order to show  that the $1/B$ terms in the first
line of eq.~(\ref{eqFQ}) drop out,
let us first consider $Q^{\mu_1\ldots\mu_{r-1}}$. 
Using eq.~(\ref{eqNEWRED}) with $N=4$ for the first term and 
eq.~(\ref{eqdef}) with $N=4$ for the second term in 
eq.~(\ref{eqONEOVERB6}), we obtain:
\begin{eqnarray}
Q^{\mu_1\ldots\mu_{r-1}} & = & Q_1^{\mu_1\ldots\mu_{r-1}}+ 
\sum_{j \in S} \Biggl( \Xi_{r-1}[(b_j \,\calv_{a_{r-1}}^{\mu_{r-1}} - 
B \, \calc_{j \, a_{r-1}}^{\mu_{r-1}}) \,
Q_{2 \, j}^{\mu_1\ldots\mu_{r-2}}]\label{eqONEOVERB7}\\
& & \mbox{} + \sum_{k \in S\setminus\{j\}} \Xi_{r-1}\, \int \diffk \;  
\frac{ (q_j^2 -m_j^2) \; (q_k^2 -m_k^2) \; 
 q_{a_1}^{\mu_{1}}\ldots q_{a_{r-2}}^{\mu_{r-2}}}{
 \prod_{i \in S}(q_i^2-m_i^2+i\delta)}  \, 
 \left[ \calc_{j \, a_{r-1}}^{\mu_{r-1}} \, b_k - b_j \, 
 \calc_{k \, a_{r-1}}^{\mu_{r-1}} \right] \Biggr) \nn
\end{eqnarray}
where 
\begin{eqnarray}
Q_1^{\mu_1\ldots\mu_{r-1}} & = & \sum_{j \in S} \Xi_{r-1}\, \Biggl[ 
 \left( b_j \, 
\calt^{\mu_{r-1} \, \nu}_{a_{r-1} \, d} - 2 \,
\calc_{j \, a_{r-1}}^{\mu_{r-1}} \,\calv^{\nu}_{d} \right)\,
\int \diffz \, \int \diffl \; \frac{ l_{\nu}\; 
\tilde{q}_{a_1}^{\mu_{1}}\ldots \tilde{q}_{a_{r-2}}^{\mu_{r-2}}}{(l^2-R_j^2)^4} 
\label{eqONEOVERB8A} \\
Q_{2 \, j}^{\mu_1\ldots\mu_{r-2}} & = &  \int \diffz \, \int \diffl \; 
\frac{\tilde{q}_{a_1}^{\mu_{1}}\ldots \tilde{q}_{a_{r-2}}^{\mu_{r-2}}}{(l^2-R_j^2)^4} 
\, \left( 2 \, \sum_{i=1}^4 z_{i}\, l\cdot\Delta_{d\,i} + l^2 + R_j^2 
\right)
\label{eqONEOVERB8}\\
R_j^2 &=& - \frac{1}{2}\,\sum_{k,l\in S \setminus \{j\}} z_{k} \,
\mcals{j}_{k \, l} \, z_{l} \;,\nn
\end{eqnarray} 
and the identities (\ref{eqONEOVERB1}), (\ref{eqONEOVERB2}) and 
(\ref{eqONEOVERB3}) have been used.
The last term of eq. (\ref{eqONEOVERB7}) vanishes due to 
antisymmetry with respect to $j$ and $k$. 
Therefore, the first line in eq.~(\ref{eqFQ}) 
can be rewritten as 
\be
\frac{\calv_{a_r}^{\mu_r}}{B}\, \left[
-\Xi_{r-1}\,F^{\mu_1\ldots\mu_{r-1}} 
+ Q_1^{\mu_1\ldots\mu_{r-1}}+
\sum_{j \in S} \, \Xi_{r-1}[(b_j \,\calv_{a_{r-1}}^{\mu_{r-1}} - 
B \, \calc_{j \, a_{r-1}}^{\mu_{r-1}}) \,
Q_{2 \, j}^{\mu_1\ldots\mu_{r-2}}]\, \right]  
\ee
The term multiplying $Q_{2 \, j}^{\mu_1\ldots\mu_{r-2}}$  involves 
$(b_j \,\calv_{a_{i}}^{\mu_{i}} - 
B \, \calc_{j \, a_{i}}^{\mu_{i}})\, \calv_{a_{r}}^{\mu_{r}}/B$,  
and due to eq. (\ref{eqEQUATION3}) we have
\begin{eqnarray}
\frac{\calv_{a_{r}}^{\mu_{r}}}{B}\,(b_j \,\calv_{a_{i}}^{\mu_{i}} - 
B \, \calc_{j \, a_{i}}^{\mu_{i}}) & = & \frac{1}{2} \, b_j\,\caltf^{\mu_{i} 
\, \mu_{r}}_{a_{i} \, a_{r}} - \calv_{a_{r}}^{\mu_{r}} \, 
\calc_{j \, a_{i}}^{\mu_{i}}\;,
\label{eqONEOVERB14}
\end{eqnarray}
explicitly free from $1/B$ terms.

\vspace{0.3cm} 

\noindent
For the remaining contribution, we will show that 
$Q_1^{\mu_1\ldots\mu_{r-1}}=\Xi_{r-1}\,
F^{\mu_1\ldots\mu_{r-1}}+{\cal O}(\eps)$ for $r<6$. 
For $r\geq 6$, the ${\cal O}(\eps)$ terms will combine with UV poles and 
therefore have to be taken into account.
For $r<6$, we will prove  that 
\begin{equation}
F^{\mu_1\ldots\mu_{r-1}} = \sum_{j \in S} \, 
\frac{\calv_{a_{r-1}}^{\mu_{r-1}}}{B} \, \Xi_{r-2} 
\left[ \left( b_j \,\calv_{a_{r-2}}^{\mu_{r-2}} - B \, 
\calc_{j \, a_{r-2}}^{\mu_{r-2}} \right) 
\,Q_{2 \, j}^{\mu_1\ldots\mu_{r-3}} \right] + {\cal O}(\eps)
\label{eqONEOVERB9}
\end{equation}
and show by direct calculation that we also have
\begin{equation}
Q_1^{\mu_1\ldots\mu_{r-1}}=\Xi_{r-1}\left[ \sum_{j \in S} 
\,\frac{\calv_{a_{r-1}}^{\mu_{r-1}}}{B} \, 
\Xi_{r-2}\left[ \left(b_j \,\calv_{a_{r-2}}^{\mu_{r-2}} - B \, 
\calc_{j \, a_{r-2}}^{\mu_{r-2}} \right) \,
Q_{2 \, j}^{\mu_1\ldots\mu_{r-3}} \right] \right] + {\cal O}(\eps) 
\label{eqONEOVERB12}
\end{equation}
Eq.~(\ref{eqONEOVERB12}) is established in subsection \ref{eqproof} 
for $r=1,\ldots,5$. 
For $r=6$, the terms of ${\cal O}(\eps)$ are not negligible because the rank 6 pentagon is UV divergent.
This fact leads to the extra term for $r=6$ in eq.~(\ref{eqONEOVERB15}) below.



\noindent
Let us now show 
eq.~(\ref{eqONEOVERB9}) for $r\leq 5$ by induction. The induction
start is $r=2$, because for $r=1$ the absence of $1/B$ terms is trivial. 
For $r=2$, we combine eqs.~(\ref{n5r1}) and (\ref{eqrm1}) to obtain
\be
F^{\mu_1}(a_1)=(4-n)\calv_{a_1}^{\mu_1}\,I^{n+2}_5(S)= {\cal O}(\eps)
\ee
From eq.~(\ref{eqONEOVERB8A}), we obtain
\begin{eqnarray}
Q_1^{\mu_1} & = & \sum_{j \in S} \,
\left( b_j \, \calt^{\mu_1 \, \nu}_{a_1 \, d} - 2 \,
\calc_{j \, a_{1}}^{\mu_{1}} \, \calv^{\nu}_{d} \right) 
\,
\int \diffz  \int \diffl \, \frac{l_{\nu}}{(l^2-R_j^2)^4}  
=0 
\end{eqnarray}
Now let us assume that (\ref{eqONEOVERB9}) is fulfilled for rank $r-1$.
To prove the step $r-1\to r$, we use eq.~(\ref{eqFQ}) for 
$I_5^{n,\,\mu_1 \ldots \mu_{r}}(a_1, \cdots,a_r\,;S)$
and replace $F^{\mu_1\ldots\mu_{r-1}}$ by (\ref{eqONEOVERB9}), 
which is true by the induction assumption, to obtain
\begin{eqnarray}
I_5^{n,\,\mu_1  \ldots \mu_{r}}(a_1, \cdots,a_r\,;S) 
& = & \frac{\calv_{a_{r}}^{\mu_{r}}}{B}\, \Biggl(
 -\,\Xi_{r-1}\left\{\sum_{j \in S} \,
 \frac{\calv_{a_{r-1}}^{\mu_{r-1}}}{B}\,
 \Xi_{r-2}\left[ \left( b_j \,\calv_{a_{r-2}}^{\mu_{r-2}} - B \, 
 \calc_{j \, a_{r-2}}^{\mu_{r-2}} \right) \,Q_{2 \, j}^{\mu_1\ldots
 \mu_{r-3}} \right] \right\} \nonumber \\
& & \mbox{} + Q_1^{\mu_1\ldots\mu_{r-1}} + \sum_{j\in S}\Xi_{r-1}
\left[ \left( b_j \,\calv_{a_{r-1}}^{\mu_{r-1}} - B \, 
\calc_{j \, a_{r-1}}^{\mu_{r-1}} \right) \,Q_{2j}^{\mu_1\ldots\mu_{r-2}} 
\right] \Biggr) \nonumber \\
& & \mbox{} - \sum_{j \in S} \, \calc_{j \, a_{r}}^{\mu_{r}} \, 
I^{n,\,\mu_1  \ldots \mu_{r-1}}_4(a_1, \cdots,a_{r-1}\,;
S\setminus\{j\}) \;.
\label{eqONEOVERB11}
\end{eqnarray}
Comparing with eq.~(\ref{eqONEOVERB12}), we see that the term 
$\Xi_{r-1}\left\{\ldots\right\}$ in the first line of 
eq.~(\ref{eqONEOVERB11}) is equal to $Q_1^{\mu_1\ldots\mu_{r-1}}$.
Comparing the remaining terms to eqs.~(\ref{eqrm1}) and (\ref{eqONEOVERB9}) 
proves our assumption. 
Therefore, using eq.~(\ref{eqONEOVERB14}),  
we see that rank $r$ 5-point integrals can be written as
\begin{eqnarray}
&&I_5^{n,\,\mu_1 \ldots \mu_{r}}(a_1, \cdots,a_r\,;S)  = \nonumber\\
&&\sum_{j \in S} \, \Biggl\{ \Xi_{r-1}\left[
\left( \frac{1}{2} \, b_j \, \caltf^{\mu_{r-1} \, \mu_{r}}_{a_{r-1} \, a_{r}} 
- \calv_{a_{r}}^{\mu_{r}} \, \calc_{j \, a_{r-1}}^{\mu_{r-1}} \right) 
\, Q_{2 \, j}^{\mu_1\ldots\mu_{r-2}} \right]
\label{eqONEOVERB15} \\ 
& & \mbox{} - \calc_{j \, a_{r}}^{\mu_{r}} \, 
I^{n,\,\mu_1  \ldots \mu_{r-1}}_4(a_1, \cdots,a_{r-1}\,;S
\setminus\{j\}) \, \Biggr\}  \nonumber\\
&&
+ \left\{\begin{array}{ll}{\cal O}(\eps)& \mbox{for }r\leq 5\\
 \frac{1}{5}\,\frac{n-4}{4}\,
   \frac{1}{48\eps}\,\sum_{j \in S}\sum_{i \in S\setminus\{j\}}
\left[ \frac{\calv_{\cdot}^{\cdot}}{B}\,{\cal D}_{j\,\cdot\cdot}^{\cdot\cdot}
(g^{\cdot\cdot}\Delta^{\cdot}_{i\cdot}+g^{\cdot\cdot}\Delta^{\cdot}_{i\cdot}+
g^{\cdot\cdot}\Delta^{\cdot}_{i\cdot}) \right]^{\mu_1\ldots\mu_6}_{a_1\ldots a_6}& \mbox{for }r=6
\end{array}
 \right.
 \;,\nn
\end{eqnarray}
where 
\begin{eqnarray}
{\cal D}^{\mu_1  \mu_2}_{j\,a_1 \, a_2} & = & 
(b_j \, \calt^{\mu_1 \, \mu_2}_{a_1 \, a_2} - 
\calc^{\mu_2}_{j \, a_2} \, \calv^{\mu_1}_{a_1} - 
\calc^{\mu_1}_{j \, a_1} \, \calv^{\mu_2}_{a_2})\;.
\label{eqDEFD} 
\end{eqnarray}
Note that 
\begin{eqnarray}
\sum_j {\cal D}^{\mu_1  \mu_2}_{j\,a_1  a_2}&=&
B\,\left({\cal T}_{a_1 a_2}^{\mu_1 \mu_2}-2\frac{\calv^{\mu_1}_{a_1}\calv^{\mu_2}_{a_2}}{B} \right)\nonumber\\
&=&B\,\left({\cal T}_{a_1 a_2\,[n]}^{\mu_1 \mu_2}-{\cal T}_{a_1 a_2\,[4]}^{\mu_1 \mu_2} \right)
=B\,g^{\mu_1 \mu_2}_{[n-4]}\;.
\end{eqnarray}
We see that eq.~(\ref{eqONEOVERB15})  
is a combination of 4-point integrals, and a rational part for $r=6$.

\subsection{Auxiliary relations} \label{eqproof}

Here we establish eq.~(\ref{eqONEOVERB12})
by direct calculation of $Q^{\mu_1\ldots\mu_{r-1}}$ 
and $Q_{2j}^{\mu_1\ldots\mu_{r-2}}$ for $r=1,\ldots,5$.  
Using the definition (\ref{eqONEOVERB8A}) and eqs.~(\ref{eqONEOVERB1}),
(\ref{eqONEOVERB2}) and (\ref{eqONEOVERB3}), one gets:
\begin{eqnarray}
Q_1^{\mu_1 \, \mu_2} & = &  - \frac{1}{2} \, \sum_{j \in S} \, 
{\cal D}^{\mu_1  \mu_2}_{j\,a_1 \, a_2} \,I^{n+2}_{4}(S\setminus\{j\})\label{eqQ1r2}\\
Q_1^{\mu_1 \, \mu_2 \, \mu_3} & = &  - \frac{1}{3} \, \sum_{j \in S}\, 
\sum_{i \in S\setminus\{j\}}\, I^{n+2}_{4}(i;S\setminus\{j\}) 
\, ({\cal D}^{\mu_2  \mu_3}_{j\,a_2 \, a_3} 
\, \Delta^{\mu_1}_{a_1 \, i} + {\cal D}^{\mu_1  \mu_3}_{j\,a_1 \, a_3} \, 
\Delta^{\mu_2}_{a_2 \, i}  + {\cal D}^{\mu_1  \mu_2}_{j\,a_1 \, a_2} \, 
\Delta^{\mu_3}_{a_3 \, i} ) \nn\\
&&\label{eqQ1r3}\\
Q_1^{\mu_1 \, \mu_2 \, \mu_3 \, \mu_4} & = &  \frac{1}{4} \, 
\sum_{j \in S} \, \Biggl[
  \frac{1}{2} \,  I^{n+4}_{4}(S\setminus\{j\}) \, 
  ( g^{\mu_1 \, \mu_2} \, {\cal D}^{\mu_3  \mu_4}_{j\,a_3 \, a_4}  
  + g^{\mu_1 \, \mu_3} \, {\cal D}^{\mu_2  \mu_4}_{j\,a_2 \, a_4} 
  + g^{\mu_1 \, \mu_4} \, {\cal D}^{\mu_2  \mu_3}_{j\,a_2 \, a_3}\nn\\
& & \mbox{} + g^{\mu_2 \, \mu_3} \, {\cal D}^{\mu_1  \mu_4}_{j\,a_1 \, a_4}  
+ g^{\mu_2 \, \mu_4} \, {\cal D}^{\mu_1  \mu_3}_{j\,a_1 \, a_3} + 
g^{\mu_3 \, \mu_4} \, {\cal D}^{\mu_1  \mu_2}_{j\,a_1 \, a_2} ) \nn \\
& & \mbox{}  - \sum_{i,k \in S\setminus\{j\}}\, I^{n+2}_{4}(i,k;S\setminus\{j\}) 
\nonumber \\ 
& & \mbox{} \times ( \Delta^{\mu_1}_{a_1 \, i} \, 
\Delta^{\mu_2}_{a_2 \, k} \, {\cal D}^{\mu_3  \mu_4}_{j\,a_3 \, a_4} + 
\Delta^{\mu_1}_{a_1 \, i} \, \Delta^{\mu_3}_{a_3 \, k} \, 
{\cal D}^{\mu_2  \mu_4}_{j\,a_2 \, a_4} + \Delta^{\mu_1}_{a_1 \, i} \, 
\Delta^{\mu_4}_{a_4 \, k} \, {\cal D}^{\mu_2  \mu_3}_{j\,a_2 \, a_3}\nonumber \\
& & \mbox{} + \Delta^{\mu_2}_{a_2 \, i} \, \Delta^{\mu_3}_{a_3 \, k} 
\, {\cal D}^{\mu_1  \mu_4}_{j\,a_1 \, a_4} + \Delta^{\mu_2}_{a_2 \, i} \, 
\Delta^{\mu_4}_{a_4 \, k} \, {\cal D}^{\mu_1  \mu_3}_{j\,a_1 \, a_3} + 
\Delta^{\mu_3}_{a_3 \, i} \, \Delta^{\mu_4}_{a_4 \, k} \, 
{\cal D}^{\mu_1  \mu_2}_{j\,a_1 \, a_2} ) \Biggr] \label{eqQ1r4}\\
Q_1^{\mu_1 \, \mu_2 \, \mu_3 \, \mu_4\, \mu_5} & = &  \frac{1}{5} \, 
\sum_{j \in S} \, \Biggl[
  \frac{1}{2} \, \sum_{i\in S\setminus\{j\}} I^{n+4}_{4}(i;S\setminus\{j\}) \nn\\
&& \times \left( g^{\mu_1 \, \mu_2} \, {\cal D}^{\mu_3  \mu_4}_{j\,a_3 \, a_4}  \,\Delta^{\mu_5}_{a_5 i}
  + g^{\mu_1 \, \mu_3} \, {\cal D}^{\mu_2  \mu_4}_{j\,a_2 \, a_4} \,\Delta^{\mu_5}_{a_5 i}
  + \ldots \mbox{(altogether 30 terms)}\right)\nn\\
& & \mbox{}  - \sum_{i,k,l \in S\setminus\{j\}}\, I^{n+2}_{4}(i,k,l;S\setminus\{j\}) 
\nonumber \\ 
& & \mbox{} \times \left( \Delta^{\mu_1}_{a_1 \, i} \, 
\Delta^{\mu_2}_{a_2 \, k} \,\Delta^{\mu_3}_{a_3 \, l}\,
{\cal D}^{\mu_4  \mu_5}_{j\,a_4 \, a_5} + \ldots \mbox{(altogether 10 terms)}\right) \Biggr]\;. \label{eqQ1r5}
\end{eqnarray}
Similarly, eq.~(\ref{eqONEOVERB8}) leads to
\begin{eqnarray}
Q_{2j}^{\mu_1} & = &  (2-n) \, \sum_{i \in S\setminus\{j\}} \, 
\Delta^{\mu_1}_{a_1 \, i} \,
I^{n+2}_{4}(i;S\setminus\{j\}) \label{eqQ2r1}\\
Q_{2j}^{\mu_1 \, \mu_2} & = &  (1-n)  \, \Biggl[ 
- \frac{1}{2} \, g^{\mu_1 \, \mu_2} \, I^{n+4}_{4}(S\setminus\{j\})  
\nonumber \\
& & \mbox{} + 
 \sum_{i,k \in S\setminus\{j\}} \, \Delta^{\mu_1}_{a_1 \, i}  \, 
 \Delta^{\mu_2}_{a_2 \, k}  \, I^{n+2}_{4}(i,k;S\setminus\{j\}) 
 \Biggr] \label{eqQ2r2} \\
Q_{2j}^{\mu_1 \, \mu_2 \, \mu_3} & = &  n \, \Biggl[ 
 \frac{1}{2} \, \sum_{i \in S\setminus\{j\}} \, I^{n+4}_{4}(i;S\setminus\{j\}) 
 \, ( g^{\mu_1 \, \mu_2} \, \Delta^{\mu_3}_{a_3 \, i} + g^{\mu_1 \, 
 \mu_3} \, \Delta^{\mu_2}_{a_2 \, i} + g^{\mu_2 \, \mu_3} \, 
 \Delta^{\mu_1}_{a_1 \, i} ) \nonumber \\
& & \mbox{}  -  \sum_{i,k,l \in S\setminus\{j\}} \, \Delta^{\mu_1}_{a_1 \, i}  \, 
\Delta^{\mu_2}_{a_2 \, k} \, \Delta^{\mu_3}_{a_3 \, l}  \, 
I^{n+2}_{4}(i,k,l;S\setminus\{j\}) \Biggr] \label{eqQ2r3}\\
Q_{2j}^{\mu_1 \, \mu_2 \, \mu_3\, \mu_4} & = &  (-1-n) \, \Biggl[ 
\frac{1}{4}\left\{ \,
g^{\mu_1 \, \mu_2} g^{\mu_3 \, \mu_4}+
g^{\mu_1 \, \mu_3} g^{\mu_2 \, \mu_4}+
g^{\mu_1 \, \mu_4} g^{\mu_2 \, \mu_3}
\,\right\} I^{n+6}_{4}(S\setminus\{j\})
\nonumber \\
& & \mbox{} 
 -\frac{1}{2} \, \sum_{i_1,i_2 \in S\setminus\{j\}} \, I^{n+4}_{4}(i_1,i_2,;S\setminus\{j\}) 
 \, \left\{\right.\nonumber\\
& & \mbox{}  \left. 
  \quad\qquad g^{\mu_1 \, \mu_2} \, \Delta^{\mu_3}_{a_3 \, i_1} \Delta^{\mu_4}_{a_4 \, i_2} 
   + g^{\mu_1 \, \mu_3} \, \Delta^{\mu_2}_{a_2 \, i_1} \Delta^{\mu_4}_{a_4 \, i_2}\right.\nonumber\\
& & \mbox{}  \left.
 \qquad + g^{\mu_2 \, \mu_3} \, \Delta^{\mu_1}_{a_1 \, i_1} \Delta^{\mu_4}_{a_4 \, i_2}
   + g^{\mu_1 \, \mu_4} \, \Delta^{\mu_2}_{a_2 \, i_1} \Delta^{\mu_3}_{a_3 \, i_2}\right.\nonumber\\
& & \mbox{}  \left.
 \qquad + g^{\mu_2 \, \mu_4} \, \Delta^{\mu_1}_{a_1 \, i_1} \Delta^{\mu_3}_{a_3 \, i_2}
  + g^{\mu_3 \, \mu_4} \, \Delta^{\mu_1}_{a_1 \, i_1} \Delta^{\mu_2}_{a_2 \, i_2}
  \right\} \nonumber \\
& & \mbox{}  +  \sum_{i_1,i_2,i_3,i_4 \in S\setminus\{j\}}  \Delta^{\mu_1}_{a_1 \, i_1}  \, 
\Delta^{\mu_2}_{a_2 \, i_2} \, \Delta^{\mu_3}_{a_3 \, i_3}  \, \Delta^{\mu_4}_{a_4 \, i_4}\,
I^{n+2}_{4}(i_1,i_2,i_3,i_4;S\setminus\{j\}) \Biggr] 
\label{eqQ2r4} 
\end{eqnarray}
If we insert eqs.~(\ref{eqQ2r1}) to (\ref{eqQ2r3}) 
for $Q_{2j}$ into the 
right-hand side of eq.~(\ref{eqONEOVERB12}) and use 
eq.~(\ref{eqONEOVERB14}), we obtain the expressions 
(\ref{eqQ1r2}) to (\ref{eqQ1r4}) for $Q_1$, up to terms of order $\eps$ stemming from 
the dimensionality $n$ in the prefactor of the 
$Q_{2j}^{\mu_1 \ldots \mu_r}$ integrals, which become important 
for rank 6 pentagons since the latter are UV divergent, as explained above.

\section{Useful relations}\label{relations}

In this appendix we give a collection of formulae
which are useful if one wants to perform algebraic simplifications 
of loop amplitudes. The relations in subsection \ref{formrel} 
can also be very useful to  
perform checks on the implementation of the form factors in a 
computer program.
Some of the relations are already given in the main text, but the purpose here
is to list them for quick reference. 

\subsection{Relations between the form factors}\label{formrel}

The following identities have been used extensively to obtain the
relations given below:
\begin{eqnarray}
q_a \cdot \Delta_{b \, c} & = & \frac{1}{2} \left( \, q_b^2-m_b^2 -
[\,q_c^2 - m_c^2\,]-
\calst_{a \, b} + \calst_{a \, c} \right)  
\label{eqQDOTDELTA} \\
\Delta_{l \, a} \cdot \Delta_{b \, c} & = & \frac{1}{2} 
\left( \, \calst_{l \, c} - \calst_{l \, b} + \calst_{a \, b} - 
\calst_{a \, c} \right)  
\label{eqDELTADOTDELTA}
\end{eqnarray}
Using relations (\ref{eqQDOTDELTA}) and (\ref{eqDELTADOTDELTA}) and 
multiplying with a vector $\Delta$ both the 
definition and the expression in terms of form factors 
of an integral, one finds 
relations between the form factors.

\subsubsection{Four-point functions}

For example, we have
\begin{eqnarray}
I_4^{n,\,\mu_1}(a;S) \, \Delta_{b \, c}^{\mu_1} & = & 
\int d\bar{k} \; 
\frac{q_{a}\cdot\Delta_{b \, c}}{\prod_{i \in S} (q_i^2-m_i^2+i\delta)} \nonumber \\
& = & \frac{1}{2} \, \Biggl[
\int d\bar{k} \; 
\frac{1}{\prod_{i \in S\setminus\{b\}} (q_i^2-m_i^2+i\delta)}
- \int d\bar{k} \; 
\frac{1}{\prod_{i \in S\setminus\{c\}}(q_i^2-m_i^2+i\delta)}\nn\\
&& -
(\calst_{a \, b} - \calst_{a \, c}) \, \int 
d\bar{k} \;\frac{1}{\prod_{i \in S} (q_i^2-m_i^2+i\delta)}
\Biggr] \nonumber \\
& = & \frac{1}{2} \, \sum_{l \in S} \,  A^{4,1}_{l}(S) \, 
\left( \, \calst_{l \, c} - \calst_{l \, b} + \calst_{a \, b} - 
\calst_{a \, c} \right) 
\label{eqD1}
\end{eqnarray}
Eq.~(\ref{eqD1}) implies
\begin{eqnarray}
\sum_{l \in S} \,  A^{4,1}_{l}(S) & = & - A^{4,0}(S) 
\label{eq1R4P1} \\
\sum_{l \in S} \,  A^{4,1}_{l}(S) \,\left( \, \calst_{l \, c} - 
\calst_{l \, b} \right) & = & A^{3,0}(S\setminus\{b\}) - 
A^{3,0}(S\setminus\{c\})\;.
\label{eq1R4P2} 
\end{eqnarray}
One can proceed in the same way for all the four-point functions and 
finds for rank 2:
\begin{eqnarray}
\sum_{l_2 \in S} \,   A^{4,2}_{l_1 \, l_2}(S) & = & 
- A^{4,1}_{l_1}(S) 
\label{eq2R4P1} \\
\sum_{l_2 \in S} \,  A^{4,2}_{l_1 \, l_2}(S) \,\left( \, 
\calst_{l_2 \, c} - \calst_{l_2 \, b} \right) & = & 2 \, 
( \delta_{l_1 \, c} -  \delta_{l_1 \, b} ) \, B^{4,2}(S) \nonumber \\
& & \mbox{} + \dbar_{l_1 \, b} \, A^{3,1}_{l_1}(S\setminus\{b\}) - 
\dbar_{l_1 \, c} \, A^{3,1}_{l_1}(S\setminus\{c\})
\;,\label{eq2R4P2} 
\end{eqnarray}
where the definition 
\begin{equation}
\dbar_{j l} = 1- \delta_{j l} =  \left\{
\begin{array}{c}
1 \; \mbox{if} \; j \neq l \\
0 \; \mbox{if} \; j = l
\end{array}
\right.
\label{eqDEFDELTABAR}
\end{equation}
has been used. 
For rank 3, one obtains:
\begin{eqnarray}
\sum_{l \in S} \,   B^{4,3}_{l}(S) & = & - B^{4,2}(S) 
\label{eq3R4P1} \\
\sum_{l \in S} \,  B^{4,3}_{l}(S) \,\left( \, \calst_{l \, c} - 
\calst_{l \, b} 
\right) & = &  B^{3,2}(S\setminus\{b\}) - B^{3,2}(S\setminus\{c\})
\label{eq3R4P2} \\
\sum_{l_3 \in S} \,   A^{4,3}_{l_1 \, l_2 \, l_3}(S) & = & 
- A^{4,2}_{l_1 \, l_2}(S) 
\label{eq3R4P3} \\
\sum_{l_3 \in S} \,  A^{4,3}_{l_1 \, l_2 \, l_3}(S) \,\left( \, 
\calst_{l_3 \, c} - 
\calst_{l_3 \, b} \right) 
& = & 2 \, ( \delta_{l_2 \, c} - \delta_{l_2 \, b} ) \, 
B^{4,3}_{l_1}(S) + 2 \, ( \delta_{l_1 \, c} - \delta_{l_1 \, b} ) 
\, B^{4,3}_{l_2}(S) \label{eq3R4P4} \\
& & \mbox{} + \dbar_{l_1 \, b} \,\dbar_{l_2 \, b} \, 
A^{3,2}_{l_1 \, l_2}(S\setminus\{b\}) - \dbar_{l_1 \, c} \, 
\dbar_{l_2 \, c} \, A^{3,2}_{l_1 \, l_2}(S\setminus\{c\})\nn 
\end{eqnarray}
For rank 4:
\begin{eqnarray}
\sum_{l_2 \in S} \,   B^{4,4}_{l_1 \, l_2}(S) & = & - B^{4,3}_{l_1}(S) 
\label{eq4R4P1} \\
\sum_{l_2 \in S} \,  B^{4,4}_{l_1 \, l_2}(S) \,\left( \, \calst_{l_2 \, c} - \calst_{l_2 \, b} \right) & = &  2 \, (\delta_{l_1 \, c} - 
\delta_{l_1 \, b}) \, C^{4,4}(S) \nonumber \\
& & \mbox{} + \dbar_{l_1 \, b} \, B^{3,3}_{l_1}(S\setminus\{b\}) - \dbar_{l_1 \, c} \, B^{3,3}_{l_1}(S\setminus\{c\})
\label{eq4R4P2} \\
\sum_{l_4 \in S} \,   A^{4,4}_{l_1 \, l_2 \, l_3 \, l_4}(S) & = & - A^{4,3}_{l_1 \, l_2 \, l_3}(S) 
\label{eq4R4P3} \\
\sum_{l_4 \in S} \,  A^{4,4}_{l_1 \, l_2 \, l_3 \, l_4}(S) \,\left( \, \calst_{l_4 \, c} - \calst_{l_4 \, b} \right) & = & 2 \, \sum_{i = l_1,l_2,l_3} \, ( \delta_{i \, c} - \delta_{i \, b} ) \, B^{4,4}_{\{l_1,l_2,l_3\} - \{i\}}(S)  \nonumber \\
& & \mbox{} + \dbar_{l_1 \, b} \, \dbar_{l_2 \, b} \, \dbar_{l_3 \, b} \, A^{3,3}_{l_1 \, l_2 \, l_3}(S\setminus\{b\}) \nonumber \\
& & \mbox{}  - \dbar_{l_1 \, c} \, \dbar_{l_2 \, c} \, \dbar_{l_3 \, c} \, A^{3,3}_{l_1 \, l_2 \, l_3}(S\setminus\{c\})
\label{eq4R4P4} 
\end{eqnarray}

\subsubsection{Five-point functions}
Below we will use the definition 
\be
H_{ij}=2\,\left(\frac{b_ib_j}{B}-S_{ij}^{-1}\right)
\ee
Rank 1:
\begin{eqnarray}
\sum_{l \in S} \,  A^{5,1}_{l}(S) & = & - A^{5,0}(S) 
\label{eq1R5P1} \\
\sum_{l \in S} \,  A^{5,1}_{l}(S) \,
\left( \, \calst_{l \, c} - \calst_{l \, b} \right) & = 
& A^{4,0}(S\setminus\{b\}) - A^{4,0}(S\setminus\{c\})
\label{eq1R5P2} 
\end{eqnarray}
Rank 2:
\begin{eqnarray}
\sum_{l_2 \in S} \,   A^{5,2}_{l_1 \, l_2}(S) & = & - A^{5,1}_{l_1}(S) 
\label{eq2R5P1} \\
\sum_{l_2 \in S} \,  A^{5,2}_{l_1 \, l_2}(S) \,\left( \, \calst_{l_2 \, c} - 
\calst_{l_2 \, b} \right) & = & 2 \, ( \delta_{l_1 \, c} -  \delta_{l_1 \, b} ) 
\, B^{5,2}(S) \nonumber \\
& & \mbox{} + \dbar_{l_1 \, b} \, A^{4,1}_{l_1}(S\setminus\{b\}) - 
\dbar_{l_1 \, c} \, A^{4,1}_{l_1}(S\setminus\{c\})
\label{eq2R5P2} 
\end{eqnarray}
Rank 3:
\begin{eqnarray}
\sum_{l \in S} \,  B^{5,3}_{l}(S)  =  - B^{5,2}(S) 
\hspace*{7cm}\label{eq3R5P1} \\
\sum_{l_3 \in S} \,  A^{5,3}_{l_1 \, l_2 \, l_3}(S)  =  - A^{5,2}_{l_1 \, l_2}(S) 
\hspace*{7cm}\label{eq3R5P3} \\
\sum_{l_3 \in S} \,  \left[A^{5,3}_{l_1 \, l_2 \, l_3 }(S) +
H_{l_1l_2}\,B^{5,3}_{l_3}(S)
\right]
\,\left( \, \calst_{l_3 \, c} -\calst_{l_3 \, b} \right) 
=  \hspace*{4cm}\nn\\
H_{l_1l_2}\,\left[ B^{4,2}(S\setminus\{b\})-
B^{4,2}(S\setminus\{c\})  \right]
+2 \sum_{i = l_1,l_2} \, ( \delta_{i \, c} - \delta_{i \, b} ) 
\, B^{5,3}_{\{l_1,l_2\} \setminus \{i\}}(S)  \nonumber \\
 \mbox{} + \dbar_{l_1 \, b} \, \dbar_{l_2 \, b} 
 \, A^{4,2}_{l_1 \, l_2 }(S\setminus\{b\})   
 - \dbar_{l_1 \, c} \,\dbar_{l_2 \, c} 
 \, A^{4,2}_{l_1 \, l_2 }(S\setminus\{c\})
\label{eq3R5P4} 
\end{eqnarray}
Rank 4:
\begin{eqnarray}
\sum_{l_2 \in S} \,   B^{5,4}_{l_1 \, l_2}(S)  =  
- B^{5,3}_{l_1}(S) 
\hspace*{7cm}\label{eq4R5P1} \\
\sum_{l_4 \in S} \,   A^{5,4}_{l_1 \, l_2 \, l_3 \, l_4}(S)  =  
- A^{5,3}_{l_1 \, l_2 \, l_3}(S) 
\hspace*{7cm}\label{eq4R5P3} \\
\sum_{l_4 \in S} \,  \left[A^{5,4}_{l_1 \, l_2 \, l_3 \, l_4}(S) +
H_{l_1l_2}\,B^{5,4}_{l_3\,l_4}(S)+
H_{l_1l_3}\,B^{5,4}_{l_2\,l_4}(S)+
H_{l_2l_3}\,B^{5,4}_{l_1\,l_4}(S)
\right]
\,\left( \, \calst_{l_4 \, c} -\calst_{l_4 \, b} \right) =\nonumber\\
  \quad H_{l_1l_2}\,\left[2\,(\delta_{l_3\, c}-\delta_{l_3 \, b})
\,C^{5,4}(S)+
\dbar_{l_3\,b}\, B^{4,3}_{l_3}(S\setminus\{b\})-\dbar_{l_3\,c}\,
B^{4,3}_{l_3}(S\setminus\{c\})  \right]\nonumber\\
+H_{l_1l_3}\,\left[2\,(\delta_{l_2\, c}-\delta_{l_2 \, b})
\,C^{5,4}(S)+
\dbar_{l_2\,b}\, B^{4,3}_{l_2}(S\setminus\{b\})-\dbar_{l_2\,c}\,
B^{4,3}_{l_2}(S\setminus\{c\})  \right]\nonumber\\
+H_{l_2l_3}\,\left[2\,(\delta_{l_1\, c}-\delta_{l_1 \, b})
\,C^{5,4}(S)+
\dbar_{l_1\,b}\, B^{4,3}_{l_1}(S\setminus\{b\})-\dbar_{l_1\,c}\,
B^{4,3}_{l_1}(S\setminus\{c\})  \right]\nonumber\\
+2 \sum_{i = l_1,l_2,l_3} \, ( \delta_{i \, c} - \delta_{i \, b} ) 
\, B^{5,4}_{\{l_1,l_2,l_3\} \setminus \{i\}}(S)  \nonumber \\
 \mbox{} + \dbar_{l_1 \, b} \, \dbar_{l_2 \, b} \,\dbar_{l_3 \, b}
 \, A^{4,3}_{l_1 \, l_2 \, l_3}(S\setminus\{b\}) 
   - \dbar_{l_1 \, c} \,\dbar_{l_2 \, c} \,\dbar_{l_3 \, c}
 \, A^{4,3}_{l_1 \, l_2 \, l_3}(S\setminus\{c\})\nn\\
\label{eq4R5P4} 
\end{eqnarray}
Rank 5:
\be
\sum_{l_5 \in S} \,   A^{5,5}_{l_1 \, l_2 \, l_3 \, l_4 \, l_5}(S) 
 =  - A^{5,4}_{l_1 \, l_2 \, l_3 \, l_4}(S) \hspace*{7cm}
\label{eq5R5P3} 
\ee
\begin{eqnarray}
&& (H_{l_1 \, l_2} \, H_{l_3 \, l_4} + H_{l_1 \, l_3} \, H_{l_2 \, l_4} + H_{l_1 \, l_4} 
 \, H_{l_2 \, l_3})
\, \sum_{l_5 \in S} C^{5,5}_{l_5}(S)  \nonumber \\
& &  + H_{l_1 \, l_2}\, \sum_{l_5 \in S} B^{5,5}_{l_3 \, l_4 \, l_5}(S)
	 + H_{l_1 \, l_3}\, \sum_{l_5 \in S} B^{5,5}_{l_2 \, l_4 \, l_5}(S)
	 + H_{l_1 \, l_4}\, \sum_{l_5 \in S} B^{5,5}_{l_2 \, l_3 \, l_5}(S) \nn \\
& & + H_{l_2 \, l_3}\, \sum_{l_5 \in S} B^{5,5}_{l_1 \, l_4 \, l_5}(S)
	 + H_{l_2 \, l_4}\, \sum_{l_5 \in S} B^{5,5}_{l_1 \, l_3 \, l_5}(S)
	 + H_{l_3 \, l_4}\, \sum_{l_5 \in S} B^{5,5}_{l_1 \, l_2 \, l_5}(S)  \nn \\
&= &  - (H_{l_1 \, l_2} \, H_{l_3 \, l_4} + H_{l_1 \, l_3} \, H_{l_2 \, l_4} + 
H_{l_1 \, l_4} \, H_{l_2 \, l_3}) \, C^{5,4}(S)\nn\\
&&	 - H_{l_1 \, l_2} \, B^{5,4}_{l_3 \, l_4 }(S)
	 - H_{l_1 \, l_3} \, B^{5,4}_{l_2 \, l_4 }(S) 
	 - H_{l_1 \, l_4} \, B^{5,4}_{l_2 \, l_3 }(S)\nonumber \\
& &	 - H_{l_2 \, l_3} \, B^{5,4}_{l_1 \, l_4 }(S)
	 - H_{l_2 \, l_4} \, B^{5,4}_{l_1 \, l_3 }(S)
	 - H_{l_3 \, l_4} \, B^{5,4}_{l_1 \, l_2 }(S)\\
&&\nn\\
&& (H_{l_1 \, l_2} \, H_{l_3 \, l_4}+H_{l_1 \, l_3} \, H_{l_2 \, l_4}+
H_{l_1 \, l_4} \, H_{l_2 \, l_3})
          \, \sum_{l_5 \in S} \,(\calst_{l_5 \, c }-\calst_{l_5 \, b }) \, 
	  C^{5,5}_{l_5}(S)   \nonumber \\
& & \mbox{} + H_{l_1 \, l_2}
	  \, \left[ 2 \, (\delta_{l_3 \, b }-\delta_{l_3 \, c }) \, C^{5,5}_{l_4}(S)
	   +2 \, (\delta_{l_4 \, b }-\delta_{l_4 \, c }) \, C^{5,5}_{l_3}(S) +
	   \sum_{l_5 \in S} \,(\calst_{l_5 \, c }-\calst_{l_5 \, b }) \, 
	   B^{5,5}_{l_3 \, l_4 \, l_5}(S) \right] \nonumber \\
& & \mbox{}+ H_{l_1 \, l_3}
	  \, \left[ 2 \, (\delta_{l_2 \, b }-\delta_{l_2 \, c }) \, C^{5,5}_{l_4}(S)
	   +2 \, (\delta_{l_4 \, b }-\delta_{l_4 \, c }) \, C^{5,5}_{l_2}(S)
	   +\sum_{l_5 \in S} \,(\calst_{l_5 \, c }-\calst_{l_5 \, b }) \, 
	   B^{5,5}_{l_2 \, l_4 \, l_5}(S) \right] \nonumber \\
& & \mbox{} + H_{l_1 \, l_4}
	  \, \left[ 2 \, (\delta_{l_2 \, b }-\delta_{l_2 \, c }) \, C^{5,5}_{l_3}(S)
	   +2 \, (\delta_{l_3 \, b }-\delta_{l_3 \, c }) \, C^{5,5}_{l_2}(S)
	   +\sum_{l_5 \in S} \,(\calst_{l_5 \, c }-\calst_{l_5 \, b }) \, 
	   B^{5,5}_{l_2 \, l_3 \, l_5}(S) \right] \nonumber \\
& & \mbox{} + H_{l_2 \, l_3}
	  \, \left[ 2 \, (\delta_{l_1 \, b }-\delta_{l_1 \, c }) \, C^{5,5}_{l_4}(S)
	   +2 \, (\delta_{l_4 \, b }-\delta_{l_4 \, c }) \, C^{5,5}_{l_1}(S)
	   +\sum_{l_5 \in S} \,(\calst_{l_5 \, c }-\calst_{l_5 \, b }) \, 
	   B^{5,5}_{l_1 \, l_4 \, l_5}(S) \right] \nonumber \\
& & \mbox{} + H_{l_2 \, l_4}
	  \, \left[ 2 \, (\delta_{l_1 \, b }-\delta_{l_1 \, c }) \, C^{5,5}_{l_3}(S)
	   +2 \, (\delta_{l_3 \, b }-\delta_{l_3 \, c }) \, C^{5,5}_{l_1}(S)
	   +\sum_{l_5 \in S} \,(\calst_{l_5 \, c }-\calst_{l_5 \, b }) \, 
	   B^{5,5}_{l_1 \, l_3 \, l_5}(S) \right] \nonumber \\
& & \mbox{} + H_{l_3 \, l_4}
	  \, \left[ 2 \, (\delta_{l_1 \, b }-\delta_{l_1 \, c }) \, C^{5,5}_{l_2}(S)
	   +2 \, (\delta_{l_2 \, b }-\delta_{l_2 \, c }) \, C^{5,5}_{l_1}(S)
	   +\sum_{l_5 \in S} \,(\calst_{l_5 \, c }-\calst_{l_5 \, b }) \, 
	   B^{5,5}_{l_1 \, l_2 \, l_5}(S) \right] \nonumber \\
& & \mbox{} + \sum_{l_5 \in S} \,(\calst_{l_5 \, c }-\calst_{l_5 \, b }) \, 
A^{5,5}_{l_1\,l_2\,l_3\,l_4\,l_5}(S)\nn\\
&& \mbox{} + 2 \, (\delta_{l_1 \, b }-\delta_{l_1 \, c }) \, 
	 B^{5,5}_{l_2 \, l_3 \, l_4}(S)
	 + 2 \, (\delta_{l_2 \, b }-\delta_{l_2 \, c }) \, 
	 B^{5,5}_{l_1 \, l_3 \, l_4}(S) \nonumber \\
& & \mbox{} + 2 \, (\delta_{l_3 \, b }-\delta_{l_3 \, c }) \, 
B^{5,5}_{l_1 \, l_2 \, l_4}(S)
	 + 2 \, (\delta_{l_4 \, b }-\delta_{l_4 \, c }) \, 
	 B^{5,5}_{l_1 \, l_2 \, l_3}(S) \nonumber \\
&= & \mbox{}   (H_{l_1 \, l_2} \, H_{l_3 \, l_4}+
H_{l_1 \, l_3} \, H_{l_2 \, l_4}+
H_{l_1 \, l_4} \, H_{l_2 \, l_3})
	  \, ( C^{4,4}(S\setminus \{b\})-C^{4,4}(S\setminus \{c\}) ) \nonumber \\
& & \mbox{}+ H_{l_1 \, l_2}
	  \, ( \dbar_{l_3 \, b } \, \dbar_{l_4 \, b } \, 
	  B^{4,4}_{l_3 \, l_4}(S\setminus \{ b \})
	   -\dbar_{l_3 \, c } \, \dbar_{l_4 \, c } \, 
	   B^{4,4}_{l_3 \, l_4}(S\setminus \{ c \}) )\nonumber \\
& & \mbox{}	 + H_{l_1 \, l_3}
	  \, ( \dbar_{l_2 \, b } \, \dbar_{l_4 \, b } \, 
	  B^{4,4}_{l_2 \, l_4}(S\setminus \{ b \})
	   -\dbar_{l_2 \, c } \, \dbar_{l_4 \, c } \, 
	   B^{4,4}_{l_2 \, l_4}(S\setminus \{ c \}) ) \nonumber \\
& & \mbox{}	 + H_{l_1 \, l_4}
	  \, ( \dbar_{l_2 \, b } \, \dbar_{l_3 \, b } \, 
	  B^{4,4}_{l_2 \, l_3}(S\setminus \{ b \})
	   -\dbar_{l_2 \, c } \, \dbar_{l_3 \, c } \, 
	   B^{4,4}_{l_2 \, l_3}(S\setminus \{ c \}) ) \nonumber \\
& & \mbox{}	 + H_{l_2 \, l_3}
	  \, ( \dbar_{l_1 \, b } \, \dbar_{l_4 \, b } \, 
	  B^{4,4}_{l_1 \, l_4}(S\setminus \{ b \})
	   -\dbar_{l_1 \, c } \, \dbar_{l_4 \, c } \, 
	   B^{4,4}_{l_1 \, l_4}(S\setminus \{ c \}) ) \nonumber \\
& & \mbox{}	 + H_{l_2 \, l_4}
	  \, ( \dbar_{l_1 \, b } \, \dbar_{l_3 \, b } \, 
	  B^{4,4}_{l_1 \, l_3}(S\setminus \{ b \})
	   -\dbar_{l_1 \, c } \, \dbar_{l_3 \, c } \, 
	   B^{4,4}_{l_1 \, l_3}(S\setminus \{ c \}) ) \nonumber \\
& & \mbox{}	 + H_{l_3 \, l_4}
	  \, ( \dbar_{l_1 \, b } \, \dbar_{l_2 \, b } \, 
	  B^{4,4}_{l_1 \, l_2}(S\setminus \{ b \})
	   -\dbar_{l_1 \, c } \, \dbar_{l_2 \, c } \, 
	   B^{4,4}_{l_1 \, l_2}(S\setminus \{ c \}) ) \nonumber \\
& & \mbox{}	 +  \dbar_{l_1 \, b } \, \dbar_{l_2 \, b } \, 
\dbar_{l_3 \, b } \, \dbar_{l_4 \, b }
	      \, A^{4,4}_{l_1 \, l_2 \, l_3 \, l_4}(S\setminus \{ b \})
	    -\dbar_{l_1 \, c } \, \dbar_{l_2 \, c } \, \dbar_{l_3 \, c } \, 
	    \dbar_{l_4 \, c }
	      \, A^{4,4}_{l_1 \, l_2 \, l_3 \, l_4}(S\setminus \{ c \}) 
\label{eq5R5P4} 
\end{eqnarray}


\subsection{Relations between reduction coefficients}\label{proof_of_eq1)}
The following relation 
is useful to cancel $1/B$ terms in the form factors 
for 5-point integrals: 
\begin{eqnarray}
{\cal S}_{c\,l}^{-1}\,\left(b_j^{\{c\}}-b_j\right)-
b_c\,\left(\micals{c}_{j\,l}-{\cal S}_{j\,l}^{-1}\right)
&=&0
\label{d117}
\end{eqnarray}
To prove eq.~(\ref{d117}), we introduce the auxiliary relation
\begin{eqnarray}
-  ({\cal S}^{\{c\}})^{-1}_{j \, l} + {\cal S}_{j\,l}^{-1} +
 {\cal S}_{c \, l}^{-1} \, 
\caly_{j \, c}=0\;,\label{eqFR1}\\
\mbox{where }\;\;\caly_{j \, c} = \sum_{l \in S \setminus \{c\}} 
({\cal S}^{\{c\}})^{-1}_{j \, l} \; \cals_{l \, c}\nn
\end{eqnarray}
To show relation~(\ref{eqFR1}), let us assume the right-hand side is not zero, 
but some tensor $\alpha_{jlc}$\,: 
\begin{equation}
-  \micals{c}_{j \, l} + \invcals_{j \, l} + \invcals_{c \, l} \, 
\caly_{j \, c}=\alpha_{jlc}
\label{eqDEFG}
\end{equation}
Now, we compute
\begin{eqnarray}
\sum_{l \in S} \, \cals_{k \, l} \, \alpha_{jlc} & = & 
- \sum_{l \in S\setminus\{c\}} \, \cals_{k \, l} \, \micals{c}_{j \, l} + 
\delta_{k \, j} + \delta_{c \, k} \, \caly_{j \, c} \nn\\
& = & \left\{ \begin{array}{ll}
  - \delta_{j \, k} + \delta_{j \, k} = 0 & \mbox{if $k \neq c$}\nn\\
  - \caly_{j \, c} + \caly_{j \, c} = 0 & \mbox{if $k = c$}
\end{array}
\right.
\end{eqnarray}
Since $\cals$ is invertible, we must have $\alpha_{jlc} = 0$.
Summing eq.~(\ref{eqFR1}) over  $l \in S $ yields
\begin{equation}
- b^{\{c\}}_{j} + b_{j} + b_{c} \, \caly_{j \, c} = 0
\label{eqFR2}
\end{equation}
Now we multiply eq.~(\ref{eqFR1}) with $b_c$ and eq.~(\ref{eqFR2})
with ${\cal S}_{c \, l}^{-1}$ and take the difference of 
the two resulting equations to obtain eq.~(\ref{d117}).\\
Multiplying eq.~(\ref{d117}) by $\Delta_{l \,a}^{\mu}$ and 
summing over $l \in  S$ leads to
\begin{equation}
\calc^{\mu}_{c \, a} \, b^{\{c\}}_j - b_c \, \mcalc{c}^{\mu}_{j \, a} =
\calc^{\mu}_{c \, a} \, b_j - b_c \, \calc^{\mu}_{j \, a}
\label{eqONEOVERB1} 
\end{equation}
Summing eq.~(\ref{eqONEOVERB1}) over $j$ in $S \setminus \{c\}$
yields
\begin{equation}
\calc^{\mu}_{c \, a} \, B^{\{c\}} - b_c \, \mcalv{c}^{\mu}_{a} =
 \calc^{\mu}_{c \, a} \, B - b_c \, \calv^{\mu}_{a}
\label{eqONEOVERB2}  
\end{equation}
and summing eq.~(\ref{eqONEOVERB1}), 
multiplied by $\Delta^{\nu}_{j b}$, over $j$ in $S \setminus \{c\}$, 
we obtain
\begin{equation}
\calc^{\mu}_{c \, a} \, \mcalv{c}^{\nu}_{b} - \frac{1}{2} \, b_c \, 
\mcaltf{c}^{\mu \, \nu}_{a \, b}  =  \calc^{\mu}_{c \, a} \, 
\calv^{\nu}_{b} - \frac{1}{2} \, b_c \, \caltf^{\mu \, \nu}_{a \, b}\;.
\label{eqONEOVERB3}
\end{equation}

\noindent
Further, one can show by direct calculation:
\bea
\sum_{i\in S} b_i\,(\Delta_{i\,a}^2-m_i^2)&=&1+B\,m_a^2\\
\sum_{i,j\in S}\,\Delta_{i\,a}^{\mu}\,{\cal S}^{-1}_{ij}\,
\Delta_{j\,c}^2&=&\Delta_{c\,a}^{\mu}+m_c^2\,{\cal V}_{a}^{\mu}+
\sum_{j\in S} m_j^2\,\calc_{j\,a}^{\mu}
\label{A2}
\eea

\subsection{Special relations for $N= 5$}\label{proof_of_eq3}

For $N=5$, the external vectors form a basis of Minkowski space, 
such that the metric (in 4 dimensions) can be expressed by 
the tensor ${\cal H}^{\mu \nu}$, constructed from external vectors 
only: 
\bea 
{\cal H}^{\mu \nu}_{ab}&=&\sum_{i,j\in S}\,
\Delta_{i\,a}^{\mu}\,\left(\frac{b_ib_j}{B}-{\cal S}^{-1}_{ij}\right)\,
\Delta_{j\,b}^{\nu}
 =\frac{g_{[4]}^{\mu \nu}}{2}\;,\label{Hmunu}
\eea
This fact implies the relation 
\begin{equation}
\caltf^{\mu \, \nu}_{a \, b} = 2 \, \frac{\calv_a^{\mu} \, 
\calv_b^{\nu}}{B}\;\;\mbox{ for } N = 5\;\;.
\label{eqPROOF5}
\end{equation}
The proof is straightforward:

\noindent
Multiply the definition 
of ${\cal H}^{\mu \nu}_{a b}$ by $\Delta_{il}^{\mu} \, 
\Delta_{mn}^{\nu}$, we obtain
\bea
{\cal H}^{\mu \nu}_{a b} \, \Delta_{i\,l}^{\mu} \,\Delta_{m\,n}^{\nu} 
&=&\sum_{j,k \in S}  \, \left[\frac{b_j \, b_k }{B}- 
\cals^{-1}_{jk} \,\right] \, \Delta_{j\,a}\cdot \Delta_{i\,l} \; 
\Delta_{k\,b}\cdot \Delta_{m\,n}\label{eqPROOF1}
\eea
Using now 
$$\Delta_{j \, a} \cdot \Delta_{i \, c}  =  \frac{1}{2} 
\left( \, \calst_{j \, c} - \calst_{j \, i} + \calst_{a \, i} - 
\calst_{a \, c} \right)  $$
we get:
\begin{equation}
{\cal H}^{\mu \nu}_{a b} \, \Delta_{il}^{\mu} \, \Delta_{mn}^{\nu} = 
\frac{1}{2} \,\Delta_{il} \cdot \Delta_{mn}
=\frac{1}{2} \, g^{\mu \nu}_{[4]} \, \Delta_{il}^{\mu} \, 
\Delta_{mn}^{\nu}
\label{eqPROOF3pre}
\end{equation}
As the vectors $\Delta$ form a basis of Minkowski space 
for $N=5$, we conclude that
\begin{equation}
{\cal H}^{\mu \nu}_{a b} 
=\frac{1}{2} \, g^{\mu \nu}_{[4]} \;.
\label{eqPROOF3}
\end{equation}
On the other hand, using the definitions 
(see eqs.~(\ref{eqTDEFN}) and (\ref{eqVDEF}))
\bea
\calt^{\mu \, \nu}_{a \, b} & = & g^{\mu \, \nu} + 
2 \, \sum_{j,k \in S} \, 
 \invcals_{j \, k} \, \Delta^{\mu}_{k \, a} \, 
\Delta^{\nu}_{j \, b}\;\nn\\
\calv_a^{\mu} & = &\sum_{k \in S} \, b_k \, \Delta^{\mu}_{k \, a}\nn
\eea
we see that
${\cal H}^{\mu \nu}_{a b}$ is equal to 
\begin{equation}
{\cal H}^{\mu \nu}_{a b}= \frac{\calv_a^{\mu} \, \calv_b^{\nu}}{B}
-\frac{1}{2}\left(\caltf^{\mu \nu}_{a \, b}-g^{\mu \nu}_{[4]}\right)
\end{equation}
such that, together with eq.~(\ref{eqPROOF3}), we must have  
\begin{equation}
\caltf^{\mu \, \nu}_{a \, b} = 2 \, \frac{\calv_a^{\mu} \, \calv_b^{\nu}}{B}
\;\;\mbox{ for } N = 5\;.
\end{equation}

\subsection{Special relations for $N=6$}
\bea
B=\sum_{i=1}^6 b_i&=&0\label{bzero}\\
\sum_{i=1}^6 b_i\,\Delta_{i\,a}^{\mu}&=&{\cal V}_a^{\mu}=0\\
\sum_{i=1}^6 b_i\,(q_i^2-m_i^2)&=&1\label{c117}\\
\sum_{i,j=1}^6\,\Delta_{i\,a}^{\mu}\,{\cal S}^{-1}_{ij} 
\Delta_{j\,b}^{\nu}&=&
\sum_{j=1}^6\,{\cal C}_{j\,a}^{\mu}\, \Delta_{j\,b}^{\nu}=
-\frac{1}{2}\,g^{\mu\nu}_{[4]}
\eea


\section{Hexagon relations from helicity decomposition}\label{chiral_hex_rels}

In the present appendix we provide very compact expressions for the reduction 
coefficients $b_{i}$ and ${\calc}^{\mu}_{ia}$ in the specific six-point case
where the particles in the loop  as well as all legs are massless.

For massless gauge theory amplitudes usually the spinor helicity formalism
\cite{bkdgw,klesti,gaswubook,xuzhch,gunkun} is used for the treatment of vector
bosons as well as for massless fermions. In this formalism the spinor degrees
of freedom are systematically projected on helicity eigenstates. This is often
essential if one wishes to write down the final result for an amplitude in a
compact way. Now, one of the technical points which had emerged in the massless
six-point calculations of \cite{yukawa,Binoth:2002qh} is that the reduction
coefficients $b_i$ can be written very compactly in terms of 
spinor traces,  
e.g.
\bea
b_1 
&=& 
{(123456)(3456)-2s_{34}s_{45}s_{56}(6123)
\over {\rm det} \, {\cals}} \label{b1spinorrep}\\
{\rm det} \, {\cals} &=& 
4s_{12}s_{23}s_{34}s_{45}s_{56}s_{61}-(123456)^2\;.
\label{detSspinorrep}
\eea
Here and in the following $(12\cdots)$ is a shorthand notation for  
$\tr( {\psla}_{1} \, {\psla}_{2} \cdots )$. The corresponding formulae for 
$b_2,\ldots,b_6$ are obtained by cyclic permutation.

In the present appendix we first show that the helicity decomposition leads to
further simplifications in eqs.~(\ref{b1spinorrep}), (\ref{detSspinorrep}) and  
yields even more compact expressions. The two main identities responsible for 
the surprising simplifications in the Yukawa model calculation of \cite{yukawa}
become quite transparent in this approach. More surprising is perhaps that,
even for the plain six-point determinant $\det{\cal S}$, the introduction of
chirality leads to a simple formula for its square root which cannot be written
in terms of the Mandelstam variables alone. We present this further
simplification of eqs.~(\ref{b1spinorrep}), (\ref{detSspinorrep}) in
eqs.~(\ref{simpdets}), (\ref{bieps}) below. The helicity representation also yields very 
compact expressions for the reduction coefficients ${\calc}^{\mu}_{ia}$.

\vspace{0.3cm}

We use here the helicity representation only as a means to get more compact 
formulae than those provided by the general method described in the main text
for the fully massless $N=6$ case. Before we proceed let us however stress that
the helicity representation naturally leads to an alternative algorithm for the
reduction of fully massless six-point rank $r$ integrals, still into a linear
combination of five-point rank $r-1$ integrals, though decomposed on a
different tensor basis. In this appendix we only sketch how this algorithm
proceeds for the six point rank one tensor integrals. We then match the
reduction thus obtained with the one derived in section \ref{subtensor}, 
which provides us
with the above mentioned compact expressions for the reduction coefficients
${\calc}^{\mu}_{ia}$. 

\vspace{0.3cm}

Let us introduce the notation\footnote{Our spinor helicity conventions follow \cite{dixonreview}; in
particular $\tr(\gamma_5ijkl) = 4i\varepsilon(ijkl)$.}
\bea
(+ ijk\cdots  s) 
&\equiv & 
\tr
\left( 
  \half (\Eins + \gamma_5)
  {\psla}_{i} \, {\psla}_{j} \, {\psla}_{k} \cdots {\psla}_{s} 
\right)
= [ij]\langle jk \rangle [k\cdots s]\langle si\rangle
\nn\\
(- ijk\cdots  s) 
&\equiv &
\tr
\left( 
  \half (\Eins - \gamma_5)
  {\psla}_{i} \, {\psla}_{j} \, {\psla}_{k} \cdots {\psla}_{s} 
\right)
= \langle ij\rangle [jk] \langle k\cdots s\rangle [si]
\label{deftracepm}
\eea
The product of two oppositely handed traces containing the same substring of
some momenta $p_{i_1} \cdots p_{i_l}$ obeys the useful identity
\bea
&& (+i_1i_2\cdots i_lj_1j_2\cdots j_s)(-i_1i_2\cdots i_lk_1k_2\cdots k_t)
\nn\\&&
\qquad \qquad=s_{i_1i_2}s_{i_2i_3}\cdots s_{i_{l-1}i_l}
((-)^lj_1j_2\cdots j_si_1k_tk_{t-1}\cdots k_1i_l)
\label{idrlprod}
\eea
We can apply this to rewrite $\det {\cals}$ as follows:
\bea
\det  {\cals}
&=& 
4s_{12}s_{23}s_{34}s_{45}s_{56}s_{61}
- (123456)^2\nn\\
&=&
4(+123456)(-123456) - \Bigl( (+123456) + (-123456)\Bigr)^2
\nn\\
&=&
- \Bigl( (+123456) - (-123456)\Bigr)^2
=
-(\gamma_5123456)^2
\label{simpdets}
\eea
Similarly, in eq.~(\ref{b1spinorrep}) the numerator of, say, $b_5$ can be
written as
\bea
(123456)(1234)-2s_{12}s_{23}s_{34}(4561)
&=&
\Bigl((+123456)+(-123456)\Bigr)
\Bigl((+1234)+(-1234)\Bigr)
\nn\\&&
-2(+123456)(-1234)-2(-123456)(+1234)
\nn\\
&=&
\Bigl((+123456)-(-123456)\Bigr)
\Bigl((+1234)-(-1234)\Bigr)\nn\\
\label{simpnum}
\eea
Combining this result with eq.(\ref{simpdets}) one factor of 
$(+123456)-(-123456)$ cancels between numerator and denominator, and
$b_5$ becomes simply
\bea
b_5 &=&-{(+1234)-(-1234)\over  (+123456) - (-123456)}
=
-{(\gamma_51234)\over  (\gamma_5123456)}
=
-4i {\varepsilon(1234)\over  (\gamma_5123456)}
\qquad\qquad
\label{b5eps}
\eea
By cyclicity,
\bea
b_i &=& (-1)^i4i \,\,{\varepsilon\bigl((i+2)(i+3)(i+4)(i+5)\bigr)\over  
(\gamma_5123456)}
\label{bieps}
\eea
The redundancy relation $\sum_{i=1}^6 b_i =0$ (eq.~(\ref{bzero})) is 
straightforwardly checked by the chiral representation (\ref{bieps}) for the 
reduction coefficients $b_i$, as it now reads:
\bea
\varepsilon(1234)-\varepsilon(2345)+ \varepsilon(3456)-\varepsilon(4561)+
\varepsilon(5612)-\varepsilon(6123) = 0
\label{sumepszero}
 \eea
which is ensured trivially by momentum conservation\footnote{It is 
{\em not} meant here that eq.~(\ref{bzero}) is the consequence
of momentum conservation. On the contrary, it expresses the extra
constraint that the Gram determinant vanishes for 
$N=6$ \cite{Bern:1993kr,tenred1}
i.e. that any five of the six external momenta $p_{i}, \, i=1,\cdots,6$ 
are linearly dependent in a four-dimensional space-time.}.

\vspace{0.3cm}

\noindent  
Next, let us recall the two main identities used in \cite{yukawa} for 
simplifying the calculation of the six-scalar Yukawa model amplitude: 
\bea
(3456)+(123456)b_1 
&=& 
-2s_{34}s_{45}s_{56}b_4 \label{mainid1}\\
(14) +b_2(1234)+b_5(4561) &=& 0  \label{mainid2}
\eea
Eq. (\ref{mainid1}) is proved as follows:
\bea
(3456)+(123456)\,b_1 
&=& \frac{\bigl[(+3456)+(-3456)\bigr]\bigl[(+123456)-(-123456)\bigr]}{(+123456)-(-123456)}
\nn\\
&&
-\frac{\bigl[(+123456)+(-123456)\bigr]\bigl[(+3456)-(-3456)\bigr]}{(+123456)-(-123456)}
\nn\\
&=&
2\frac{(+345612)(-3456)-(-345612)(+3456)}{(+123456)-(-123456)}
\nn\\
&=&
2s_{34}s_{45}s_{56}\frac{(+1236)-(-1236)}{(+123456)-(-123456)}\nn\\
&=& -2s_{34}s_{45}s_{56}\,b_4
\label{proofmainid1}
\eea 
Here (\ref{idrlprod}) was used in the third step. The proof of (\ref{mainid2}) 
is similar.

The representation (\ref{bieps}) for $b_i$  further suggests that, for the
reduction of  six-point tensor integrals, it might be natural to expand one of
the numerator momenta in terms of vectors dual to the external momenta along
the lines of \cite{vanver}.  Let us define a dual basis 
$\lbrace v^{(i)}_1,v^{(i)}_2, v^{(i)}_3, v^{(i)}_4\rbrace$, $i=1,\ldots, 6,$ 
for each of the six choices of four consecutive momenta\footnote{Note that
contrary to \cite{vanver} we use the $p_i$ rather than the $r_i$ in building
the dual basis vectors.}:
\bea
v^{(i)}_{1\mu} &=& (-1)^i\varepsilon(\mu,i+3,i+4,i+5),\nn\\
v^{(i)}_{2\mu} &=& (-1)^i\varepsilon(i+2,\mu,i+4,i+5),\nn\\
v^{(i)}_{3\mu} &=& (-1)^i\varepsilon(i+2,i+3,\mu,i+5),\nn\\
v^{(i)}_{4\mu} &=& (-1)^i\varepsilon(i+2,i+3,i+4,\mu)\nn
\label{defvi}
\eea
For generic external momenta, each of these six bases can be used to expand the
loop momentum $k$. Namely, further defining
\bea
\varepsilon^{(i)} \equiv (-1)^i\varepsilon(i+2,i+3,i+4,i+5)
\label{defepsi}
\eea
one has
\bea
v^{(i)}_a \cdot p_k = \varepsilon^{(i)} \delta_{k,i+a+1} \qquad (k=i+2,\ldots, i+5)
\label{vidotpk}
\eea 
Therefore,
\bea
q_{a_{m}}^{\mu} &=& {1\over\varepsilon^{(i)}}\sum_{a=1}^4 v^{(i)\mu}_a\, 
p_{i+a+1}\cdot q_{a_{m}}
\qquad (i=1,\ldots,6)
\label{kexpand}
\eea
The only caveat here is that we have used a projection of the momentum
$q^{\mu}_{a_m}$ in the loop on four-dimensional space, while the loop integration
requires dimensional continuation. However, for rank one integrals this is
safe, since at one loop the index $\mu$ will be contracted with some external
momentum or polarisation only. We will comment briefly on the case of higher
ranks below.

\vspace{0.3cm}
 
Using this equation together with eq.~(\ref{c117}) and rewriting
\bea
p_{i+a+1}\cdot q_{a_m} 
=
{1\over 2} 
\Bigl[
 q_{i+a+1}^2 - q_{i+a}^2 - {\cals}_{i+a+1,a_{m}} + {\cals}_{i+a,a_{m}} 
\Bigr]
\label{rewritepk}
\eea
we obtain
\bea
q_{a_{m}}^{\mu} &=& \sum_{i=1}^6 b_iq_i^2 \,  
{1\over\varepsilon^{(i)}}\sum_{a=1}^4 v^{(i)\mu}_a\, p_{i+a+1}\cdot q_{a_{m}}
\nn\\
&=&
{2i\over (\gamma_5 123456)}
\sum_{i=1}^6
q_i^2
\sum_{a=1}^4
 \Bigl[
  q_{i+a+1}^2 - q_{i+a}^2 -{\cals}_{i+a+1,a_{m}} + {\cals}_{i+a,a_{m}}
 \Bigr] \, 
v^{(i)\mu}_a
\label{qexpfinredund}
\eea
Note that the $\varepsilon^{(i)}$ from the numerator of $b_i$ in eq.
(\ref{bieps}) has cancelled against the same $\varepsilon^{(i)}$ from the
denominator of eq.(\ref{kexpand}); this is the main point of the algorithm.

The $v_a^{(i)}$ are not independent since
\bea
v^{(i)}_1 = v^{(i+1)}_4
\label{vvtriv}
\eea
and, by momentum conservation, also
\bea
v^{(i)}_1-v^{(i)}_2 &=& -(v^{(i+2)}_3-v^{(i+2)}_4) \label{vv1}\\
v^{(i)}_2-v^{(i)}_3 &=& -(v^{(i+3)}_2-v^{(i+3)}_3) \label{vv2}
\eea
Remarkably, the three identities (\ref{vvtriv}),(\ref{vv1}),(\ref{vv2}) 
together just imply that the 48 products of $q_i^2$'s in (\ref{qexpfinredund})
cancel out in pairs. This leaves us with
\bea
q_{a_m}^{\mu} &=& {2i\over (\gamma_5 123456)}
\sum_{i=1}^6
q_i^2
\sum_{a=1}^4
 \Bigl({\cals}_{i+a,a_m}-{\cals}_{i+a+1,a_m}\Bigr)\,
v^{(i)\mu}_a
\label{qexpfin}
\eea
Using the decomposition (\ref{qexpfin}) in the rank one massless six-point 
integral $I^{n,\,\mu}_6$ we obtain the following reduction to five-point rank zero 
integrals:
\bea
I^{n,\,\mu}_6(a_m;S) &=&
{2i\over (\gamma_5 123456)} 
\sum_{i=1}^6
\sum_{a=1}^4
\Bigl( 
 {\cals}_{i+a,a_{m}} - {\cals}_{i+a+1,a_{m}}
\Bigr) \, 
v^{(i)\mu}_a \, I_5^n(S \setminus \{i\})
\label{I6mufin}
\eea
We can write this result more compactly by
introducing the convention that $v^{(i)\mu}_5=v^{(i)\mu}_6 = 0$, and
defining
\bea
d_a^{(i)\mu} &\equiv& v_{a-i-1}^{(i)\mu}-v_{a-i}^{(i)\mu} 
\qquad (i,a = 1,\ldots,6)
\label{defd}
\eea
In terms of the vectors $d_a^{(i)\mu}$, eqs.~(\ref{vvtriv}),(\ref{vv1}),(\ref{vv2}) can be neatly combined to
\bea
d^{(i)}_j &=& - \,d^{(j)}_i  
\qquad (i,j = 1,\ldots,6)
\label{reld}
\eea
We also note that
\bea
\sum_{a=1}^6d^{(i)\mu}_a &=& 0 \qquad (i=1,\ldots,6)
\label{sumd}\\
\sum_{j=1}^6r_j^{\mu}d^{(i)\nu}_j
&=&\varepsilon^{(i)}\,
g^{\mu\nu}_{[4]}
\qquad (i = 1,\ldots,6)
\label{sumrd}
\eea
Now, further defining
\bea
{\cal D}_{i a_{m}}^{\mu} 
&\equiv & 
\sum_{a=1}^6 \, {\cals}_{a\, a_{m}} \, d^{(i)\mu}_a 
\qquad (i,a_m = 1,\ldots,6)
\label{defDgen}
\eea
one has
\bea
I_6^{n,\,\mu} (a_m;S) 
&=&
-{2i \over (\gamma_5 123456)} \sum_{i=1}^6
{\cal D}^{\mu}_{i a_{m}} \, I_5^n(S \setminus \{i\})
\label{I6mufindm}
\eea
Comparing eq.~(\ref{I6mufindm}) with eq.~(\ref{eqNEWIDEF2}) of section
\ref{subtensor} for rank one
-- in which the first term on the r.h.s. vanishes as there is no higher
dimensional integral for $N=6$ -- we identify
\begin{equation}\label{identif}
{\cal C}^{\mu}_{ia} = {2i \over (\gamma_5 123456)} \, {\cal D}^{\mu}_{ia}
\end{equation}

\vspace{0.3cm}


Let us add one comment on the application of this algorithm in the case of
arbitrary rank. Applying the decomposition (\ref{qexpfin}) to the six-point 
integral of arbitrary rank we obtain
\bea
I_6^{n,\,\mu_1, \ldots, \mu_r}(a_1, \ldots, a_r;S) &=&
-{2i\over (\gamma_5 123456)}
\sum_{i=1}^6
{\cal D}^{\mu_1}_{i a_{1}} \, 
I_5^{n,\,\mu_2, \cdots, \mu_r}(a_2, \ldots ,a_r; S \setminus \{i\})
\label{I6higher}
\eea
In the case of arbitrary rank the implied projection of the loop momentum $q^{\mu}_{a_m}$ on 
four-dimensional space requires a more careful consideration. This trick could fail if a contraction of two indices with a metric tensor $g_{\mn}$ occurs, as can be the case at ranks $r \geq 2$.
However, if $g_{\mn}$ is the full $n$-dimensional metric, then the integral
breaks down to a lower rank integral anyway, while if $g_{\mn}$ is the 
($n-4$)-dimensional metric then $g_{[n-4]} = g_{[n]} - g_{[4]}$ can be used before
applying (\ref{I6higher}).

In any case the index ``$\mu$'' gets absorbed into the vectors $d^{(i)\mu}_a$,
and ultimately will be contracted either (i) with an external momentum 
$p_{l\mu}$, (ii) with an $\varepsilon$ - tensor, or (iii) with a polarisation 
vector $\varepsilon_{l\mu}^{\pm}(p_k,q)$. Let us consider these cases in turn.\\
(i) Contraction with external momenta produces $\varepsilon$-tensors. These can, using
momentum conservation, be reduced to the six  $\varepsilon^{(i)}$'s.\\
(ii) Here we encounter the contraction of two 
$\varepsilon$-tensors involving arbitrary momenta,
\bea
\varepsilon({\mu,a,b,c}) \, \varepsilon({\mu,i,j,k})
&=& 
{\rm det} 
\left(
\matrix{p_a\cdot p_i & p_a\cdot p_j & p_a\cdot p_k \cr
        p_b\cdot p_i & p_b\cdot p_j & p_b\cdot p_k \cr
        p_c\cdot p_i & p_c\cdot p_j & p_c\cdot p_k}
\right)
\nonumber \\
& = &
{1\over 8} \Bigl[(abckji)-(abcijk)\Bigr]
\label{defgabcijk}
\eea
(iii) As usual we take the reference momentum for $\varepsilon_l$
to be some other external momentum $p_s$, so that
\bea
\varepsilon_{l\mu}^{\pm}(p_s) &=&
\pm {\langle s^{\mp}\vert \gamma_{\mu} \vert l^{\mp}\rangle\over
\sqrt{2}\langle s^{\mp}\vert l^{\pm}\rangle }
\label{defpol}
\eea
Then it is easily shown by using the Fierz identity that
\bea
\varepsilon_{l\mu}^+(p_s) \, \varepsilon(\mu,i,j,k) 
&=& 
{i\over 2\sqrt{2}\langle ls\rangle}
\biggl[
 {\langle ks\rangle\over\langle kl\rangle}(+lijk) -
 {\langle is\rangle\over\langle il\rangle}(-lijk)
\biggr] 
\nn\\
\varepsilon_{l\mu}^-(p_s) \, \varepsilon(\mu,i,j,k) 
&=& 
{i\over 2\sqrt{2}[ls]}
\biggl[
 {[ks]\over[kl]}(-lijk) - {[is]\over[il]}(+lijk)
\biggr] 
\label{epscontract1}
\eea
In writing eqs.\,(\ref{epscontract1}) we have assumed $i,k\ne l$, which is not 
a restriction.

\end{appendix}


\begin{thebibliography}{999}

\bibitem{lhcrep}
{\it Proceedings of the Workshop on Standard Model Physics 
(and more) at the LHC}, 
eds. G. Altarelli \& M.L. Mangano, CERN 2000-004,
  [hep-ph/0005025];\\
{\it Proceedings of the 3rd Les Houches Workshop: 
Physics at TeV Colliders}, eds. 
G.~B\'elanger, F.~Boudjema, J.~P.~Guillet and E.~Pilon, 
Les Houches, France, May 26-June 3, 2003.
%
\bibitem{Draggiotis:2002hm}
P.~D.~Draggiotis, R.~H.~P.~Kleiss and C.~G.~Papadopoulos,
Eur.\ Phys.\ J.\ C {\bf 24} (2002) 447
[hep-ph/0202201].

\bibitem{Giele:1991vf}
W.~T.~Giele and E.~W.~N.~Glover,
Phys.\ Rev.\ D {\bf 46} (1992) 1980;\\
W.~T.~Giele, E.~W.~N.~Glover and D.~A.~Kosower,
Nucl.\ Phys.\ B {\bf 403} (1993) 633
[hep-ph/9302225].

\bibitem{Frixione:1995ms}
S.~Frixione, Z.~Kunszt and A.~Signer,
Nucl.\ Phys.\ B {\bf 467} (1996) 399
[hep-ph/9512328].

\bibitem{Nagy:1996bz}
Z.~Nagy and Z.~Tr\'ocs\'anyi,
Nucl.\ Phys.\ B {\bf 486} (1997) 189
[hep-ph/9610498].

\bibitem{Catani:1996vz}
S.~Catani and M.~H.~Seymour,
Nucl.\ Phys.\ B {\bf 485} (1997) 291
[Erratum-ibid.\ B {\bf 510} (1997) 291]
[hep-ph/9605323].

\bibitem{Hahn:2000kx}
  T.~Hahn,
  Comput.\ Phys.\ Commun.\  {\bf 140} (2001) 418
  [hep-ph/0012260].


\bibitem{Belanger:2003sd}
  G.~Belanger, F.~Boudjema, J.~Fujimoto, T.~Ishikawa, T.~Kaneko, K.~Kato and Y.~Shimizu,
  Phys.\ Rept.\  {\bf 430} (2006) 117
  [hep-ph/0308080].

\bibitem{Kaneko:1994fd}
  T.~Kaneko,
  Comput.\ Phys.\ Commun.\  {\bf 92} (1995) 127
  [hep-th/9408107].

\bibitem{Nogueira:1991ex}
  P.~Nogueira,
  J.\ Comput.\ Phys.\  {\bf 105} (1993) 279.

\bibitem{Bern:1993mq}
Z.~Bern, L.~J.~Dixon and D.~A.~Kosower,
Phys.\ Rev.\ Lett.\  {\bf 70} (1993) 2677
[hep-ph/9302280].

\bibitem{Kunszt:1994tq}
Z.~Kunszt, A.~Signer and Z.~Trocsanyi,
Phys.\ Lett.\ B {\bf 336} (1994) 529
[hep-ph/9405386].

\bibitem{Kilgore:1996sq}
  W.~B.~Kilgore and W.~T.~Giele,
  Phys.\ Rev.\ D {\bf 55} (1997) 7183
  [hep-ph/9610433].
\bibitem{Campbell:1997tv}
J.~M.~Campbell, E.~W.~N.~Glover and D.~J.~Miller,
Phys.\ Lett.\ B {\bf 409} (1997) 503
[hep-ph/9706297].

\bibitem{Bern:1997sc}
Z.~Bern, L.~J.~Dixon and D.~A.~Kosower,
Nucl.\ Phys.\ B {\bf 513} (1998) 3
[hep-ph/9708239].

\bibitem{Reina:2001sf}
L.~Reina and S.~Dawson,
Phys.\ Rev.\ Lett.\  {\bf 87} (2001) 201804
[hep-ph/0107101].

\bibitem{Beenakker:2002nc}
W.~Beenakker, S.~Dittmaier, M.~Kr\"amer, B.~Pl\"umper, M.~Spira and P.~M.~Zerwas,
Nucl.\ Phys.\ B {\bf 653} (2003) 151
[hep-ph/0211352].

\bibitem{Belanger:2002ik}
G.~B\'elanger, F.~Boudjema, J.~Fujimoto, T.~Ishikawa, T.~Kaneko, K.~Kato and Y.~Shimizu,
Phys.\ Lett.\ B {\bf 559} (2003) 252
[hep-ph/0212261].

\bibitem{Denner:2003iy}
A.~Denner, S.~Dittmaier, M.~Roth and M.~M.~Weber,
Nucl.\ Phys.\ B {\bf 660} (2003) 289
[hep-ph/0302198].

\bibitem{Belanger:2003ya}
G.~B\'elanger {\it et al.},
Phys.\ Lett.\ B {\bf 576} (2003) 152
[hep-ph/0309010].

\bibitem{Belanger:2003nm}
G.~B\'elanger {\it et al.},
Phys.\ Lett.\ B {\bf 571} (2003) 163
[hep-ph/0307029].

\bibitem{Denner:2003ri}
A.~Denner, S.~Dittmaier, M.~Roth and M.~M.~Weber,
Phys.\ Lett.\ B {\bf 575} (2003) 290
[hep-ph/0307193].

\bibitem{Denner:2003yg}
A.~Denner, S.~Dittmaier, M.~Roth and M.~M.~Weber,
Phys.\ Lett.\ B {\bf 560} (2003) 196
[hep-ph/0301189].

\bibitem{Chen:2003yd}
  H.~Chen, W.~G.~Ma, R.~Y.~Zhang, P.~J.~Zhou, H.~S.~Hou and Y.~B.~Sun,
  Nucl.\ Phys.\ B {\bf 683} (2004) 196
  [hep-ph/0309106].

\bibitem{Denner:2003zp}
A.~Denner, S.~Dittmaier, M.~Roth and M.~M.~Weber,
Nucl.\ Phys.\ B {\bf 680} (2004) 85
[hep-ph/0309274].


\bibitem{Nagy:2001fj}
  Z.~Nagy,
  Phys.\ Rev.\ Lett.\  {\bf 88} (2002) 122003
  [hep-ph/0110315].

\bibitem{Nagy:2003tz}
  Z.~Nagy,
  Phys.\ Rev.\ D {\bf 68} (2003) 094002
  [hep-ph/0307268].

\bibitem{Campbell:2002tg}
  J.~Campbell and R.~K.~Ellis,
  Phys.\ Rev.\ D {\bf 65} (2002) 113007
  [hep-ph/0202176].
\bibitem{Yasui:2002bn}
Y.~Yasui,
Phys.\ Rev.\ D {\bf 66} (2002) 094012
[hep-ph/0203163].

\bibitem{DelDuca:2001fn}
  V.~Del Duca, W.~Kilgore, C.~Oleari, C.~Schmidt and D.~Zeppenfeld,
  Nucl.\ Phys.\ B {\bf 616} (2001) 367
  [hep-ph/0108030].

\bibitem{Binoth:2003xk}
T.~Binoth, J.~P.~Guillet and F.~Mahmoudi,
JHEP {\bf 0402} (2004) 057
[hep-ph/0312334].


\bibitem{Boudjema:2004id}
F.~Boudjema, J.~Fujimoto, T.~Ishikawa, T.~Kaneko, K.~Kato, Y.~Kurihara and Y.~Shimizu,
Nucl.\ Phys.\ Proc.\ Suppl.\  {\bf 135} (2004) 323
[hep-ph/0407079].

\bibitem{Denner:2005es}
  A.~Denner, S.~Dittmaier, M.~Roth and L.~H.~Wieders,
  Phys.\ Lett.\ B {\bf 612} (2005) 223
  [hep-ph/0502063];\,
  Nucl.\ Phys.\ B {\bf 724} (2005) 247
  [hep-ph/0505042].
  

\bibitem{Frixione:2005gz}
  S.~Frixione and B.~R.~Webber,
  hep-ph/0506182;\\
  S.~Frixione, P.~Nason and B.~R.~Webber,
  JHEP {\bf 0308} (2003) 007
  [hep-ph/0305252];\\
  S.~Frixione and B.~R.~Webber,
  JHEP {\bf 0206} (2002) 029
  [hep-ph/0204244].

\bibitem{Nagy:2005aa}
  Z.~Nagy and D.~E.~Soper,
  JHEP {\bf 0510} (2005) 024
  [hep-ph/0503053].

  

\bibitem{Soper:1999xk}
D.~E.~Soper,
Phys.\ Rev.\ D {\bf 62} (2000) 014009
[hep-ph/9910292].

\bibitem{Kramer:2002cd}
M.~Kr\"amer and D.~E.~Soper,
Phys.\ Rev.\ D {\bf 66} (2002) 054017
[hep-ph/0204113].

  
\bibitem{melrose}
D.~B.~Melrose,
Nuovo Cim.\  {\bf 40}, (1965) 181.

\bibitem{passarinoveltman}
G.~Passarino and M.~J.~G.~Veltman,
Nucl.\ Phys.\ B {\bf 160} (1979) 151.
\bibitem{vanver}
W.~L.~van Neerven and J.~A.~M.~Vermaseren,
Phys.\ Lett.\ B {\bf 137}, 241 (1984).
\bibitem{Kotikov:1991pm}
  A.~V.~Kotikov,
  Phys.\ Lett.\ B {\bf 267}, (1991) 123 \\
  A.~V.~Kotikov,
  Mod.\ Phys.\ Lett.\ A {\bf 6}, (1991) 3133.
\bibitem{tarasov}
O.~V.~Tarasov,
Phys.\ Rev.\ D {\bf 54}, 6479 (1996)
[hep-th/9606018];\\
J.~Fleischer, F.~Jegerlehner and O.V.~Tarasov,
Nucl.\ Phys.\ B {\bf 566}, 423 (2000)
[hep-ph/9907327].
\bibitem{oldenb}
G.J. van Oldenborgh and J.A.M. Vermaseren,
Z.\ Phys.\ C {\bf 46} 425 (1990).
\bibitem{davydychev}
A.~I.~Davydychev,
Phys.\ Lett.\ B {\bf 263}, 107 (1991).
  
\bibitem{Bern:1992em}
Z.~Bern, L.~J.~Dixon and D.~A.~Kosower,
Phys.\ Lett.\ B {\bf 302} (1993) 299
[Erratum-ibid.\ B {\bf 318} (1993) 649]
[hep-ph/9212308].
\bibitem{Bern:1993kr}
Z.~Bern, L.~J.~Dixon and D.~A.~Kosower,
Nucl.\ Phys.\ B {\bf 412} (1994) 751
[hep-ph/9306240].

\bibitem{Campbell:1996zw}
J.~M.~Campbell, E.~W.~N.~Glover and D.~J.~Miller,
Nucl.\ Phys.\ B {\bf 498} (1997) 397
[hep-ph/9612413].
\bibitem{Pittau:1997mv}
R.~Pittau,
Comput.\ Phys.\ Commun.\  {\bf 111} (1998) 48
[hep-ph/9712418];\\
R.~Pittau,
Comput.\ Phys.\ Commun.\  {\bf 104} (1997) 23
[hep-ph/9607309].
\bibitem{Weinzierl:1998we}
  S.~Weinzierl,
  Phys.\ Lett.\ B {\bf 450} (1999) 234
  [hep-ph/9811365].
  
  
\bibitem{tenred1}
 T.~Binoth, J.Ph.~Guillet, G.~Heinrich,
 Nucl. Phys. {\bf B572} (2000) 361
 [hep-ph/9911342].

\bibitem{Denner:2002ii}
A.~Denner and S.~Dittmaier,
Nucl.\ Phys.\ B {\bf 658} (2003) 175
[hep-ph/0212259].
\bibitem{Dittmaier:2003bc}
S.~Dittmaier,
Nucl.\ Phys.\ B {\bf 675} (2003) 447
[hep-ph/0308246].

\bibitem{Duplancic:2003tv}
  G.~Duplancic and B.~Nizic,
  Eur.\ Phys.\ J.\ C {\bf 35} (2004) 105
  [hep-ph/0303184].
  
\bibitem{Giele:2004iy}
W.~T.~Giele and E.~W.~N.~Glover,
JHEP {\bf 0404} (2004) 029
[hep-ph/0402152].

\bibitem{Giele:2004ub}
W.~Giele, E.~W.~N.~Glover and G.~Zanderighi,
Nucl.\ Phys.\ Proc.\ Suppl.\  {\bf 135} (2004) 275
[hep-ph/0407016].

\bibitem{delAguila:2004nf}
F.~del Aguila and R.~Pittau,
JHEP {\bf 0407} (2004) 017
[hep-ph/0404120].

\bibitem{vanHameren:2005ed}
  A.~van Hameren, J.~Vollinga and S.~Weinzierl,
  Eur.\ Phys.\ J.\ C {\bf 41} (2005) 361
  [arXiv:hep-ph/0502165].



\bibitem{passarino} 
A. Ferroglia, M. Passera, G. Passarino and S. Uccirati,
Nucl.\ Phys.\ B {\bf 650}, 162 (2003)
[hep-ph/0209219];\\
G. Passarino and S. Uccirati,
Nucl.\ Phys.\ B {\bf 629}, 97 (2002)
[hep-ph/0112004];\\
G. Passarino,
Nucl.\ Phys.\ B {\bf 619}, 257 (2001)
[hep-ph/0108252].


\bibitem{Kurihara:2005ja}
  Y.~Kurihara and T.~Kaneko,
  Comput.\ Phys.\ Commun.\  {\bf 174} (2006) 530
  [hep-ph/0503003].

  
\bibitem{Nagy:2003qn}
Z.~Nagy and D.~E.~Soper,
JHEP {\bf 0309} (2003) 055
[hep-ph/0308127].

\bibitem{nikolas} 
T.~Binoth, G.~Heinrich and N.~Kauer,
Nucl.\ Phys.\ B {\bf 654}, 277 (2003)
[hep-ph/0210023].


\bibitem{Neq4Oneloop}
Z.~Bern, L.~J.~Dixon, D.~C.~Dunbar and D.~A.~Kosower,
Nucl.\ Phys.\ B {\bf 425} (1994) 217 
[hep-ph/9403226].

\bibitem{Neq1Oneloop}
Z.\ Bern, L.\ J.\ Dixon, D.\ C.\ Dunbar and D.\ A.\ Kosower,
Nucl.\ Phys.\ B {\bf 435} (1995) 59 [hep-ph/9409265].
%
\bibitem{UnitarityMachinery}
Z.\ Bern and A.\ G.\ Morgan,
Nucl.\ Phys.\ B {\bf 467} (1996) 479 [hep-ph/9511336];\\
%
Z.\ Bern, L.\ J.\ Dixon and D.\ A.\ Kosower,
Ann.\ Rev.\ Nucl.\ Part.\ Sci.\  {\bf 46} (1996) 109 
[hep-ph/9602280];\\
%
Z.\ Bern, L.\ J.\ Dixon and D.\ A.\ Kosower,
Nucl.\ Phys.\ Proc.\ Suppl.\  {\bf 51C} (1996) 243 
[hep-ph/9606378].

\bibitem{Cachazo:2004zb}
F.~Cachazo, P.~Svrcek and E.~Witten,
JHEP {\bf 0410} (2004) 074
[hep-th/0406177];
A.~Brandhuber, B.~Spence and G.~Travaglini,
Nucl.\ Phys.\ B {\bf 706} (2005) 150 
[hep-th/0407214];
J.~Bedford, A.~Brandhuber, B.~Spence and G.~Travaglini,
Nucl.\ Phys.\ B {\bf 706} (2005) 100
[hep-th/0410280];
R.~Britto, F.~Cachazo and B.~Feng,
Phys.\ Lett.\ B {\bf 611} (2005) 167
[hep-th/0411107];
Phys.\ Rev.\ D {\bf 71} (2005) 025012 [hep-th/0410179];
Nucl.\ Phys.\ B {\bf 725} (2005) 275 [hep-th/0412103];
R.~Britto, E.~Buchbinder, F.~Cachazo and B.~Feng,
Phys.\ Rev.\ D {\bf 72} (2005) 065012 [hep-ph/0503132].

\bibitem{BBKR}
I.~Bena, Z.~Bern, D.~A.~Kosower and R.~Roiban,
Phys.\ Rev.\ D {\bf 71} (2005) 106010
[hep-th/0410054].

\bibitem{Bern:2004ky}
Z.~Bern, V.~Del Duca, L.~J.~Dixon and D.~A.~Kosower,
Phys.\ Rev.\ D {\bf 71} (2005) 045006
[hep-th/0410224].

\bibitem{Bern:2004bt}
Z.~Bern, L.~J.~Dixon and D.~A.~Kosower,
Phys.\ Rev.\ D {\bf 72} (2005) 045014 [hep-th/0412210].

\bibitem{Bern:2005hs}
Z.~Bern, L.~J.~Dixon and D.~A.~Kosower,
Phys.\ Rev.\ D {\bf 71} (2005) 105013 [hep-th/0501240].

\bibitem{Badger:2005zh}
  S.~D.~Badger, E.~W.~N.~Glover, V.~V.~Khoze and P.~Svrcek,
 JHEP {\bf 0507} (2005) 025 [hep-th/0504159].

\bibitem{yukawa}
 T.~Binoth, J.~P.~Guillet, G.~Heinrich and C.~Schubert,
Nucl.\ Phys.\ B {\bf 615} (2001) 385
[hep-ph/0106243].

\bibitem{Binoth:2002qh}
  T.~Binoth,
  Nucl.\ Phys.\ Proc.\ Suppl.\  {\bf 116} (2003) 387
  [hep-ph/0211125].

\bibitem{Binoth:2005ua}
T.~Binoth, M.~Ciccolini, N.~Kauer and M.~Kr\"amer,
JHEP {\bf 0503} (2005) 065
[hep-ph/0503094].

\bibitem{Davydychev:1999mq}
  A.~I.~Davydychev,
  Phys.\ Rev.\ D {\bf 61} (2000) 087701
  [hep-ph/9910224].
\bibitem{Davydychev:2000na}
  A.~I.~Davydychev and M.~Y.~Kalmykov,
  Nucl.\ Phys.\ B {\bf 605} (2001) 266
  [hep-th/0012189].

  
 \bibitem{watson}
  J.~Papavassiliou, E.~de Rafael and N.~J.~Watson,
  Nucl.\ Phys.\ B {\bf 503} (1997) 79
  [hep-ph/9612237].
 
\bibitem{barnett}
S. Barnett, {\em Matrices: Methods and Applications}, Oxford Applied
Mathematics and Computing Sciences Series, Clarendon Press, Oxford 1990,
chapter 10, p.~250.
 
\bibitem{swedesbook}
G.~Dahlquist \& \AA.~Bj\"ork, {\em Numerical Methods} (trad. N.~Anderson),
Prentice-Hall series in automatic computations (1974), section 5.2.5,
p.~143-144.

\bibitem{Kurihara:2005at}
  Y.~Kurihara,
   ``Dimensionally regularized one-loop tensor-integrals with massless  internal
  Eur.\ Phys.\ J.\ C {\bf 45} (2006) 427
  [hep-ph/0504251].

\bibitem{bjorkendrell}
J.D. Bjorken, S.D. Drell, Relativistic Quantum Fields, McGraw-Hill (1965).
\bibitem{edenlandshoff} R. J. Eden, P. V. Landshoff, D.I. Olive, J.C. Polkinghorne,
The analytic S-matrix, Cambridge University Press (1966).

\bibitem{secdec_papers}
  T.~Binoth and G.~Heinrich,
  Nucl.\ Phys.\ B {\bf 585} (2000) 741
  [hep-ph/0004013];\\
  T.~Binoth and G.~Heinrich,
  Nucl.\ Phys.\ B {\bf 680} (2004) 375
  [hep-ph/0305234].


\bibitem{bkdgw}
F.A. Berends, R. Kleiss, P. De Causmaecker, R. Gastmans, and T.T. Wu,
Phys. Lett. {\bf B 103} (1981) 124.

\bibitem{klesti}
R. Kleiss and W.J. Stirling, Nucl. Phys, {\bf B 262} (1985) 235.

\bibitem{gaswubook}
R. Gastmans and T.T. Wu, {\it The Ubiquitous Photon: Helicity Method for QED and
QCD}, Clarendon Press 1990.

\bibitem{xuzhch}
Z. Xu, D.-H. Zhang, and L. Chang, Nucl. Phys. {\bf B 291} (1987) 392.

\bibitem{gunkun}
J.F. Gunion and Z. Kunszt, Phys. Lett. {\bf B 161} (1985) 333.

\bibitem{dixonreview}
L. Dixon, {\it Calculating scattering amplitudes efficiently}, TASI Lectures, Boulder TASI 95,
539 [hep-ph/9601359].



\end{thebibliography}
\end{document}